\documentclass[a4paper,11pt]{article}
\usepackage{graphicx,amssymb,amstext,amsmath}
\usepackage{color}

\definecolor{cbl}{rgb}{0,0,1}                

\newcommand{\bc}{\begin{center}}
\newcommand{\ec}{\end{center}}
\def\ba#1{\begin{array}{#1}\displaystyle}
\newcommand{\ea}{\end{array}}

\newcommand{\beq}{\begin{equation}}
\newcommand{\eeq}{\end{equation}}
\newcommand{\beqa}{\begin{eqnarray}}
\newcommand{\eeqa}{\end{eqnarray}}

\newcommand{\n}{\nonumber\\}
\newcommand{\bi}{\begin{itemize}}
\newcommand{\ei}{\end{itemize}}

\def\lt#1{\left#1}
\def\rt#1{\right#1}
\def\t#1{\tilde{#1}}

\def\b#1{\bar{#1}}
\def\frc#1#2{\frac{#1}{#2}}

\newcommand{\p}{\partial}

\newcommand{\bra}{\langle}
\newcommand{\ket}{\rangle}
\newcommand{\Z}{{\mathbb{Z}}}

\newcommand{\R}{{\mathbb{R}}}

\newcommand{\hC}{{\hat{\mathbb{C}}}}

\newcommand{\Or}{{\cal O}}

\newcommand{\ep}{\epsilon}
\newcommand{\varep}{\varepsilon}

\newcommand{\Tr}{{\rm Tr}}

\newcommand{\TT}{{\cal T}}
\newcommand{\Tp}{:{\cal T}\phi:}
\newcommand{\bTp}{:\tilde{\cal T}\phi:}

\begin{document}

\begin{titlepage}
\vspace{0.2cm}
\begin{center}

{\large{\bf{Entanglement Entropy of Non-Unitary Integrable Quantum Field Theory}}}

\vspace{0.8cm} {\large \text{Davide Bianchini${}^{\bullet}$, Olalla A. Castro-Alvaredo${}^{\bullet}$ and Benjamin Doyon${}^{\circ}$}}

\vspace{0.2cm}
{{\small ${}^{\bullet}$} Department of Mathematics, City University London, Northampton Square EC1V 0HB, UK }\\

{{\small ${}^{\circ}$} Department of Mathematics, King's College London, Strand WC2R 2LS, UK}
\end{center}

\vspace{1cm}
In this paper we study the simplest massive 1+1 dimensional integrable quantum field theory which can be described as a perturbation of a non-unitary minimal conformal field theory: the Lee-Yang model. We are particularly interested in the features of the bi-partite entanglement entropy for this model and on building blocks thereof, namely twist field form factors. Non-unitarity selects out a new type of twist field as the operator whose two-point function (appropriately normalized) yields the entanglement entropy. We compute this two-point function both from a form factor expansion and by means of perturbed conformal field theory. We find good agreement with CFT predictions put forward in a recent work involving the present authors. In particular, our results are consistent with a scaling of the entanglement entropy given by $\frac{c_{\text{eff}}}{3}\log \ell$ where $c_{\text{eff}}$ is the effective central charge of the theory (a positive number related to the central charge) and $\ell$ is the size of the region.
Furthermore the form factor expansion of twist fields allows us to explore the large region limit of the entanglement entropy and find the next-to-leading order correction to saturation. We find that this correction is very different from its counterpart in unitary models. Whereas in the latter case, it had a form depending only on few parameters of the model (the particle spectrum), it appears to be much more model-dependent for non-unitary models.

\medskip
\medskip
\medskip
\medskip
\medskip
\medskip
\medskip
\medskip
\medskip
\medskip
\medskip
\medskip
\medskip
\medskip
\medskip
\medskip

\hfill \today

\end{titlepage}

\section{Introduction}

Entanglement is a fundamental property of quantum systems which relates to the outcomes of local measurements: performing a local measurement may affect
the outcome of local measurements far away. This property represents the 
single main difference between quantum and classical systems.
Technological advances have taken entanglement from a strange quantum phenomenon to a valuable resource at the heart of various fields of research such as quantum computation and quantum cryptography.
There has also been great interest in developing efficient
(theoretical) measures of entanglement, not just in view of the applications above but also as a means to extract valuable information about emergent properties of quantum states of extended systems. One such measure for many-body quantum systems is the bi-partite entanglement entropy (EE) \cite{bennet},
which we will consider here. Other
measures of entanglement exist, see e.g.
\cite{bennet,Osterloh,Osborne,Barnum,Verstraete}, which occur in
the context of quantum computing, for instance. In its most general
understanding, the EE is a measure of the amount of quantum
entanglement, in a pure quantum state, between the degrees of
freedom associated to two sets of independent observables whose
union is complete on the Hilbert space.  In the present paper, the two sets of observables
correspond to the local observables in two complementary connected regions, $A$ and $\bar{A}$, of a 1+1-dimensional (1 space + 1 time
dimension) extended quantum model, and we will consider cases where the quantum state is the ground state
of a {\it non-unitary}, near-critical model. 

Prominent examples of extended one-dimensional quantum systems are quantum spin chains. Their entanglement has been
extensively studied in the literature
\cite{Eisert,Latorre1,Latorre2,Latorre3,Jin,Lambert,Casini,KeatingM05,Weston}. These examples however all refer to unitary quantum spin chains. Interesting examples of non-unitary spin chain systems exist, for instance the famous quantum group invariant integrable XXZ spin chain, with generically non-Hermitian boundary terms; in the thermodynamic limit it has critical points associated with the minimal models of conformal field theory (CFT), including the non-unitary series \cite{PS1,PS2,PS3,PS4}. Another example is provided by the Hamiltonian studied by von Gehlen in \cite{gehlen1,gehlen2}: the Ising model in the presence of a longitudinal imaginary magnetic field. This Hamiltonian has a critical line (in the phase space of its two couplings) which has been identified with the Lee-Yang non-unitary minimal model of CFT, with central charge $c=-22/5$ \cite{Fisher, LYCardy}. In all these examples, the local, extended Hamiltonians are non-Hermitian, yet have {\em real and bounded energy spectra}. Their critical points are described by CFT models containing non-unitary representations of the Virasoro algebra with real weights, and whose ground states are not the conformal vacua, but negative-weight modules.

Non-Hermitian Hamiltonians with real spectra are the subject of much current research especially in connection with {\tt PT}-symmetry or pseudo/quasi Hermiticity  \cite{sgh,pt} (see \cite{Bender1, Bender2, Bender3} for reviews and \cite{revpt} for the interplay with integrability). For instance the critical line of von Gehlen's system \cite{gehlen1,gehlen2} described above, can be related to ${\tt PT}$--symmetry breaking in that it separates the phase space into two regions, one where only real eigenvalues occur, and another where pairs of complex conjugated eigenvalues arise \cite{meandreas}. Experimental studies and theoretical descriptions of new physical phenomena connected to non-Hermitian Hamiltonians have recently emerged, including optical effects \cite{wow1,wow2,wow3}, transitions from ballistic to diffusive transport \cite{exper}, and dynamical phase transitions \cite{nott,timeint}. Non-Hermitian quantum mechanics is also used in the description of non-equilibrium systems \cite{bw}, quantum Hall transitions \cite{huck}, and quantum annealing \cite{nest}.

At quantum critical points, the scaling limit of the EE has been widely studied within unitary models of CFT \cite{CallanW94,HolzheyLW94,Latorre1, Latorre2,Calabrese:2004eu,Calabrese:2005in}. In particular, the combination of a geometric description, Riemann uniformization techniques and standard expressions for CFT partition functions is very fruitful. Recently \cite{BCDLR}, this was generalized to non-unitary CFT, where a general formula was obtained using such techniques. Near critical points, the scaling limit is instead described by massive quantum field theory (QFT), and geometric techniques relying on conformal mappings break down. As was found in \cite{entropy,nexttonext,next}, the most powerful way of studying the EE in unitary models of QFT is using an approach based on local {\em branch-point twist fields}. However, the question of the EE in non-unitary near-critical models is much more delicate, and standard arguments give little indications as to how to modify the field-theoretical approach. Importantly, the rigorous derivation presented in \cite{BCDLR} provided a precise local-field description of the EE involving composite fields in the branch-point twist family, thus opening the door to its study in non-unitary QFT. In the present paper, using techniques of integrable QFT, we will study the scaling limit of the EE in the near-critical region of von Gehlen's model, described by the Lee-Yang QFT model.

This paper is organized as follows. In section \ref{1} we recall the main definitions and techniques, and provide a summary of our main results. In section \ref{2} we introduce the Lee-Yang model and some general results on the form factor expansion of correlation functions, their logarithms and expectation values of local fields. In section \ref{33} we review the twist field form factor equations and present solutions for the branch-point twist fields fields $\TT$ and $\Tp$ in the Lee-Yang model. In section \ref{4} we test our form factor solutions by performing a form factor expansion of the functions $\log\left(\bra \Tp \ket^{-2} \bra \Tp(r) \bTp (0) \ket\right)$ and $\log\left(\bra \TT \ket^{-2} \bra \TT(r) \tilde{\TT} (0) \ket\right)$ and recovering the behaviours $-4x_{\Tp} \log(mr)$ and $-4 x_{\TT} \log(mr)$ for some constants $x_{\TT}$, $x_{\Tp}$ which we compare to CFT predictions. In section \ref{PCFT} we compare a form factor computation of the two-point functions above with a computation in zeroth order conformal perturbation theory. As a byproduct, we find general formulae for some of the CFT structure constants entering the OPEs of $\TT$ and $\tilde{\TT}$ and of $\Tp$ with $\bTp$. In section \ref{6} we present numerical results for the R\'enyi entropy near criticality and a detailed computation of the first three leading corrections to saturation of the EE. We find that the next-to-leading order correction to saturation is non-universal. In section \ref{7} we present our conclusions and outlook. In  appendix \ref{appendixc} we explain how the normalization and conformal dimension of the field $\Tp$ are fixed by CFT. In appendix \ref{appendixa} we present a detailed analysis of the one-particle form factor contribution to the two-point functions of $\TT$ and $\Tp$. A large $n$ expansion of this function demonstrates that it provides a very substantial contribution to the power law behaviour of the two-point functions at short distances. In appendix \ref{appendixe} we present a computation of the three particle form factor of Lee-Yang twist fields. In appendix \ref{appendixd} we perform a computation of some of the structure constants entering the OPE of fields $\TT$ and $\tilde{\TT}$ and of fields $\Tp$ and $\bTp$ in CFT. In appendix \ref{appendixb} we present a computation of the numerical coefficient of the next-to-leading order correction to saturation of the EE in the Lee-Yang model.

\section{General aspects and summary of main results}\label{1}

In order to provide a formal definition of the
EE, the Hilbert space of an extended
quantum system, such as a spin chain, is decomposed into a tensor product of
local Hilbert spaces associated to its sites. Grouping together sites associated to the regions
$A$ and $\bar{A}$, this gives: \begin{equation} \label{Hdecomp} {\cal H} = {\cal A} \otimes
\b{{\cal A}}.
\end{equation} The EE in a state $|\psi\ket$ is the von
Neumann entropy of the reduced density matrix $\rho_A$ associated
to $A$: \beq\label{defeegs}
    S_A = -\Tr_{{\cal A}} \rho_A\log \rho_A ~,\quad \rho_A = \Tr_{\b{{\cal A}}} |\psi\ket \bra \psi|.
\eeq
Another frequently used measure of entanglement is the R\'enyi entropy,
\beq\label{renyi}
    S_A^{(n)} = \frac{\log \Tr_{{\cal A}} \rho_A^n}{1-n},
\eeq
which specializes to the von Neumann entropy at $n=1$,
\beq\label{formulan1}
\lim_{n\rightarrow 1} S_A^{(n)} = -\lim_{n\to 1}\frc{d}{dn} \Tr_{{\cal A}} \rho_A^n =S_A.
\eeq
We will study the ground state entanglement entropy in the scaling limit of infinite-length quantum chains. The scaling limit gives the universal part of the quantum chain behaviour near quantum critical points, described by 1+1-dimensional QFT. It is obtained
by approaching the critical point while letting the length $\ell$ of the region $A$ go to infinity in a fixed proportion with the correlation length $\xi$ (measured in number of lattice sites). If $\xi=\infty$ from the start, the system is exactly at its critical point, and the scaling limit is described by CFT. In this case the entanglement entropy of unitary critical systems, as a function of $\ell$, is divergent in a way which was first understood in \cite{CallanW94,HolzheyLW94}, numerically confirmed in \cite{Latorre1, Latorre2} and  generalized and reinterpreted in \cite{Calabrese:2004eu,Calabrese:2005in}. The divergency is logarithmic with a proportionality constant depending on the central charge $c$ of the CFT,
\beq\label{divSA}
  S_A^{(n)}(\ell)= \frac{c (n+1)}{6n} \log \frac{\ell}{\varepsilon} + o(1),\qquad S_A(\ell) = \frc{c}3 \log \frac{\ell}{\varepsilon} + o(1) \qquad\mbox{(CFT)},
\eeq
and where $\varepsilon$ is a non-universal ultraviolet cut-off (proportional to the lattice spacing) which is chosen so as to encode all $o(1)$ corrections. The formulae above are easily adapted to the case of an infinite region $\ell=\infty$ near criticality $\xi<\infty$, where $\ell$ is simply replaced by $\xi$ in \eqref{divSA} \cite{Calabrese:2004eu,Calabrese:2005in}. In the full scaling limit, where $\ell$ and $\xi$ are both large and in proportion to each other, there is a universal scaling function $f(\ell/\xi)$ which interpolates between the two results,
\beq\label{divSAmassive}
  S_A^{(n)}(\ell)= \frac{c (n+1)}{6n} \log \frac{\ell}{\varepsilon} + f(\ell/\xi) + o(1) \qquad \mbox{(QFT)}.
\eeq
In this case the result is much less trivial and has been studied in unitary integrable \cite{entropy,nexttonext} and non-integrable \cite{next} models using massive QFT techniques.

We may ask how (if at all) the entanglement entropy is affected by non-unitarity. At criticality, it was shown in \cite{BCDLR}  that the entanglement entropy scales instead as
\beq\label{divSA2}
   S_A^{(n)}(\ell) = \frac{c_{\text{eff}}(n+1)}{6n} \log \frac{\ell}{\varepsilon} + o(1), \qquad S_A(\ell) = \frc{c_{\text{eff}}}3 \log \frac{\ell}{\varepsilon} + o(1)\qquad
   \mbox{(non-unitary CFT)},
\eeq
where $c_{\text{eff}}:=c-24 \Delta$, and $\Delta$ is the smallest (often negative in non-unitary models) scaling dimension of a primary field in the CFT. For the Lee-Yang model, for example, $c_{\text{eff}}=\frac{4}{5}$ as $\Delta=-\frac{1}{5}$. This result is not entirely surprising as the work of Itzykson, Saleur and Zuber \cite{ceff} had previously shown that the {\em effective central charge} $c_{\rm eff}$ also replaces $c$ in the expression of the ground state free energy found by Affleck \cite{affleck} and Bl\"ote, Cardy and Nightingale \cite{BCN}. However, the question of the entanglement entropy in non-unitary near-critical models is much more delicate. Importantly, the rigorous derivation of (\ref{divSA2}) presented in \cite{BCDLR} has lead to new insights into the computation of entanglement entropy
in non-unitary theories and its field theoretical interpretation, opening the door to its study away from criticality in QFT.

It is known since some time
\cite{CallanW94,HolzheyLW94,Calabrese:2004eu,Calabrese:2005in}
that the bi-partite entanglement entropy
in the scaling limit can be re-written in
terms of more geometric quantities, using a method
known as the ``replica trick''. The essence of the method is to
``replace" the original QFT model by a new model consisting of
$n$ copies (replicas) of the original one. These are used to represent $\rho_A^n$ when $n$ is an integer, and then to evaluate $\Tr_{\cal A} \rho_A^n$. The quantities $S_A$ and $S_A^{(n)}$ for general $n$ are then obtained by ``analytic continuation'' in $n$.
The matrix multiplications in $\rho_A^n$ and the trace operation give rise to the condition that
the copies be connected cyclically through a finite cut on the
region $A$. As a consequence, this trace is proportional to the partition function
$Z_n(x_1,x_2)$ of the original (euclidean) QFT model on a Riemann
surface ${\cal M}_{n,x_1,x_2}$ with two branch points, at the
points $x_1$ and $x_2$ in $\R^2$, and $n$ sheets cyclically
connected. The positions $x_1$ and $x_2$ of the branch points
are dimensionful positions in the QFT model corresponding to the end-points of
the region $A$ in the scaling limit.
This gives: \beq\label{iden}
    S_A(r) = -\lim_{n\to 1}\frc{d}{dn} \frc{Z_n(x_1,x_2)}{Z_1^n}.
\eeq Here, $r:=|x_1-x_2|$ is the euclidean distance between $x_1$ and
$x_2$. The above concepts hold, in principle, for any QFT model, unitary or not. In CFT, one may evaluate this by using the uniformization theorem: the Riemann surface ${\cal M}_{n,x_1,x_2}$ can be conformally mapped to the Riemann sphere with two punctures (or the cylinder) by using the map $g$ reproduced in appendix \ref{appendixc}. 

In the EE context, it was first noticed in \cite{Calabrese:2004eu,Calabrese:2005in} that the ratio of partition functions above can be reinterpreted as correlation functions of certain fields, which were not otherwise specified, in unitary CFT. This idea was then generalized to unitary massive theories in \cite{entropy} and the fields where identified as {\it branch-point twist fields}
${\mathcal{T}}(x_1), \tilde{\mathcal{T}}(x_2)$ characterized by their non-trivial exchange relations with other fields of the $n$-copy theory. These twist fields
are defined only in the replica model (e.g.~they become the identity field when $n=1$),
and are primary fields arising from the extra permutation
symmetry present in the replica theory; they are associated to the $\Z_n$ symmetry generators $j\mapsto j+1\;{\rm mod}\; n$ and $j\mapsto j-1 \;{\rm mod}\; n$ respectively. In CFT, such twist fields and their relation to partition functions on Riemann surfaces were in fact studied much before their use in the computation of the EE was emphasized, see for instance \cite{kniz}. In terms of these fields, the replica partition function is given by
\begin{equation}
\frac{Z_n(x_1,x_2)}{Z_1^n}=\mathcal{Z}_n \varepsilon^{4\Delta_{\TT}}
\langle {\mathcal{T}}(x_1) \tilde{\mathcal{T}}(x_2)\rangle, 
\label{result}
\end{equation}
where $\langle {\mathcal{T}}(x_1) \tilde{\mathcal{T}}(x_2)\rangle$ is a two-point function in the ground state of the replica theory. The branch-point twist fields are chosen so as to have the CFT
normalisation (e.g.~the leading term in their OPE has coefficient 1). 
The constant ${\cal
Z}_n$, with ${\cal Z}_1=1$, is an $n$-dependent non-universal
constant, $\varepsilon$ is a short-distance cut-off which is
scaled in such a way that $d {\cal Z}_n/dn=0$ at $n=1$, and,
finally, $\Delta_{\TT}$ is the conformal dimension of the counter parts of
the fields $\mathcal{T}, \tilde{\mathcal{T}}$ in the underlying
$n$-copy conformal field theory,
\begin{equation}\label{dn}
    \Delta_{\TT}=\frac{c}{24}\left(n-\frac{1}{n}\right),
\end{equation}
which can be obtained by CFT arguments
\cite{kniz,Calabrese:2005in,entropy}. It is easy to show that the formula (\ref{result}) when inserted in (\ref{formulan1}) indeed reproduces (\ref{divSA}) for CFT.

The derivation above assumes unitarity of the theories under consideration. In such case $\Delta_\TT$ is by construction the lowest conformal dimension of any field in the replica theory which has the twist property.  The CFT derivation of (\ref{divSA}) has been generalized to the non-unitary case in \cite{BCDLR} leading to the expressions (\ref{divSA2}). In this work it was also observed that the EE could be computed from a representation of the $\mathbb{Z}_n$-orbifold partition function of the theory via correlation functions involving certain $\mathbb{Z}_n$ twist fields of the $n$-copy replica theory. This representation requires new twist fields $\Tp$ and $\bTp$, obtained from the primary twist fields $\mathcal{T}$ and $\tilde{\mathcal{T}}$ as leading descendants in the product with the lowest-dimension field $\phi$ (of conformal dimension $\Delta$). More precisely:
\beq
	\Tp(y) = n^{2\Delta-1} \lim_{x\to y} |x-y|^{2\Delta(1-\frac{1}{n})} \sum_{j=1}^n \TT (y) \phi_j(x), \label{newfield}
\eeq
and similarly for $\bTp$.
These composite fields were first introduced in \cite{ctheorem} and further studied in \cite{Levi}. The constant $n^{2\Delta-1}$ ensures conformal normalization, namely
\beq
	\bra \Tp(x_1) \bTp(x_2) \ket =|x_1-x_2|^{-4\Delta_{\Tp}}
\eeq
in CFT, where  
\beq
\Delta_{\Tp}=\Delta_{\bTp}=\Delta_{\TT}+\frac{\Delta}{n}
\eeq
are the conformal dimensions of $\Tp$ and $\bTp$. In the context of the study of the EE they were first obtained in \cite{ctheorem}. However, as for many other quantities in this context, they had emerged previously in the study of orbifold CFT, see e.g.~\cite{KacWakimoto,Bouwknegt,Borisov}. 

A detailed derivation of both the normalization constant and the power law in (\ref{newfield}) is given in appendix \ref{appendixc}.
\begin{figure}
\begin{center}
\includegraphics[width=7.92cm]{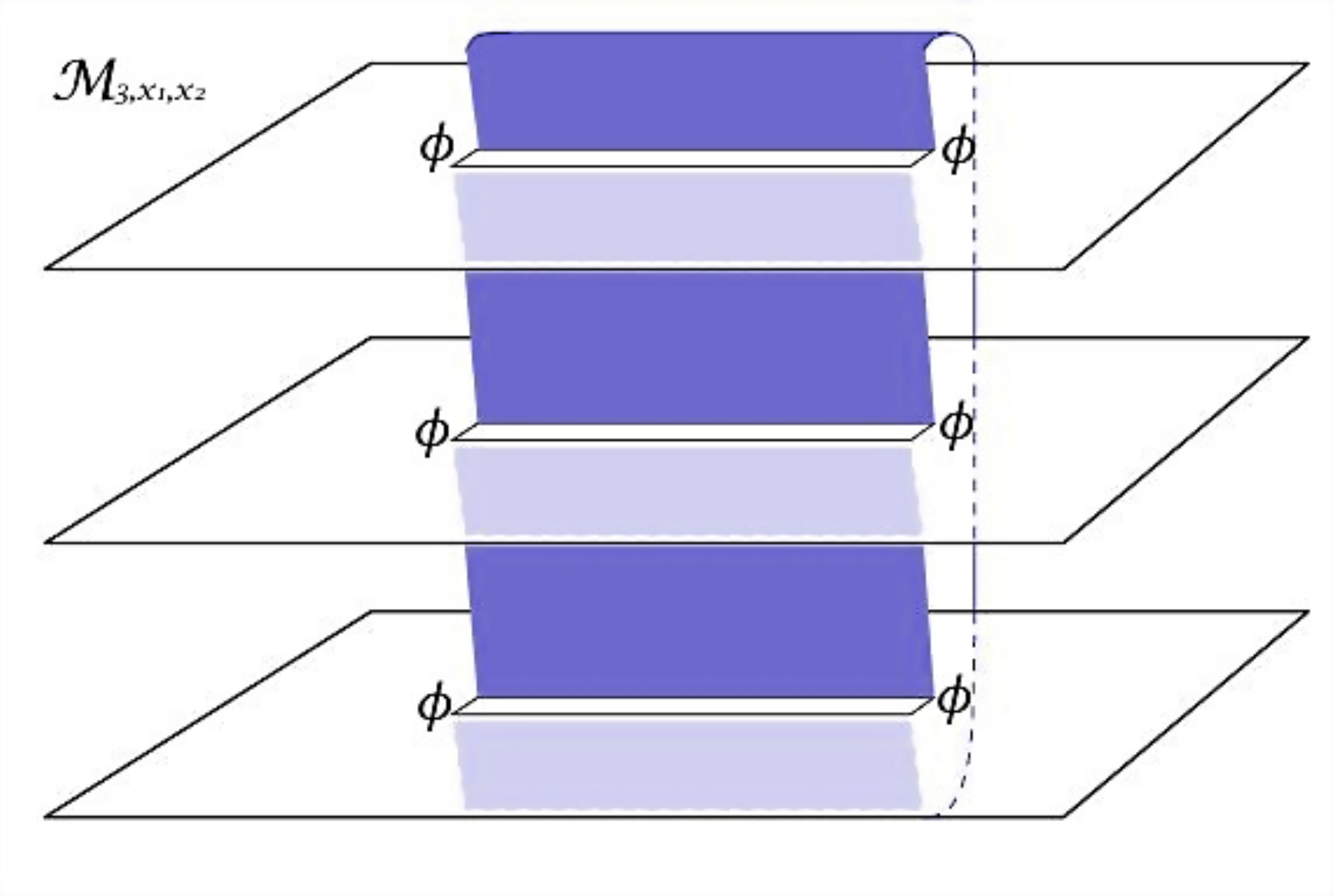} 
\caption{An artist's impression of the Riemann surface $\mathcal{M}_{n,x_1,x_2}$ for $n=3$ with field insertions $\phi$ at the branch points $x_1$, $x_2$.}
\end{center}
\end{figure}
The dimension $\Delta_{\Tp}$ arises naturally in computations of the entanglement entropy in non-unitary CFT, and, as noticed in \cite{BCDLR}, suggests that for such theories, the partition function of the $n$-copy theory may be written instead as:
\beq
\Tr_{\cal A}(\rho_A^n) = \frac{Z_n(x_1,x_2)}{Z_1^n}
= \mathcal{Z}_n\varepsilon^{4(\Delta_{\Tp}-n\Delta)} \frac{\langle\Tp(x_1) \bTp(x_2)\rangle}{\langle \phi(x_1)\phi(x_2)\rangle^n}, \label{newt}
\eeq
where again $\mathcal{Z}_n$ is such that it and its derivative at $n=1$ are 1 and $\varepsilon$ is a short distance cut-off. Compared to (\ref{result}) the expression (\ref{newt}) involves not only a different twist field but also a normalization given by $\langle \phi(x_1)\phi(x_2)\rangle^n$. For CFT it is easy to interpret this normalization as simply the norm of the ground state which in radial quantization is created by the action of the field $\phi$ on the conformal vacuum. As for the unitary case,  it is easy to show that the formula (\ref{newt}) when inserted in (\ref{formulan1}) reproduces (\ref{divSA2}) for CFT. It is natural to assume that the same expression will hold beyond criticality. This paper is a first step towards putting this assumption to the test beyond criticality.

\subsection{Summary of main results}

From the formulae above it is clear that a study of the EE in massive QFT is in principle only possible by studying correlation functions of twist fields. This approach has been pursued successfully in several works \cite{entropy,other,nexttonext} where the ratio of partition functions (\ref{result}) at large distances $r=|x_1-x_2|$ (the infrared
(IR) region) has been studied for unitary 
1+1-dimensional integrable QFTs. Integrability
means that in these models there is no particle production in any
scattering process and that the scattering ($S$) matrix factorizes
into products of two-particle $S$-matrices which can be calculated
exactly (for reviews see e.g.
\cite{Karowski:1978eg,ZZ,abdalla,Mussardo:1992uc,Dorey:1996gd}). 
Although most of the integrable theories studied in this framework are unitary, well-known examples of non-unitary integrable QFTs exist. Best known among those examples is the Lee-Yang model whose exact $S$-matrix was first given in \cite{CardyMuss2}.

Taking the known $S$-matrix of an integrable model as input it is possible to compute the matrix elements of local operators (also called form
factors). This is done by solving a set of consistency equations
\cite{KW,Smirnovbook}, also known as the form factor bootstrap
program for integrable QFTs. It is this particular feature of integrable models which makes them  interesting, as it means there is a systematic, non-perturbative way of computing multi-point functions of local fields. These computations are by no means easy, but often provide good numerical accuracy.  In \cite{entropy}, the form factor program was
generalised to branch point twist fields leading to the evaluation of
(\ref{result}) for various unitary models. In this paper we will pursue this program for the Lee-Yang model employing the formula (\ref{newt}). 
Our main results can be summarized as follows:
\begin{itemize}
\item[1)] {\it Form factor program for twist fields:} We have found that the twist field form factor equations together with the requirement of form factor clustering are sufficient conditions to entirely fix all form factor solutions for any particle numbers. In particular, these constraints immediately give rise to two form factor families, naturally identifiable with the fields $\TT$ and $\Tp$. We have carried out a zeroth order perturbed CFT computation of the twist field two-point function for several values of $n$ and compared this to a truncated form factor expansion of the same correlator. The former is expected to be accurate at short distances, the latter at large distances. Nevertheless, the agreement is relatively good, thus confirming the validity of the form factors found.

\item[2)] {\it Saturation of the EE at large subsystem size}:
Let us absorb all non-universal $o(1)$ constants of the short-distance behaviour of the R\'enyi EE into a short-distance cutoff $\ep_n$. Subtracting this non-universal contribution, the EE at large distances then saturates to a universal constant which can be calculated using QFT. More precisely, we find
\beqa
	S^{(n)}_A(r)&\sim& -\frc{c_{\rm eff}(n+1)}{6n} \log(m\ep_n) + U_n +o(1) \qquad (mr\to\infty)\n
    &\sim& \frc{c_{\rm eff}(n+1)}{6n} \log(r/\ep_n) + o(1) \qquad\qquad (mr\to0)
    \label{Snsl}
\eeqa
where the universal saturation $U_n$ is given by
\beq\label{Un}
U_n=\frc1{1-n}\,
\frc{K_{\phi}^n}{K_{\Tp}}.
\eeq
The constants $K_{\Or}$ are fundamental properties of QFT fields $\Or$, defined by
\beq\label{Kointro}
	K_\Or = \frc{\lim_{mr\to0}\,
    (mr)^{4x_\Or}\bra\Or(r)\widetilde\Or(0)\ket}{
    \lim_{mr\to\infty}\,
    \bra\Or(r)\widetilde\Or(0)\ket}
\eeq
where $x_\Or$ is the unique exponent making the limit finite and nonzero, and where $\widetilde\Or$ is the ``conjugate'' under internal symmetries ($\widetilde\phi = \phi$ and $\widetilde{\Tp} = \bTp$). In the unitary case, $x_\Or$ is the conformal dimension of $\Or$, and $K_\Or = m^{2\Delta_\Or}\bra\Or\ket^{-1}$ under the CFT normalization of $\Or$. In the non-unitary case, both of these statements are modified. In particular, in the Lee-Yang model, these constants can be expressed in terms of massive QFT and of CFT data as
\beq
	K_\phi = m^{2\Delta}\,\frc{\tilde{C}_{\phi\phi}^\phi}{\bra\phi\ket}
    ,\quad
    K_{\Tp} = m^{4\Delta_{\Tp}-2n\Delta}\,
    \frc{\tilde{C}_{\Tp\Tp}^{\phi_1\cdots\phi_n}\,
    \bra\phi\ket^n}
    {\bra\Tp\ket^2}
\eeq
where the vacuum expectation values are under CFT normalization, and $\tilde{C}_{\phi\phi}^\phi$ and $\tilde{C}_{\Tp\Tp}^{\phi_1\cdots\phi_n}$ are structure constants of conformal OPEs. Constants $K_{\Or}$ can also be expressed solely in terms of form factors of the massive model, as in \eqref{vev}.

\item[3)] {\it Leading order correction to saturation}: For unitary theories, one of the most interesting results
\cite{entropy,other} has been the identification of a {\it universal}
leading order correction to the large-distance (large-$r$)
saturation of the entropy of all unitary integrable theories. This exponentially decaying correction, of order $o(e^{-2mr})$, has a higher degree of universality than usual QFT quantities, as it only
depends on the particle spectrum of the model. In \cite{next} it was
shown that, even more strikingly,
this feature holds beyond integrability. This, however, seems to be broken in non-unitary models. For the Lee-Yang model, we found
\beq
S_A(r)\sim-\frac{2}{15}\log(m\ep) +U -a K_0(mr)+ {O}(e^{-2mr}),
\eeq
where $a=-0.0769782...$ is a constant that is (a priori) model-dependent. This suggests that we may use this feature of the EE as a means to identify non-unitary critical points. Indeed, given a spin chain model whose critical point is not known, a study of entanglement at criticality will reveal the value of $c_{\text{eff}}$. However, this does not say if $c_{\rm eff}=c$ or not. Considering large size corrections away from criticality will reveal different types of exponential decay depending on whether or not the theory is unitary.
\end{itemize}

\section{$S$-matrix and form factors in the Lee-Yang model}
\label{2}

\subsection{$S$-matrix}

The Lee-Yang model is one of the simplest 1+1 dimensional integrable QFTs. From the CFT point of view, it may be regarded as a perturbation of the non-unitary minimal model associated with central charge $c=-\frac{22}{5}$. The primary operator content of the theory is very simple, consisting of the identity and a scalar field $\phi$ of conformal dimension $\Delta=-\frac{1}{5}$. Perturbing this CFT by the scalar field we obtain the massive Lee-Yang model. This theory has a single particle spectrum. The scattering amplitude corresponding to the scattering of two particles of the same type was found by Cardy and Mussardo \cite{CardyMuss2} and can be written as
$$
S(\theta)=\frac{\tanh\frac{1}{2}\left(\theta+\frac{2\pi i}{3}\right)}{\tanh\frac{1}{2}\left(\theta-\frac{2 \pi i}{3}\right)}.
$$
It has a pole in the physical sheet at $\theta=\frac{2\pi i}{3}$ corresponding to the formation of a bound state, which in this case is the same fundamental particle of the theory. We note that the non-unitarity is manifested by the fact that the associated residue has the wrong sign. Nevertheless, the corresponding integrable massive model is well defined. The $n$-copy model, where the $\Z_n$ twist fields live, possesses $n$ particle species $\mu=1,\ldots,n$, and a two-particle scattering matrix given by $S_{\mu_1 \mu_2}(\theta) = S(\theta)^{\delta_{\mu_1,\mu_2}}$.

\subsection{Form factor expansions of two-point functions}
\label{ssff}

In this paper we study the correlators $\langle \mathcal{T}(r)\tilde{\mathcal{T}}(0) \rangle$ and, especially, $\langle \Tp(r)\bTp(0) \rangle$ and $\langle \phi(r)\phi(0) \rangle$, as well as the associated entanglement entropy obtained via \eqref{newt}. It is well known that two-point functions of local operators in QFT can be expressed as infinite sums involving matrix elements of these operators. The matrix elements of relevance, also known as form factors, are  defined as 
\begin{equation}
F_{k}^{\mathcal{O}|\mu _{1}\ldots \mu _{k}}(\theta _{1},\ldots
,\theta _{k}):=\left\langle
0|\mathcal{O}(0)|\theta_1,\ldots,\theta_k\right\rangle_{\mu_1,\ldots,\mu_k}^{\text{in}}
~,\label{ff}
\end{equation}
for a local field $\mathcal{O}$.
Here $|0\rangle$ represents the vacuum state and
$|\theta_1,\ldots,\theta_k\rangle_{\mu_1,\ldots,\mu_k}^{\text{in}}$
are the physical ``in'' asymptotic states of massive QFT. They
carry indices $\mu_i$, which are quantum numbers characterizing
the various particle species, and depend on the real parameters
$\theta_i$, which are called rapidities. The energy and momentum of a particle of mass $m_i$ are expressed in terms of its rapidity $\theta_i$ as $m_i  \cosh \theta_i$ and $m_i \sinh \theta_i$, respectively. In terms of form factors, two-point correlation functions (in unitary models) may be expanded as
\begin{eqnarray}
\langle \mathcal{O}(r)\mathcal{O}^\dagger(0)\rangle =
\sum_{k=0}^\infty\frac{1}{k!} \sum_{\mu_1,\ldots,\mu_k=1}^n \left(
\prod_{j=1}^k \int \limits_{-\infty }^{\infty } \frac{d\theta
_{j}}{(2\pi )}\right) \left|F_{k}^{\mathcal{O} |\mu_1 \ldots
\mu_k}(\theta_1,\ldots, \theta_{k})\right|^2\,e^{-rm
\sum\limits_{j=1}^k \cosh\theta_j}.\label{ent1}
\end{eqnarray}

As mentioned, the Lee-Yang model is non-unitary. As was noted in \cite{Smir,Z}, non-unitarity affects the form factor expansion. A consequence of this is that many fields appear to be non-Hermitian under the Hilbert structure of asymptotic states. In the Lee-Yang model, an exact calculation of the form factors of the field $\phi$ shows that $\lt(\bra0|\phi(0)|\theta_1,\ldots,\theta_k\ket^{\rm in}\rt)^* \neq {}^{\rm in}\bra \theta_1,\ldots,\theta_k|\phi(0)|0\ket$, where the right-hand side can be obtained by crossing symmetry. However, it turns out that the relation is surprisingly simple:
\beq\label{phiadd}
\lt(\bra\phi\ket^{-1}\,\bra0|\phi(0)|\theta_1,\ldots,\theta_k\ket^{\rm in}\rt)^* = (-1)^k\,\bra\phi\ket^{-1}\, {}^{\rm in}\bra \theta_1,\ldots,\theta_k|\phi(0)|0\ket.
\eeq
As a consequence, the form factor expansion of the two-point function of the field $\phi$, normalized by the square of the VEV, is a modification of \eqref{ent1} where sign factors $(-1)^k$ are included for the terms involving the $k$-particle form factors. This gives rise, in the single-copy Lee-Yang model, to:
\begin{eqnarray}
\frc{\langle \phi(r) \phi(0)\rangle}{\bra \phi\ket^2} =
\sum_{k=0}^\infty\frac{(-1)^k}{k!} \left(
\prod_{j=1}^k \int \limits_{-\infty }^{\infty } \frac{d\theta
_{j}}{(2\pi )}\right) \left|\bra\phi\ket^{-1}\,F_{k}^{\phi}(\theta_1,\ldots, \theta_{k})\right|^2\,e^{-rm
\sum\limits_{j=1}^k \cosh\theta_j}.\label{entphi1}
\end{eqnarray}

A natural way to understand this modification is through a discussion of the bound-state singularity occurring in the form factors. The additional $(-1)^k$ guarantees that the bound-state residue of the analytic continuation of the $k$-particle integrand, which, like that of the scattering matrix, has the wrong sign, is related to the $k-1$-particle integrand in a way that would guarantee locality properties. As we will see below, form factors of twist fields $\Tp$, $\bTp$, $\mathcal{T}$ and $\tilde{\mathcal{T}}$ are subject to similar bound-state residue equations as those of $\phi$. Hence, this interpretation suggests that a similar modification of \eqref{ent1} occurs for the form factor expansion of $\bra\Tp(r)\bTp(0)$ and of $\bra\mathcal{T}(r)\t\TT(0)\ket$. That is, in the $n$-copy model,
\beq
\frc{\langle \Or(r) \t\Or(0)\rangle}{\bra \Or\ket^2} =
\sum_{k=0}^\infty\frac{(-1)^k}{k!} \sum_{\mu_1,\ldots,\mu_k=1}^n \left(
\prod_{j=1}^k \int \limits_{-\infty }^{\infty } \frac{d\theta
_{j}}{(2\pi )}\right) \left|\bra\Or\ket^{-1}\,F_{k}^{\Or |\mu_1 \ldots
\mu_k}(\theta_1,\ldots, \theta_{k})\right|^2\,e^{-rm
\sum\limits_{j=1}^k \cosh\theta_j}\label{entOr}
\eeq
for both $\Or=\TT$, $\t\Or = \t\TT$ and $\Or = \Tp$, $\t\Or=\bTp$. Here we have used the fact that by symmetry under inversion of copies, $\bra\t\TT\ket = \bra\TT\ket$ and $\bra\bTp\ket = \bra\Tp\ket$, and the form factor expansion includes sums over the copy numbers $\mu_j$. That this is the correct expansion follows from an equation similar to \eqref{phiadd} for twist fields, see subsection \ref{ssfftt}

Finally, since the field $\phi$ is no longer Hermitian, its VEV is no longer expected to be real. As was shown by Zamolodchikov, $\bra\phi
\ket$ is in fact purely imaginary -- this can be explained by the fact that it occurs in the formal massive Lee-Yang action (written as a perturbation of the CFT action) with a purely imaginary coupling constant. A similar phenomenon makes the VEVs $\bra\TT\ket$ and $\bra\Tp\ket$ not necessarily real. We will determine their phases (up to multiples of $\pi$) by evaluating analytically the normalization of their leading short-distance power-law, and by observing numerically that the right-hand side of \eqref{entOr} is positive for all $mr$.

\subsection{Short-distance behaviour from form factors}

Form factor expansions \eqref{ent1} and \eqref{entphi1} are naturally large-distance expansions, in that they converge very rapidly for large values of $rm$. However, in many cases we want to explore small values of $rm$. In such cases two-point functions generally develop power-law behaviours in $rm$ and it is very difficult to extract the precise power from a form factor expansion such as those above. 

It was realized a long time ago \cite{Smir} (see also \cite{karo} for a nice derivation and application to various models and \cite{takacs} for a generalization to boundary theories) that if one is interested in the short-distance behaviour of correlators then an expansion of the logarithm of two-point function is more appropriate:
\begin{eqnarray}
\log\left(\frac{\langle \mathcal{O}(r)\t{\mathcal{O}}(0)\rangle}{\langle\mathcal{O} \rangle^2} \right)=
\sum_{k=1}^\infty\frac{(-1)^k}{k!} \sum_{\mu_1,\ldots,\mu_k=1}^n \left(
\prod_{j=1}^k \int \limits_{-\infty }^{\infty } \frac{d\theta
_{j}}{(2\pi )}\right) H_k^{\mathcal{O}|\mu_1,\ldots, \mu_k}(\theta_1, \cdots, \theta_k)e^{-rm
\sum\limits_{j=1}^k \cosh\theta_j}.\label{ent2}
\end{eqnarray}
The functions $H_k^{\mathcal{O}|\mu_1,\ldots, \mu_k}(\theta_1,\cdots, \theta_n)$ must of course be chosen so that the expansion (\ref{ent1}) is recovered when exponentiating (\ref{ent2}). This condition automatically implies for example that
\beqa
H_1^{\mathcal{O}|\mu_1}(\theta)&=&{\langle\mathcal{O} \rangle^{-2}}|F_1^{\mathcal{O}|\mu_1}(\theta)|^2,\\
H_2^{\mathcal{O}|\mu_1 \mu_2}(\theta_1,\theta_2)&=&{\langle\mathcal{O} \rangle^{-2}} |F_2^{\mathcal{O}|\mu_1 \mu_2}(\theta_1,\theta_2)|^2- H_1^{\mathcal{O}|\mu_1}(\theta_1)H_1^{\mathcal{O}|\mu_2}(\theta_2),\\
H_3^{\mathcal{O}|\mu_1 \mu_2 \mu_3}(\theta_1,\theta_2, \theta_3)&=&{\langle\mathcal{O} \rangle^{-2}} |F_3^{\mathcal{O}|\mu_1 \mu_2 \mu_3}(\theta_1,\theta_2,\theta_3)|^2- H_1^{\mathcal{O}|\mu_1}(\theta_1)H_1^{\mathcal{O}|\mu_2}(\theta_2)H_1^{\mathcal{O}|\mu_3}(\theta_3)\nonumber \\ 
&& - H_2^{\mathcal{O}|\mu_1 \mu_2}(\theta_1,\theta_2)H_1^{\mathcal{O}|\mu_3}(\theta_3)- H_2^{\mathcal{O}|\mu_2 \mu_3}(\theta_2,\theta_3)H_1^{\mathcal{O}|\mu_1}(\theta_1)\nonumber\\
&&
-H_2^{\mathcal{O}|\mu_1 \mu_3}(\theta_1,\theta_3)H_1^{\mathcal{O}|\mu_2}(\theta_2).
\eeqa
In general the $H_k$ functions can be interpreted as the ``connected parts'' of the $F_k$ functions (they are ``cumulants'' with respect to the rapidities). These are such that, if the clustering decomposition holds for the $F_{k}$'s at large rapidities for all $k$, that is
\beq
\lim_{\theta_1, \ldots, \theta_k \rightarrow \infty} F_{k+\ell}^{\mathcal{O} |\mu_1 \ldots \mu_{k+\ell}}(\theta_1,\ldots, \theta_{k+\ell})=\frac{
F_{k}^{\mathcal{O} |\mu_1 \ldots\mu_k}(\theta_1,\ldots, \theta_{k}) F_{\ell}^{\mathcal{O} |\mu_{k+1} \ldots
\mu_{k+\ell}}(\theta_{k+1},\ldots, \theta_{k+\ell})}{\bra \mathcal{O} \ket},
\eeq
$\forall k,\ell \in \mathbb{N}$, then the $H_{k}$'s vanish at large rapidities for all $k$.

Thanks to this vanishing, for $mr \ll 1$ we now expect {\em each summand} in the sum over $k$ in the expression above to be dominated by a leading term proportional to $\log(mr)$. The constant coefficient of this term, summed over all particle contributions, will then give the power which governs the short-distance behaviour of the two point function. Let us call this power $-4 x_{\mathcal{O}}$. Then, carrying out one integral in (\ref{ent2}) and expanding the result for small $mr$ we find \cite{Smir,karo,takacs}
\beqa
x_{\mathcal{O}}=\frac{1}{4 \pi}
\sum_{k=0}^\infty \frac{(-1)^k}{k!
}\sum_{\mu_1,\ldots,\mu_k=1}^n \left(
\prod_{j=2}^k \int \limits_{-\infty }^{\infty } \frac{d\theta
_{j}}{(2\pi )}\right) H_k^{\mathcal{O}|\mu_1,\ldots, \mu_k}(0,\theta_2, \cdots, \theta_k).\label{del}
\eeqa
This was used for the field $\phi$ in \cite{Z} and shown to agree well with conformal field theory results.

In addition, the proportionality constant \eqref{Kointro} of the power law behaviour at short distances,
\beq\label{Ko}
\frc{\langle \mathcal{O}(r)\t{\mathcal{O}}(0)\rangle}{\bra \mathcal{O} \ket^2}\sim K_{\mathcal{O}}\, (mr)^{-4 x_{\mathcal{O}}}\quad (mr\to0)
\eeq
can also be extracted from the form factor expansion. It was shown in \cite{karo} that by considering the leading correction to the $\log(mr)$ term in (\ref{ent2}) one may also find a form factor expansion for the constant $K_{\mathcal{O}}$ which is given by:
\beqa
K_{\mathcal{O}}=\exp\left(-\frac{1}{\pi}
\sum_{k=0}^\infty\frac{(-1)^k}{k!} \sum_{\mu_1,\ldots,\mu_k=1}^n \left(
\prod_{j=2}^k \int \limits_{-\infty }^{\infty } \frac{d\theta
_{j}}{(2\pi )}\right) H_k^{\mathcal{O}|\mu_1,\ldots, \mu_k}(0,\theta_2, \cdots, \theta_k)(\ln\frac{\xi}{2}+\gamma)\right), \label{vev}
\eeqa
with $\xi^2 =\left(\left(\sum_{j=2}^k \cosh \theta_i +1\right)^2-\left(\sum_{j=2}^k \sinh \theta_i\right)^2\right)$ and where $\gamma=0.5772157...$ is the Euler-Mascheroni constant.

\section{Twist field form factors}
\label{33}

\subsection{Form factor equations and minimal form factors}

The form factor equations for $\Z_n$ twist fields were derived in \cite{entropy}. Details of the solutions procedure for two-particle form factors appeared there, and higher particle form factors of various models were computed in \cite{nexttonext} and \cite{higher}. Interestingly, a very similar set of form factor equations had been derived much earlier \cite{nieder} in a rather different context (e.g.~the study of the response of an integrable QFT to a variation of the Unruh temperature).
The details of the computation for the Lee-Yang model are very similar to those described in these works, with the only difference that the presence of the bound state pole in the $S$-matrix imposes further conditions on the form factors. In particular, bound state poles are present in addition to kinematic poles. The form factor equations only encode locality properties of fields, hence they are unchanged for form factors of any field in a the same $\Z_n$ twist sector. In order to distinguish for form factors of $\mathcal{T}$ and $\Tp$, we will impose additional conditions, and verify the correctness of the solutions by numerical comparisons with CFT predictions.

In what follows we will consider $k$-particle form factors $F^{\mathcal{O}|\mu_1\cdots \mu_k}(\theta_1, \theta_2, \ldots, \theta_k)$ for a generic twist field $\mathcal{O}$. We will later identify this field with $\mathcal{T}$ or $:\mathcal{T}\phi:$ depending on various properties of the form factor solutions we obtain.

The two-particle form factor must satisfy (in the two-particle case we use the single argument $\theta=\theta_1-\theta_2$) 
\begin{equation}\label{ff1}
    F^{\mathcal{O}|11}(\theta)=S(\theta)  F^{\mathcal{O}|11}(-\theta)= F^{\mathcal{O}|11}(-\theta + 2 \pi i n),
\end{equation}
and the kinematic residue equations
\begin{eqnarray}
 \begin{array}{l}
\\
  \text{Res}  \\
 {\footnotesize \theta=0}
\end{array}\!\!\!\!
 F_{2}^{\mathcal{O}|\b{\mu} \mu }(\theta+i\pi)
  &=&
  i \,\langle\mathcal{O}\rangle, \label{ni3}
  \\
\begin{array}{l}
\\
  \text{Res}  \\
 {\footnotesize \theta=0}
\end{array}\!\!\!\!
 F_{2}^{\mathcal{O}|\b\mu \hat{\mu } }(\theta+i\pi)
  &=&-i\,\langle \mathcal{O}\rangle. \label{nikre}
\end{eqnarray}
Here and below we use $\hat\mu = \mu-1\; {\rm mod}\; n$. Higher particle versions of these equations read
\begin{eqnarray}
 \begin{array}{l}
\\
  \text{Res}  \\
 {\footnotesize \bar{\theta}_{0}={\theta}_{0}}
\end{array}\!\!\!\!
 F_{k+2}^{\mathcal{O}|\b{\mu} \mu  \mu_1 \ldots \mu_k}(\bar{\theta}_0+i\pi,{\theta}_{0}, \theta_1 \ldots, \theta_k)
  &=&
  i \,F_{k}^{\mathcal{O}| \mu_1 \ldots \mu_k}(\theta_1, \ldots,\theta_k), \label{3}
  \\
\begin{array}{l}
\\
  \text{Res}  \\
 {\footnotesize \bar{\theta}_{0}={\theta}_{0}}
\end{array}\!\!\!\!
 F_{k+2}^{\mathcal{O}|\b\mu \hat{\mu } \mu_1 \ldots \mu_k}(\bar{\theta}_0+i\pi,{\theta}_{0}, \theta_1 \ldots, \theta_k)
  &=&-i\prod_{i=1}^{k} S_{\hat{\mu}\mu_i}^{(n)}(\theta_{0i})
  F_{k}^{\mathcal{O}| \mu_1 \ldots \mu_k}(\theta_1, \ldots,\theta_k).\label{kre}
\end{eqnarray}

For this model there is the added difficulty of having to solve also the bound state residue equation associated to the scattering process $a+a \rightarrow a$ where $a$ is the Lee-Yang particle on copy $a$. This takes the form
\begin{eqnarray}
\begin{array}{l}
\\
  \text{Res}  \\
 {\footnotesize \theta=\bar{\theta}}
\end{array}\!\!\!
    F_{n+1}^{{\mathcal{O}}|aa \mu_1\ldots\mu_{n-1}}
        (\theta+\frc{i\pi}3,\bar{\theta}-\frc{i\pi}3,\theta_1,\ldots,\theta_{n-1})
        = i\Gamma F_{n}^{{\mathcal{O}}|a\mu_1\ldots\mu_{n-1}}(\theta,\theta_1,\ldots,\theta_{n-1})\label{bse}
\end{eqnarray}
where the so-called three-point coupling is fixed by
\beq
    \Gamma^2 = -i\lim_{\theta\to \frac{2 \pi i}{3}} (\theta-\frac{2\pi i}{3}) S(\theta)=-2 \sqrt{3}
\eeq
and by choosing the negative imaginary direction: $\Gamma = -i2^{1/2}3^{1/4}$. For $n=1$ this equation fixes the one particle form factor (which for spinless fields must be rapidity independent) through the equation
\begin{eqnarray}
\begin{array}{l}
\\
  \text{Res}  \\
 {\footnotesize \theta=\bar{\theta}}
\end{array}\!\!\!
    F_{2}^{{\mathcal{O}}|aa}
        (\theta-\bar{\theta}+\frc{2i\pi}3)
        = i\Gamma F_{1}^{{\mathcal{O}}|a}. \label{fs}
\end{eqnarray}

These equations imply that the two-particle form factor solution given in \cite{entropy} must be generalized to include the bound state pole. As discussed in \cite{Z} this may be done by defining a minimal form factor. A minimal form factor is a solution of (\ref{ff1}) which has no poles in the (extended) physical sheet $\theta\in[0,2\pi n)$ except possibly for bound state poles, and which tends to unity as $|\theta|\to\infty$. It turns out that this particular Riemann-Hilbert problem has a unique solution, and this solution possesses bound state poles with nonzero residues:
\begin{equation}
    F_{\text{min}}(\theta)=a(\theta, n) f(\theta, n),
\end{equation}
where $a(\theta, n)$ encodes the bound state pole
\begin{equation}\label{a}
    a(\theta, n)=\frac{\cosh\frac{\theta}{n}-1}{\cosh\frac{\theta}{n}-\cos \frac{2\pi }{3n}},
\end{equation}
and $f(\theta)$ is given by the integral representation
\begin{equation}\label{eq}
    f(\theta, n)=\exp \left(2\int_{0}^\infty \frac{\sinh\frac{t}{3}\sinh\frac{t}{6}}{t \sinh(nt) \cosh\frac{t}{2}} \cosh t \left(n+\frac{i\theta}{\pi}\right)\right).
\end{equation}
The latter function admits also a representation as an infinite product of gamma functions which was already given in \cite{entropy} for the sinh-Gordon model (it suffices to take $B=2/3$ and to invert the formula).

The expression \eqref{eq} may be obtained as a solution to \eqref{ff1} using a similar integral representation of the two-particle scattering amplitude. In the absence of bound state poles, the resulting $f(\theta,n)$ would directly be the minimal two-particle form factor. In the present case, however, the function tends to 1 as $|\theta|\to\infty$ but has a simple pole at $\theta=0$. The factor $a(\theta,n)$ is the unique one that shifts this pole towards the position of the allowed bound-state singularity in the physical sheet, without affecting the large-$|\theta|$ behaviour.

Using the integral or Gamma-function representation, it may be shown that
\begin{equation}\label{rationfs}
    \frac{f(i\pi,n)}{f(\frac{2\pi i}{3},n)^2}=\frac{n}{\sqrt{3}}\frac{\sin^3 \frac{\pi}{3n}}{\sin\frac{\pi}{6n} \sin \frac{\pi}{2n}}.
\end{equation}
In order to compute higher particle form factors the following more general identities are important
\begin{equation}\label{plmo}
    F_{\text{min}}(\theta+ \frac{i \pi}{3})F_{\text{min}}(\theta-\frac{i \pi}{3})=\frac{\cosh\frac{\theta}{n}-\cos \frac{2\pi }{3n}}{\cosh\frac{\theta}{n}-\cos\frac{\pi}{n}}F_{\text{min}}(\theta),
\end{equation}
\begin{equation}\label{plusipi}
F_{\text{min}}(\theta+ i\pi)F_{\text{min}}(\theta)=\frac{\sinh\frac{\theta}{2n} \sinh\left(\frac{\theta}{2n}+\frac{i\pi}{2n}\right)}{\sinh\left(\frac{\theta}{2n}-\frac{i\pi}{3n}\right)\sinh\left(\frac{\theta}{2n}+\frac{5i\pi}{6n}\right)}.
\end{equation}

\subsection{Twist field one- and two-particle form factors}

One expects that primary twist fields, with direct geometric meaning, would occur as solutions to \eqref{ff1} and to conditions of bound state and kinematical singularities with the additional requirement of convergence as $|\theta|\to\infty$. This additional requirement is expected to implement, in a path-integral picture, the least singular asymptotic condition possible at small distances near the position of the field. With these conditions, the most general form the two-particle form factor can take is
\begin{equation}
F_{2}^{\Or|11}(\theta)=\frac{ \langle \mathcal{O}\rangle
\sin\left(\frac{\pi}{n}\right)}{2 n \sinh\left(\frac{i\pi
- \theta}{2n}\right)\sinh\left(\frac{i\pi+\theta}{2n}\right)}
\frac{F_{\text{min}}(\theta)}{F_{\text{min}}(i
\pi)} + \kappa F_{\text{min}}(\theta),\label{full}
\end{equation}
where the first term is of the form required to solve the kinematic residue equation (and of the same form as for other theories previously studied \cite{entropy}) and the second term is what is commonly termed a ``kernel" solution of the kinematic residue equation (that is a solution without kinematic poles).

In general $\kappa$ is an arbitrary constant, but it may be fixed by imposing the cluster decomposition property, namely
\beq
\lim_{\theta\rightarrow \infty}F_{2}^{\mathcal{O}|11}(\theta)=\kappa:=\frac{(F_{1}^{\mathcal{O}|1})^2}{\langle \mathcal{O}\rangle},
\eeq
where we have used the fact that $\lim_{\theta \rightarrow \infty} F_{\text{min}}^{11}(\theta)=1$. Then, the one-particle form factor on copy $a$ may be fixed by combining this with equation (\ref{fs}), which translates into the following quadratic equation for $F_1^{\mathcal{O}|1}$
\begin{equation}\label{f1}
    F_1^{\mathcal{O}|1}=-\frac{1}{\Gamma}\frac{\tan \frac{\pi}{3n}}{\tan\frac{\pi}{2n}} \frac{f(\frac{2\pi i}{3},n)}{f(i\pi,n)} \langle \mathcal{O}\rangle+\frac{(F_{1}^{\mathcal{O}|1})^2}{\langle \mathcal{O}\rangle}\frac{n}{\Gamma}\tan\left(\frac{\pi}{3n}\right)f(\frac{2\pi i}{3},n).
\end{equation}
This leads to two possible solutions:
\beq
F_1^{\mathcal{O}|1}=-\langle \mathcal{O}\rangle\Gamma\frac{\cos \left(\frac{\pi }{3 n}\right)\pm 2 \sin ^2\left(\frac{\pi }{6 n}\right)}{2 n\sin\left(\frac{\pi}{3n}\right)f(\frac{2\pi i}{3},n)},
\eeq
where we have used the identity (\ref{rationfs}).

The presence of two solutions immediately suggests the existence of two different least-singular twist fields, by contrast to other models studied in the past. It is natural to conjecture that these are $\mathcal{T}$ and $\Tp$, and given this, it is a simple matter to identify their respective form factor solutions. Indeed, the former specializes to the identity at $n=1$, and the latter, to $\phi$. We note that the solution with the negative sign specializes to 0 at $n=1$, and that that with the positive sign specializes to the one-particle form factor of the field $\phi$
\beq
\frac{F_1^{\phi}}{F_0^{\phi}}=\frac{i 2^{1/2}} {3^{1/4}f(\frac{2\pi i}{3},1)} \quad \text{with} \quad F_0^{\phi}=\langle \phi\rangle = \frac{5 i m^{-\frac{2}{5}}}{24 h \sqrt{3}}\quad \text{and} \quad h=0.09704845636... \label{ratio01}
\eeq
found in \cite{Z} (note that the constant $v(0)$ in \cite{Z} is $v(0)=f(i \pi,1)^{1/2}=\frac{\sqrt{3}}{2}f(\frac{2\pi i}{3},1)$ with our present notation, and that the coupling $h$ was computed in \cite{tba1}). These properties suggest the identifications
\beq
 \frac{F_1^{\mathcal{T}|1}}{\langle \mathcal{\mathcal{T}}\rangle}=-\Gamma\frac{2 \cos \left(\frac{\pi }{3 n}\right)-1}{2 n\sin\left(\frac{\pi}{3n}\right)f(\frac{2\pi i}{3},n)},\quad
 \frac{F_1^{:\mathcal{T}\phi:|1}}{\langle :\mathcal{T}\phi:\rangle}=-\frac{\Gamma}{2 n\sin\left(\frac{\pi}{3n}\right)f(\frac{2\pi i}{3},n)}. \label{identification}
\eeq
The numerical results of sections 4, 5 and 6 provide further support for these identifications.

\subsection{Higher particle form factors}

Let us now consider only form factors of the form $F_{k}^{\mathcal{O}|11\ldots 1}(x_1,\ldots,x_k):=F_{k}^{\mathcal{O}}(x_1,\ldots,x_k)$, that is form factors involving only one particle type. This is sufficient as form factors involving other particles may be obtained from these by using the twist field form factor equations \cite{entropy}.

The higher particle form factors may be obtained by making the ansatz
\begin{equation}
 F_{k}^{\mathcal{O}}(x_1,\ldots,x_k)=Q_k(x_{1},...,x_{k}) \prod_{i<j}^{k}
\frac{F_{\text{min}}(\theta_i-\theta_j)}{(x_i-\alpha
x_j)(x_j-\alpha x_i)},\label{ansatz2}
\end{equation}
where $x_i=e^{\theta_i/n}$ and $\alpha=e^{i\pi/n}$. The functions
$Q_{k}(x_{1},...,x_{k})$ are symmetric in all variables and have
no poles on the physical sheet.

This ansatz, as usual in the context of the computation of form factors of local fields (see e.g.~\cite{Z,FMS}), expresses the form factors in such a way as to explicitly separate the part containing the poles from the part which has no singularities. In addition, the explicit presence of the minimal form factor and the symmetry in the variables $x_i$ automatically gives form factors which exhibit the correct monodromy properties in the rapidities. In the context of twist fields, this ansatz was used for the first time in \cite{higher}.

\subsubsection{Kinematic and bound state residue equations}

Using \eqref{plusipi}, the kinematic residue equation with the ansatz (\ref{ansatz2}) can be rewritten as ($k\geq 0$):
\begin{equation}\label{qq}
    Q_{k+2}(\alpha x_0,x_0,x_1,\ldots,x_k)=x_0^2 P_k(x_0,x_1,\ldots,x_k) Q_k(x_1,\ldots,x_k),
\end{equation}
where $P_k$ is the polynomial
\begin{eqnarray}
 P_k(x_0,x_1,\ldots,x_k) = C_k(n)\prod_{b=1}^k\Big(
 (x_b-\alpha^2 x_0)(x_b-\alpha^{-1} x_0)
 (x_b- \beta x_0)(x_b-\alpha \beta^{-1} x_0)\Big),
\end{eqnarray}
where $\beta=e^{-\frac{2\pi i}{3n}}$ and
\begin{equation}\label{cn}
    C_k(n)=\frac{2 \sin\frac{\pi}{n}}{n F_{\text{min}}(i\pi)}\, \alpha^{2(k+1)} = C_0(n)\alpha^{2k}.
\end{equation}
Denoting $\sigma_i^{(k)}$ the $i$-th elementary symmetric
polynomial on $k$ variables $x_1,\ldots,x_k$, which can be defined
by means of the generating function,
\begin{equation}\label{sigma}
   \sum_{i=0}^k x^{k-i} \sigma_i^{(k)}= \prod_{i=1}^k(x_i + x),
\end{equation}
we can rewrite $P_k(x_0,x_1,\ldots,x_k)$ as
\begin{eqnarray}
C_k(n)\sum^{k}_{a,b,c,d=0}(-\alpha^{2}x_0)^{k-a}(-\alpha^{-1}{x_0})^{k-b}
(-{\alpha}{\beta^{-1}}x_0)^{k-c}(-\beta x_0)^{k-d}\sigma_a^{(k)}
\sigma_b^{(k)} \sigma_c^{(k)} \sigma_d^{(k)}.\label{qqC}
\end{eqnarray}
In the following we will omit the upper index $^{(k)}$ when there is no confusion possible.

Besides \eqref{qq}, another equation that arises from the ansatz (\ref{ansatz2}) is that using the bound state residue equation (\ref{bse}). The simplest case of this equation was given in (\ref{fs}) and this allowed us, in combination with the clustering property, to fix the one-particle form factor (\ref{f1}). For higher particles, using \eqref{plmo} we find ($k\geq 1$)
\begin{equation}\label{q2}
    Q_{k+1}(x_0 \beta^{-\frac{1}{2}},{x}_0 \beta^{\frac{1}{2}},x_1,\ldots,x_{k-1})= x_0^2 U_k(x_0,x_1,\ldots,x_{k-1}) Q_k(x_0,x_1,\ldots,x_{k-1}),
\end{equation}
with
\begin{eqnarray}
    U_k(x_0,x_1,\ldots,x_{k-1})&=&H_k(n) \prod_{i=1}^{k-1}(x_i-\beta^{-2}x_0)(x_i-\beta^2 x_0)\\\nonumber
&=& H_k(n)\sum_{a,b=0}^{k-1} (-\beta^{-2}x_0)^{k-1-a} (-\beta^2 x_0)^{k-1-b} \sigma_{a}^{(k-1)}\sigma_b^{(k-1)},
\end{eqnarray}
and
\begin{equation}\label{hh2}
    H_{k}(n)= \frac{4 \Gamma \sin^2 \left(\frac{\pi}{2n}\right)}{n  \tan\left(\frac{\pi}{3n}\right) a(i\pi) f(\frac{2 \pi i}{3})}(-\alpha)^k = H_1(n) (-\alpha)^{k-1}.
\end{equation}
From the ansatz (\ref{ansatz2}) it follows that  $Q_1=F_1^{\mathcal{O}|1}$. 

\subsubsection{Three-particle form factors}

First let us analyze the two-particle case. We have by definition that $F_0^{\mathcal{O}}=Q_0=\langle\mathcal{O}\rangle$, and comparing to (\ref{full}), we obtain the polynomial
\begin{equation}
    Q_2(x_1,x_2)=\langle\mathcal{O}\rangle C_0(n) \alpha^{-1} \sigma_2 + \frac{(F_1^{\mathcal{O}|1})^2}{\langle\mathcal{O}\rangle}\left((1+\alpha)^{2}\sigma_2-\alpha\sigma_1^2\right).\label{q22}
\end{equation}
It is a simple matter to verify that this is indeed in agreement with the kinematic residue equation \eqref{qq}; given $Q_0$ this is the most general solution to \eqref{qq} ($k=0$), as was shown in \cite{higher}  (in particular, the second term vanishes at $x_1=\alpha x_2$). Further, replacing $(F_1^{\mathcal{O}|1})^2/\langle\mathcal{O}\rangle$ by the linear combination of the zero- and one-particle form factors, $Q_0$ and $Q_1(x_1)$, occurring via the quadratic equation \eqref{f1}, one can check that \eqref{q22} is in agreement with \eqref{q2}. In fact, given arbitrary $Q_0$ and $Q_1(x_1)$, the resulting expression is the {\em unique} solution to \eqref{qq} ($k=0$) and \eqref{q2} ($k=1$).

As was shown above, the additional condition of clustering imposes the one-particle form factor to take only two possible values (proportional to the vacuum expectation value), according to \eqref{identification}. For $n=1$ the solution \eqref{q22} is either zero (if we take the first solution in (\ref{identification})) or  it reduces to Zamolodchikov's two particle solution for the Lee-Yang field \cite{Z} (if we take instead the second solution in (\ref{identification})). This is in accordance with identifying the two-particle form factors with those of ${\cal T}$ and $\Tp$, respectively.

Interestingly, it turns out that the above structure subsists to higher particles: given $Q_2(x_1,x_2)$ and $Q_1(x_1)$, there is a unique solution to the kinematic and bound state residue equations \eqref{qq} ($k=1$) and \eqref{q2} ($k=2$) for the polynomial $Q_3(x_1,x_2,x_3)$. The solution has the following structure:
\begin{equation}
Q_3(x_1,x_2,x_3) = A_1 \sigma_1^3\sigma_3 + A_2 \sigma_1^2 \sigma_2^2 + A_3\sigma_1\sigma_2\sigma_3 +A_4\sigma_2^3 + A_5 \sigma_3^2, \label{q3}
\end{equation}
where the parameters $A_i$ are complicated functions of $n$ but rapidity-independent. The detailed computation of $Q_3(x_1,x_2,x_3)$ and the values of $A_i$ are reported in appendix \ref{appendixe}, and note in particular that the polynomials $\sigma_1^6$ and $\sigma_1^4\sigma_2$ have vanishing coefficients.

Again it is interesting to consider the limit $n\rightarrow 1$ of the functions $A_i$ above. Using the two solutions (\ref{identification}), we now note that all constants vanish, $A_i=0$, when we consider that corresponding to the operator $\mathcal{T}$ (where $F_1^{\mathcal{T}|1}=0$ for $n=1$), thus the three particle form factor also vanishes. On the other hand, if we consider the other solution in (\ref{identification}), which at $n=1$ should correspond to the field $\phi$, we find
\beqa
A_1=A_4=A_5=0, \quad A_2 =-A_3=\frac{(F_1^{\phi})^2 H_1(1)}{\langle\phi\rangle}=\frac{i \pi m^2 3^{1/4}}{2^{7/2}f(i\pi,1)^{3/2}}, 
\eeqa
and a simple computation shows that our three-particle form factor, as expected, reduces to Zamolodchikov's solution \cite{Z}.

It is tempting to use this benchmark (agreement with Zamolodchikov's solutions) to try and find the general solution for higher particle numbers. However, as the three-particle case shows, the reduction to $n=1$ occurs thanks to great simplifications. At this stage, it is unfortunately not obvious at all how high-particle solutions may be constructed other than by brute force computation. The main reason for this is the presence of two (rather than one) kinematic pole in the form factor ansatz (\ref{ansatz2}). This leads to polynomials $Q_k^{\mathcal{O}}(x_1, \ldots, x_k)$ of much higher degree than is the case in the standard form factor program.

Despite the complexity of the expression (\ref{q3}), there are certain simplifications that can be used to rewrite the three-particle form factor in a form which is more suitable for numerical computations. It turns out that 
\begin{eqnarray}
F_3^{\mathcal{O}|111}(\theta_1,\theta_2,\theta_3)&=&f_3(x_{1},...,x_{k}) \prod_{i<j}^{3}
\frac{F_{\text{min}}(\theta_i-\theta_j)}{(x_i-\alpha
x_j)(x_j-\alpha x_i)} \nonumber\\
&& - \frac{(F_1^{\mathcal{O}|1})^2\langle \mathcal{O}\rangle^{-1}H_1(n)}{4 \alpha\sin
   \left(\frac{\pi }{6 n}\right) \sin \left(\frac{5 \pi }{6
   n}\right)}\prod_{i<j}^{3}F_{\text{min}}(\theta_i-\theta_j),
\end{eqnarray}
where $f_3(\theta_1,\theta_2,\theta_3)$ is the function that is obtained from $Q_3(\theta_1,\theta_2,\theta_3)$ in (\ref{q3}) 
by setting all terms proportional to $\langle \mathcal{O}\rangle^{-1}$ to zero. In other words, when divided by $\prod_{i<j}(x_i-\alpha
x_j)(x_j-\alpha x_i)$, all those terms simplify giving just the second summand in the formula above. This summand represents a kernel solution to the form factor equations, in the sense already described in subsection 3.2.

Finally, note that
\beqa
\lim_{\theta_1 \rightarrow \infty}F_3^{\mathcal{O}|111}(\theta_1,\theta_2,\theta_3) &=& \frac{\alpha^{-1} F_1^{\mathcal{O}|1} C_0(n)}{4\sin
   \left(\frac{\pi }{6 n}\right) \sin \left(\frac{5 \pi }{6
   n}\right)} \frac{\left( 4\cos ^2\left(\frac{\pi }{3 n}\right) x_2 x_3 -  (x_2+x_3)^2 \right)F_{\text{min}}(\theta_2-\theta_3)}{(x_2-\alpha x_3)(x_3-\alpha x_2)}\nonumber\\
   && - \frac{(F_1^{\mathcal{O}|1})^2\langle \mathcal{O}\rangle^{-1}H_1(n)}{4 \alpha\sin
   \left(\frac{\pi }{6 n}\right) \sin \left(\frac{5 \pi }{6
   n}\right)}F_{\text{min}}(\theta_2-\theta_3)=\frac{F_1^{\mathcal{O}|1} F_2^{\mathcal{O}|11}(\theta_2-\theta_3)}{\langle \mathcal{O}\rangle}, \label{clus1}\eeqa 
where we have used the property
\beq
F_1^{\mathcal{O}|1}H_1(n)=\langle \mathcal{O}\rangle\alpha^{-1}C_0(n)-4\alpha \langle \mathcal{O}\rangle^{-1}(F_1^{\mathcal{O}|1})^2 \sin\left( \frac{\pi}{6n}\right)\sin\left( \frac{5\pi}{6n}\right),
\eeq
which can easily be derived from (\ref{f1}), (\ref{cn}) and (\ref{hh2}). In other words, the three-particle solution automatically satisfies the clustering property. This is an extremely nontrivial check of the validity of the three-particle solution. This situation is in contrast to that of the sinh-Gordon model \cite{higher}, where at each particle number, the clustering property has to be imposed in order to uniquely fix the solution. It also follows from the result above and the cluster property of the two-particle form factor that
\beq
\lim_{\theta_1,\theta_2 \rightarrow \infty} F_3^{\mathcal{O}|111}(\theta_1,\theta_2,\theta_3)=\frac{(F_1^{\mathcal{O}|1})^3}{\langle \mathcal{O}\rangle^2}. \label{clus2}
\eeq
Properties (\ref{clus1}) and (\ref{clus2}) are very important as they insure the convergence of the integrals (\ref{ent2}) for $k=3$.

\subsection{Form factors of the fields $\tilde{\TT}$ and $\bTp$}
\label{ssfftt}

In the previous subsections we have concentrated our analysis on computing the form factors of the fields $\TT$ and $\Tp$. However, the correlators we are interested in also involve the fields $\tilde{\TT}$ and $\bTp$ thus their form factors are also required. In fact the form factors of all these fields are not independent from each other. We may think of $\TT$ and ${\Tp}$ and of $\tilde{\TT}$ and $\bTp$ as twist fields associated to the two opposite cyclic permutation symmetries $ i \mapsto i + 1$ and $ i + 1 \mapsto i$ ($i=1,\ldots,n, n+1\equiv 1$). From the additional symmetry under the inversion of copy numbers it follows that
\beq
F_k^{\TT|\mu_1\ldots \mu_k}(\theta_1,\cdots,\theta_k)=
F_k^{\tilde{\TT}|(n-\mu_1)\ldots (n-\mu_k)}(\theta_1,\cdots,\theta_k),
\eeq
and similarly for $\Tp$ and $\bTp$. At the same time, as already explained in subsection \ref{ssff}, from the non-unitarity of the theory we would expect that
\beqa
\left[\bra\TT\ket^{-1}\,F_k^{\TT|\mu_1\ldots \mu_k}(\theta_1,\cdots,\theta_k)\right]^*&=&(-1)^k\, \bra\TT\ket^{-1}\,F_k^{\tilde{\TT}|\mu_1\ldots \mu_k}(\theta_k,\cdots,\theta_1)\nonumber\\
&=& (-1)^k \,\bra\TT\ket^{-1}\,F_k^{\TT|(n-\mu_1)\ldots (n-\mu_k)}(\theta_k,\cdots,\theta_1)
\eeqa
(note that $\bra\TT\ket = \bra \t\TT\ket$).
These equations both define the form factors of $\tilde{\TT}$ and impose the condition expressed by the last equality above on the form factors of $\TT$. We have verified that this is satisfied for all our solutions, and that similar equations hold for $\Tp$. These equations are the counter-part of \eqref{phiadd} for twist fields, and show that the form factor expansion \eqref{entOr} is correct.

Finally, another important relation which we have used in subsequent computations is the following identity
\beq 
F_k^{\TT|\mu_1\ldots \mu_k}(\theta_1,\cdots,\theta_k)=F_k^{\TT|1\ldots 1}(\theta_1+2\pi i (\mu_1-1),\cdots,\theta_k+2\pi i(\mu_k-1)) \quad \mu_1<\ldots<\mu_k,
\eeq
which allows us to express any form factor in terms of form factors involving only the particle living in copy 1. The same equation holds for the field $\Tp$.

\section{Identification of twist field operators: numerical results}
\label{4}

In previous sections we have provided compelling evidence for the identification of the two families of form factor solutions that we have obtained with the twist fields $\mathcal{T}$ and $:\mathcal{T}\phi:$. This evidence is based on the (highly non-trivial) fact that the one-particle and higher form factors of the field we identified as $\mathcal{T}$ vanish at $n=1$ whereas those of $:\mathcal{T}\phi:$ reduce to the form factors of $\phi$ obtained in \cite{Z}. Further evidence may be gathered by, for example, examining the short distance behaviour of the correlators $\langle \mathcal{T}(r)\tilde{\mathcal{T}}(0) \rangle$ and $\langle :\mathcal{T}\phi:(r):\tilde{\mathcal{T}}\phi:(0) \rangle$. We must therefore first understand what the expected behaviour of such correlators should be for the theory at hand. 

Let us first consider the conformal field theory. In CFT such correlators are expected to {\textit{converge}} at small distances as
\beq
\langle \mathcal{T}(r)\tilde{\mathcal{T}}(0) \rangle_{\rm CFT} = r^{-4 \Delta_{\mathcal{T}}},
\eeq
and 
\beq
\langle :\mathcal{T}\phi:(r):\tilde{\mathcal{T}}\phi:(0) \rangle_{\rm CFT} = r^{-4 \Delta_{:\mathcal{T}\phi:}}.
\eeq
Indeed note that the powers above are positive for the Lee-Yang model as both $c$ and $\Delta$ are negative (see section \ref{2}). This is of course a consequence of non-unitarity. 

In the massive theory, however, we expect that the leading short distance behaviours of these correlators should be described by a {\rm different} power law:
\beq
\langle \mathcal{T}(r)\tilde{\mathcal{T}}(0) \rangle \propto r^{-4 \Delta_{\mathcal{T}}{+}2n \Delta},\label{b1}
\eeq
and 
\beq
\langle :\mathcal{T}\phi:(r):\tilde{\mathcal{T}}\phi:(0) \rangle \propto r^{-4 \Delta_{:\mathcal{T}\phi:}{+}2 n \Delta}.\label{b2}
\eeq

The reason for this is entirely analogous to the observation made in \cite{Z} regarding the correlator $\langle \phi(r) \phi(0) \rangle$. It was found that for short distance in the massive theory the leading behaviour of this correlator was $r^{-2 \Delta}$ rather than the conformal behaviour $r^{-4 \Delta}$. Zamolodchikov argued that this was due to the fact that the leading behaviour of the conformal OPE comes from the field $\phi$ rather than the identity. In the massive theory the expectation value $\langle \phi \rangle \neq 0$ and therefore the contribution to the OPE from the field $\phi$ itself becomes the dominating term in the short distance expansion of the two-point function.

Similarly, it is possible to argue that the leading contribution to the OPEs of $\mathcal{T}$ and $\tilde{\mathcal{T}}$ and of $:\mathcal{T}\phi:$ and $:\tilde{\mathcal{T}}\phi:$ corresponds to the field $\phi_1 \phi_2 \ldots \phi_n$ where $\phi_i$ represents the field $\phi$ on copy $i$. This field has dimension $n \Delta$, and it is the field of smallest (most negative) conformal dimension that can be constructed in the $n$-copy Lee-Yang model. Since its expectation value is nonzero, it thus gives the leading contribution at short distances. Massive OPEs of twist fields will be discussed in more detail in section~\ref{PCFT}.

Thus, by employing a form factor expansion we may check whether the expected behaviours are indeed recovered from our form factor solutions. We will include up to three particle form factors as done in \cite{Z}. We have performed a numerical evaluation of the formula (\ref{del}) including up to three particle form factors for the twist fields $\mathcal{T}$ and $:\mathcal{T}\phi:$. We confirm with good accuracy that the twist fields exhibit the behaviours (\ref{b1}) and (\ref{b2}) for $mr \ll 1$. This means that (\ref{del}) holds with $x_{\TT}=\Delta_{\mathcal{T}}- n \Delta/2$ and $x_{\Tp}=\Delta_{\mathcal{\Tp}}- n \Delta/2$. The tables and plots below show our numerical results for various $n$ and a comparison to the exact CFT values.

\begin{table}[h!]
\begin{center}
\begin{tabular}{|l|c|c|c|c|c|c|}\hline
  n   & 2  &  3 & 4 & 5 & 8 & 10\\ \hline
  CFT ($-4x_{\mathcal{T}}$) &  $\frac{3}{10}=0.3$  & $\frac{34}{45}=0.756$ &$\frac{23}{20}=1.15$ &$\frac{38}{25}=1.52$ & $\frac{103}{40}=2.575$ &$\frac{163}{50}=3.26$\\ \hline
  1-particle & $0.209643$  & $0.442562$ & $0.656773$ & $0.861066$ & $1.44896$ & $1.83206$\\ \hline
  1+2-particles & $0.259028$  & $0.564549$ & $0.842992$ & $1.10754$ &$1.86697$ & $2.3611$ \\ \hline
  1+2+3-particles & $0.279487$ & $0.625075$ & $0.937636$ & $1.23376$ &$2.13554$ &$2.70666$\\ \hline
\end{tabular}
\caption{Study of the two-point function $\langle \mathcal{T}(r)\tilde{\mathcal{T}}(0) \rangle $ of the $n$-copy Lee-Yang theory at short distances. Near the critical point we expect this correlator to exhibit a power-law behaviour of the form $r^{-4 x_{\TT}}$ where $x_{\TT}=\Delta_{\mathcal{T}} - \frac{n\Delta}{2}=-\frac{n}{12}+\frac{11}{60 n}$. This value should be best reproduced in the massive theory the more form factor contributions are added. The data above show that this expectation is indeed met by considering up to three-particle form factors.}
\end{center}
\end{table}

\begin{table}[h!]
\begin{center}
\begin{tabular}{|l|c|c|c|c|c|c|}\hline
  n   & 2  &  3 & 4 & 5 & 8 & 10\\ \hline
  CFT ($-4x_{:\mathcal{T}\phi:}$) & $\frac{7}{10}=0.7 $  & $\frac{46}{45}=1.022$& $\frac{27}{20}=1.35$&$\frac{42}{25}=1.68$ & $\frac{107}{40}=2.675$ & $\frac{167}{50}=3.34$\\ \hline
  1-particle & $0.391185$  & $0.572281$ & $0.756341$ & $0.941564$ & $1.499823$ & $1.87287$\\ \hline
  1+2-particles & $0.505165$ & $0.737822$ & $0.974720$& $1.213628$ &$1.931704$ & $2.41539$ \\ \hline
  1+2+3-particles &$0.575841$ &$0.843472$ & $1.11533$ &$1.38907$ &$2.21646$ &$2.77169$\\ \hline
\end{tabular}
\caption{Study of the two-point function $\langle :\mathcal{T}\phi:(r):\tilde{\mathcal{T}}\phi:(0) \rangle $ of the $n$-copy Lee-Yang theory at short distances. Near the critical point we expect this correlator to exhibit a power-law behaviour of the form $r^{-4 x_{\Tp}}$ where $x_{\Tp}=\Delta_{:\mathcal{T}\phi:} - \frac{n\Delta}{2}=-\frac{n}{12}-\frac{1}{60 n}$. The data above show good agreement with CFT by considering up to three-particle form factors.}
\end{center}
\end{table}
\begin{figure}[h!]
 \begin{center} 
 \includegraphics[width=7.92cm]{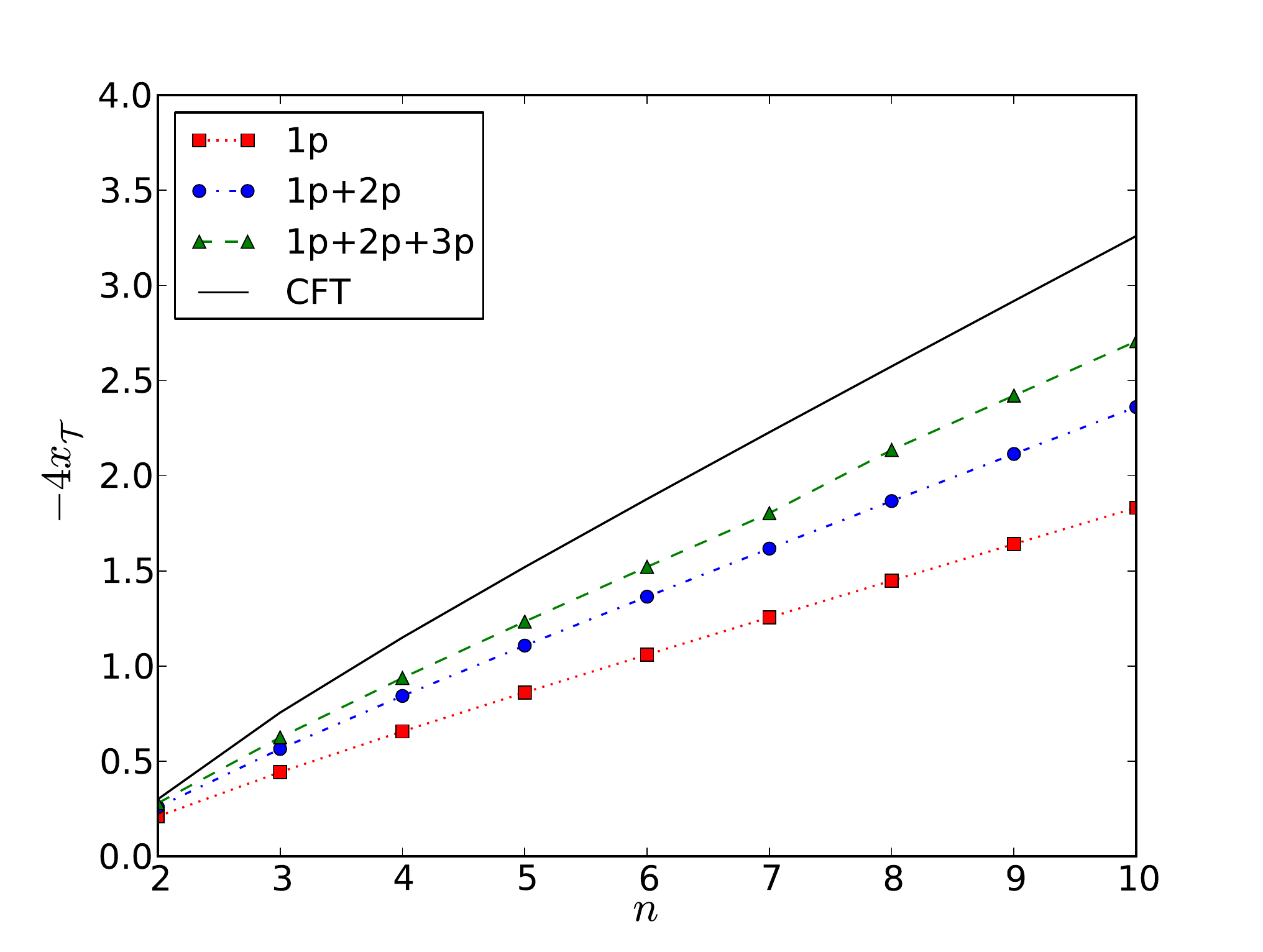} 
 \includegraphics[width=7.92cm]{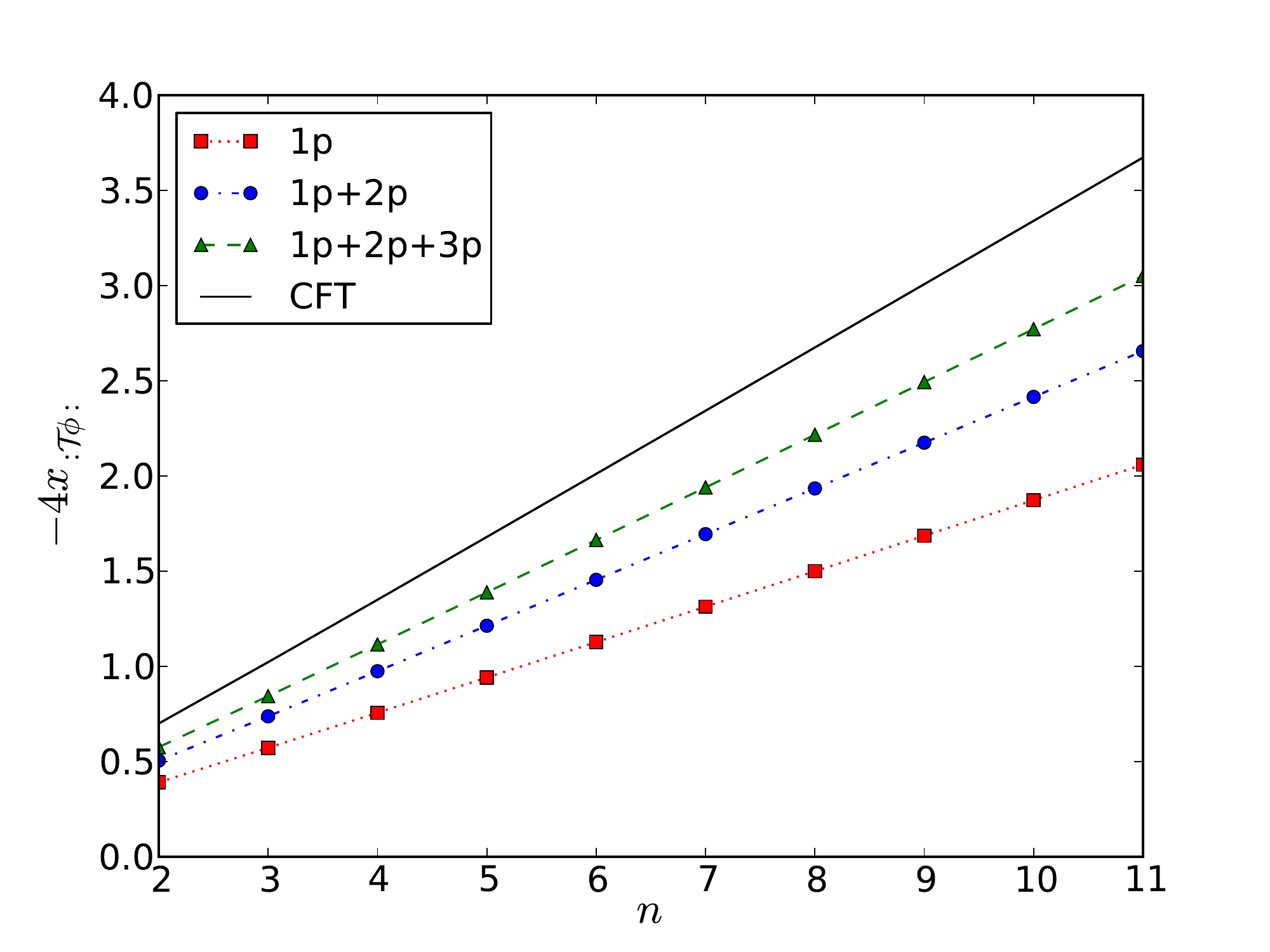} 
 \end{center} 
 \caption{Graphical representation of $-4x_{\TT}$ and $-4 x_{\Tp}$ for $n\leq 11$. The squares, circles and triangles, represent the up to one-, two- and three-particle form factor contributions. The black solid line represents the exact values at criticality. All curves clearly show strong linearity in $n$ which is consistent with the CFT behaviour, where the coefficient of $n$ (e.g.~slope of the curves) approaches the CFT value as more form factor contributions are added. The agreement with CFT gets worse as $n$ increases. This is also to be expected as the larger $n$ is, the larger the contribution of higher particle form factors becomes (all form factor contributions are in fact proportional to $n$).} 
 \label{deltan} 
 \end{figure}

\section{Comparison with perturbed conformal field theory results}\label{PCFT}

A further consistency check of our form factor solutions may be carried out by comparing a form factor expansion of the correlators $\bra\TT(x_1)\tilde{\TT}(x_2) \ket$ and $\bra \Tp (x_1) \bTp (x_2)\ket$ to its counterpart in perturbed conformal field theory. 

\subsection{Conformal perturbation theory and twist fields structure constants}

We may regard the action of the integrable quantum field theory as a perturbation of Lee-Yang CFT action by a term proportional to a coupling constant $\lambda$ and the CFT field $\phi(x,\bar{x})$ of conformal dimension $\Delta$,
\beq
S_{IQFT}=S_{CFT} + i\lambda \int d^2x \,\phi(x,\bar{x}),
\eeq
and compute correlators by performing perturbation theory about the conformal critical point on the coupling $\lambda$ \cite{Zamolodchikov:1989zs}. As is well-known, in the massive theory this coupling is related to the mass scale $m$ as $\lambda \propto m^{2-2\Delta}$ (with a known proportionality factor \cite{tba1}). The massive correlators can then be obtained by using OPEs where conformal structure constants are modified to structure functions of $mr$ that can be evaluated perturbatively in $\lambda$ (they are convergent series in integer powers of $\lambda$), and where vacuum expectation values, which are non-perturbative in $\lambda$, are nonzero. The same type of comparison between a form factor and a perturbed CFT computation was carried out in \cite{Z} for the two-point function of the field $\phi$ in Lee-Yang.

Let us now consider the OPEs of $\TT$ with $\tilde{\TT}$ and of $\Tp$ with $\bTp$. They involve only fields in the non-twisted sector (we mean by this all fields constructed by considering $n$ non-interacting copies of the fields of the original theory) and by construction they must be invariant under cyclic permutation of the copies. Let us consider the following primary, cyclically invariant, homogeneous fields, composed of multilinears in the fields $\phi_i$ on the various copies: we label them by sets $\{k_1,\ldots,k_J\}$ of $J$ different integers in $[1,n]$ for $J=1,2,\ldots,n$ (we may take $k_1<\cdots<k_J$), and take them to be
\beq
	\Phi_{k_1,\ldots,k_J} := \frc{\phi_{k_1}\cdots \phi_{k_J} + \mbox{cyclic permutations}}{{\cal S}_{k_1,\ldots,k_J}}.
\eeq
The symmetry factor ${\cal S}_{k_1,\ldots,k_J}$ is equal to the order of the subgroup of the cyclic replica permutations which preserve the sequence $k_1,k_2,\ldots,k_J$ of replica indices. That is, $\Phi_{k_1,\ldots,k_J}$ is the sum, over all  elements $\sigma\in\Z_n$ in the cyclic replica permutation group $\Z_n$, of $\sigma(\phi_{k_1}\cdots \phi_{k_J})$, divided by the order of the stabilizer, in $\Z_n$, of $\phi_{k_1}\cdots \phi_{k_J}$. This definition guarantees that in $\Phi_{k_1,\ldots,k_J}$, every independent multilinear term, including the initial term $\phi_{k_1}\cdots\phi_{k_J}$ itself, appears with coefficient 1. The number of independent multilinears in $\Phi_{k_1,\ldots,k_J}$ is $n/{\cal S}_{k_1,\ldots,k_J}$. The symmetry factors for low values of $J$ can be written explicitly:
\beqa
	{\cal S}_{1,k}&=&
\lt\{\ba{ll} 2 & (n\ \mbox{even},\ k=n/2+1) \\ 1 & (\mbox{otherwise})\ea\rt.\n
	{\cal S}_{1,k,j} &=& \lt\{\ba{ll}3 & (k-1 = j-k = n+1-j)\\
1 & (\mbox{otherwise}). \ea\rt.\n
	{\cal S}_{1,k,j,p} &=& \lt\{\ba{ll}
    4 & (k-1=j-k=p-j=n+1-p) \\
    2 & (k-1=p-j\neq j-k=n+1-p) \\
    1 & (\mbox{otherwise})
    \ea\rt.
\eeqa

The fields $\Phi_{k_1,\ldots,k_J}$ have conformal dimensions $J\Delta$. In order to have a basis of primary, cyclically invariant homogeneous fields of dimension $J\Delta$, we need to further restrict the indices $k_1,\ldots,k_J$. We may certainly fix $k_1=1$, and further restrictions hold due to the residual equivalence relation generated by $\{1,\ldots,k_J\} \sim \{1,n+2-k_J,n+1+k_2-k_J,\ldots,n+1+k_{J-1}-k_J\}$. More generally, in the set of all replica-index sets $\{k_1,\ldots,k_J\}$, there is a foliation by $\Z_n$ orbits, and a basis of fields $\Phi_{k_1,\ldots,k_J}$ can be taken as fields parametrised by single representatives of each $\Z_n$ orbit.

Let us give simple examples. For $J=1$, we have $\Phi_1 = \sum_{j=1}^n \phi_j$. For $J=2$ the basis is $\Phi_{1,2}=\phi_1\phi_2$ + all $n-1$ cyclic permutations, $\Phi_{1,3}=\phi_1\phi_3$ + all $n-1$ cyclic permutations, etc. until  $\Phi_{1,[n/2]+1}=
\phi_1\phi_{[n/2]+1}$ + all $n-1$ cyclic permutations (if $n$ is odd), or until $\Phi_{1,n/2+1}=
\phi_1\phi_{n/2+1}$ + all cyclic permutations up to $\phi_{n/2}\phi_n$ (if $n$ is even). In particular for $n=3$, we have $\phi_1\phi_2 + \phi_2\phi_3 + \phi_3\phi_1$ only; for $n=4$, we have $\phi_1\phi_2 + \phi_2\phi_3 + \phi_3\phi_4 + \phi_4\phi_1$ and $\phi_1\phi_3 + \phi_2\phi_4$; etc. There is a unique field at $J=n$: $\Phi_{1,\ldots,n}=\phi_1 \phi_2 \ldots \phi_n$, which has dimension $n\Delta$. As mentioned, this field is very important in non-unitary models since for $\Delta<0$ it provides the leading contribution (for small $r$) to the OPE, as it is the field of lowest conformal dimension.

The OPEs in the massive theory can be regarded as ``deformations'' of the conformal OPEs such that the structure constants are replaced by functions of $mr$. Denoting by $\mathcal{O}$ and $\tilde{\mathcal{O}}$ any given pair of conjugate (i.e. whose twist actions cancel out) twist fields, it takes the form
\beqa
	\mathcal{O}(x_1) \tilde{\mathcal{O}}(x_2) &\sim & r^{-4\Delta_{\mathcal{O}}}\left(C_{\mathcal{O} \tilde{\mathcal{O}}}^{\bf 1}(mr)  {\bf 1}
	+ C_{\mathcal{O}\tilde{\mathcal{O}}}^{\Phi_1}(mr)  r^{2\Delta} \Phi_1(x_2) \right.\nonumber\\
    && \left.+
	\sum_{k=2}^{[n/2]+1} C_{\mathcal{O}\tilde{\mathcal{O}}}^{\Phi_{1,k}}( mr) r^{4\Delta}
	\Phi_{1,k}(x_2) + \ldots + C_{\mathcal{O}\tilde{\mathcal{O}}}^{\Phi_{1,\ldots,n }}(mr) r^{2n\Delta}
	\Phi_{1,\dots,n} (x_2)\right)\nonumber\\
    && + \text{Virasoro descendants},
\eeqa
where $r:=|x_1-x_2|$, $m$ is the physical mass of the Lee-Yang model. The functions 
\beq
 C_{\mathcal{O}\tilde{\mathcal{O}}}^{\Phi_{k_1,\ldots,k_p}}(mr)=  \tilde{C}_{\mathcal{O}\tilde{\mathcal{O}}}^{\Phi_{k_1,\ldots,k_p}}\left(1+ C_1^{\Phi_{k_1,\ldots,k_p}}(mr)^{2-2\Delta}+ C_2^{\Phi_{k_1,\ldots,k_p}} (mr)^{2(2-2\Delta)} + \cdots \right),
 \eeq
admit an expansion in integer powers of the coupling $\lambda$, hence in powers of $(mr)^{2-2\Delta}$, and the constants $\tilde{C}_{\mathcal{O}\tilde{\mathcal{O}}}^{\Phi_{k_1,\ldots,k_p}}$ are the structure constants of the CFT. In our analysis we will in fact only consider the leading term (the CFT contribution) to these structure functions, that is, we will only carry out zeroth order perturbation theory whereby the mass dependence is introduced through the non-vanishing expectation values of OPE fields. The analysis is still non-trivial because of the presence of nonzero expectation values. Note that with the definition (\ref{newfield}) and the standard definition of $\TT$ and $\tilde{\TT}$ we have the conformal normalization
\beq
 \tilde{C}_{\TT \tilde{\TT}}^{\bf 1}=\tilde{C}_{\Tp \bTp}^{\bf 1}=1.
\eeq

The conformal OPEs (and structure constants) of the branch point twist field $\TT$ have been studied in several places in the literature. The most general study can be found in Appendix A of \cite{Headrick} where general formulae for the structure constants associated to the OPE of $\TT$ with $\tilde{\TT}$ in general (unitary) CFT are given. Structure constants have also played an important role within the study of the entanglement of disconnected regions \cite{CardyMultiple2, Gliozzi}. More recently the structure constants of other types of twist fields which arise naturally within the study of the negativity have been studied in \cite{FTneg,Hoogeveen:2014bqa}. However we do not know of any studies of the OPE and structure constants of composite fields such as $\Tp$. 
Here we provide explicit step-by-step computations of the conformal structure constants $\tilde{C}_{\TT \tilde{\TT}}^{\Phi_1}$, $\tilde{C}_{\Tp \bTp}^{\Phi_1}$, $\tilde{C}_{\TT \tilde{\TT}}^{\Phi_{1,k}}$, $\tilde{C}_{\Tp \bTp}^{\Phi_{1,k}}$, $\tilde{C}_{\TT \tilde{\TT}}^{\Phi_{1,k,j}}$ and $\tilde{C}_{\TT \tilde{\TT}}^{\Phi_{1,k,j,p}}$ (see appendix \ref{appendixd} for details) which are proportional to one-, two-, three- and four-point functions of the field $\phi$ (other structure constants would involve higher-point functions, which are harder to access). Our computations focus on the Lee-Yang model but could be easily generalized to other minimal models (for the field $\TT$ the ingredients needed for such generalization are already provided in \cite{Headrick}). Other structure constants and massive corrections thereof will involve higher point functions. The results are:
\beqa
\tilde{C}_{\TT \tilde{\TT}}^{\Phi_{1}}&=&0, \quad \tilde{C}_{\TT \tilde{\TT}}^{\Phi_{1,k}}= n^{-4\Delta}|1-e^{\frac{2\pi i (k-1)}{n}}|^{-4\Delta} \quad \text{for} \quad k>1,\nonumber\\
\tilde{C}_{\TT \tilde{\TT}}^{\Phi_{1,k,j}}&=&n^{-6\Delta} \tilde{C}_{\phi \phi}^{\phi}|(1-e^{\frac{2\pi i (k-1)}{n}})(1-e^{\frac{2\pi i (j-1)}{n}})(1-e^{\frac{2\pi i (j-k)}{n}})|^{-2\Delta} \quad \text{for} \quad j>k>1,\nonumber\\
\tilde{C}_{\TT \tilde{\TT}}^{\Phi_{1,k,j,p}}&=& n^{-8\Delta}\,
\bra \phi(e^{\frac{2\pi i}{n}}) \phi(e^{\frac{2\pi i k}{n}}) \phi(e^{\frac{2\pi i j}{n}})\phi(e^{\frac{2\pi i p}{n}}) \ket\quad \text{for} \quad p>j>k>1,\nonumber\\
	\tilde{C}_{\Tp\bTp}^{\Phi_1} &=& n^{-2\Delta} \tilde{C}_{\phi\phi}^{\phi}, \quad \tilde{C}_{\Tp\bTp}^{\Phi_{1,k}} = n^{-4\Delta}\, \kappa\lt(1-e^{\frac{2\pi i (k-1)}{n}}\rt) \quad \text{for} \quad k>1, \label{struc}
\eeqa
where $\kappa$ is a model-dependent function which characterizes the four-point function of fields $\phi$
\beq
	\bra\phi(x_1)\phi(x_2)\phi(x_3)\phi(x_4)\ket
	= \kappa(\eta) |x_1-x_4|^{-4\Delta} |x_2-x_3|^{-4\Delta},
	\quad \eta = \frc{x_{12}x_{34}}{x_{13}x_{24}}. \label{fourpoint}
\eeq
Other structure constants may be computed in terms of higher-point functions so that in general we expect
\beq
	\tilde{C}_{\TT \tilde{\TT}}^{\Phi_{1,k_2,\ldots,k_{J}}} =
	n^{-2J\Delta} \bra \phi(e^{\frac{2\pi i}{n}}) \phi(e^{\frac{2 \pi i k_2}{n}}) \ldots \phi(e^{\frac{2\pi i k_J}{n}})\ket,
\eeq
and
\beq
	\tilde{C}_{\Tp\bTp}^{\Phi_{1,k_2,\ldots,k_J}} =
	n^{-2J\Delta} \kappa(e^{\frac{2\pi i k_2}{n}},\ldots,e^{\frac{2\pi i k_J}{n}}), \label{nstruc}
\eeq
with
\beq
	\kappa(x_1,x_2,\ldots) = \lim_{y\to\infty}
	|y|^{4\Delta} \bra \phi(0)\phi(1)\phi(y)\phi(x_1)\phi(x_2)\cdots\ket.
\eeq
But for the last line in (\ref{struc}) and for (\ref{nstruc}), all formulae above are particular cases of those given in \cite{Headrick}.

The difficulty of calculating such terms is then reduced to the difficulty of obtaining higher-point functions in CFT. Such higher-point functions will also be required in order to obtain most massive corrections to the CFT structure constants, a problem which we will not be addressing in this work.

\subsection{The case $n=2$}

As explained earlier, obtaining the CFT structure constants becomes a difficult problem for the field $\Tp$ as soon as we consider OPE terms involving products of more than two fields and for the field $\TT$ when we consider products involving more than four fields. For this reason, the case $n=2$ is particularly interesting as in this case the leading contribution to the OPE is given by the bilinear fields $\Phi_{1,2}=2\phi_1 \phi_2$ defined earlier. The leading expansions in the massive theory are
\beqa
{\bra\TT(r) \tilde{\TT}(0)\ket} = 
	r^{-4\Delta_{\TT}}\left(1
	+ 2 \tilde{C}_{\TT \tilde{\TT}}^{\Phi_1} r^{2\Delta} \bra\phi\ket+
	\tilde{C}_{\TT \tilde{\TT}}^{\Phi_{1,2}} r^{4\Delta}
	\bra\phi\ket^2\right)+\cdots
\eeqa
\beqa
{\bra\Tp(r) \bTp(0)\ket} = 
	r^{-4\Delta_{\Tp}}\left(1
	+ 2 \tilde{C}_{\Tp \bTp}^{\Phi_1} r^{2\Delta} \bra\phi\ket+
	\tilde{C}_{\Tp \bTp}^{\Phi_{1,2}} r^{4\Delta}
	\bra\phi\ket^2\right)+\cdots
\eeqa
Here the numerical coefficients arise from the total numbers of independent multilinears in $\Phi_1$ and in $\Phi_{1,2}$ in the case with $n=2$, which are $n/{\cal S}_{1}=2$ and $n/{\cal S}_{1,2}=1$ respectively.  All subleading terms correspond to Virasoro descendants and massive corrections to the structure constants, hence are suppressed by positive powers.

The structure constants are
\beq
\tilde{C}_{\TT \tilde{\TT}}^{\Phi_1} =0, \quad \tilde{C}_{\TT \tilde{\TT}}^{\Phi_{1,2}}=2^{-8 \Delta},\quad
	\tilde{C}_{\Tp\bTp}^{\Phi_1}  =  2^{-2\Delta} \tilde{C}_{\phi\phi}^{\phi},\quad
	\tilde{C}_{\Tp\bTp}^{\Phi_{1,2}} = 2^{-4\Delta} \,\kappa(2).\label{constants}
\eeq
In the Lee-Yang model all the constants (\ref{constants}) can be computed. The CFT  structure constant $\tilde{C}_{\phi\phi}^{\phi}$ can be found for instance in \cite{Z},
\beq
\tilde{C}_{\phi\phi}^{\phi}=\frac{i}{5} \frac{\Gamma(\frac{1}{5})^{\frac{3}{2}}\Gamma(\frac{2}{5})^{\frac{1}{2}}}{\Gamma(\frac{4}{5})^{\frac{3}{2}}\Gamma(\frac{3}{5})^{\frac{1}{2}}}=i (1.91131...).
\eeq
\begin{figure}[h!]
 \begin{center} 
 \includegraphics[width=7.5cm]{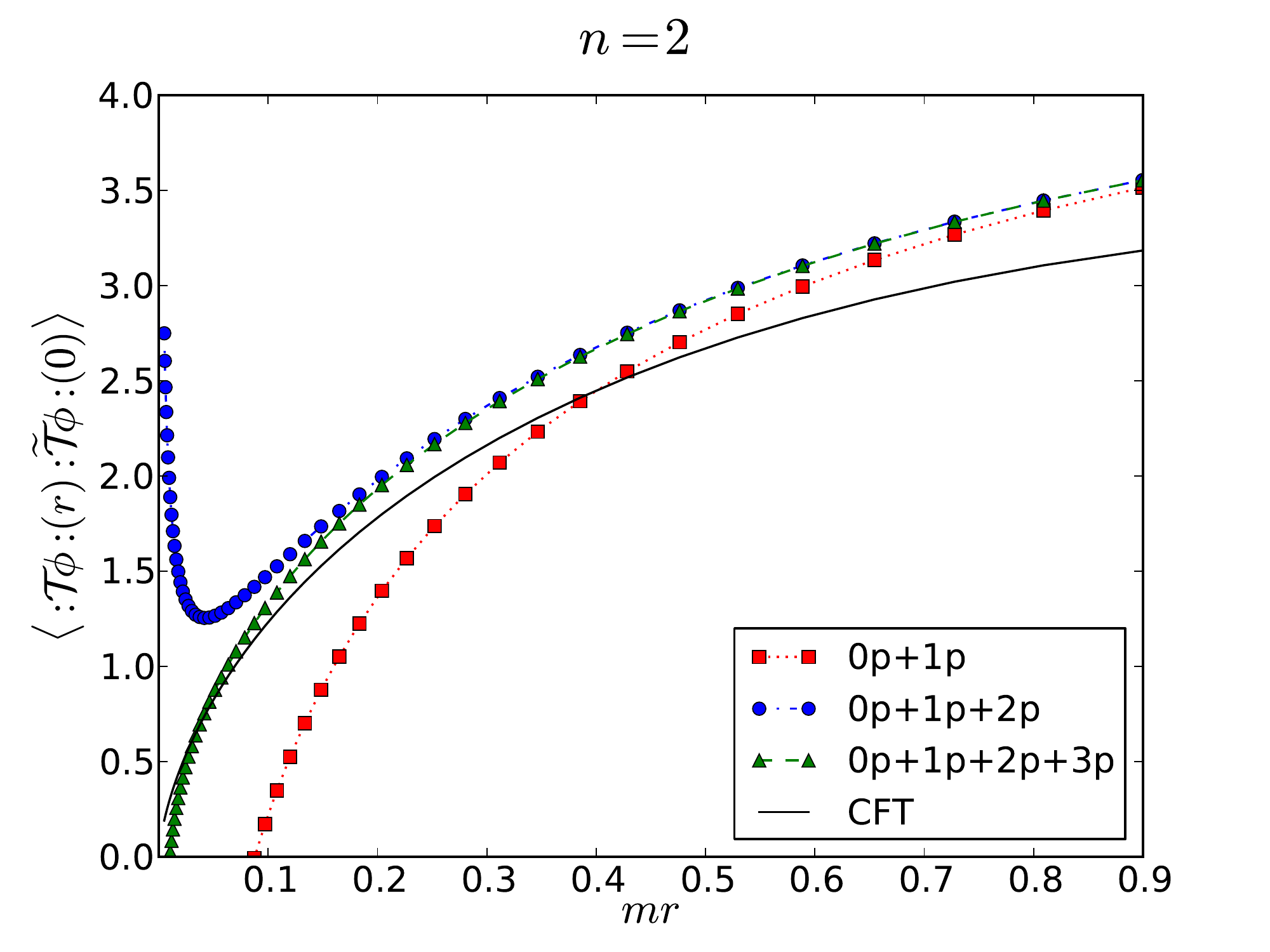} 
 \includegraphics[width=7.5cm]{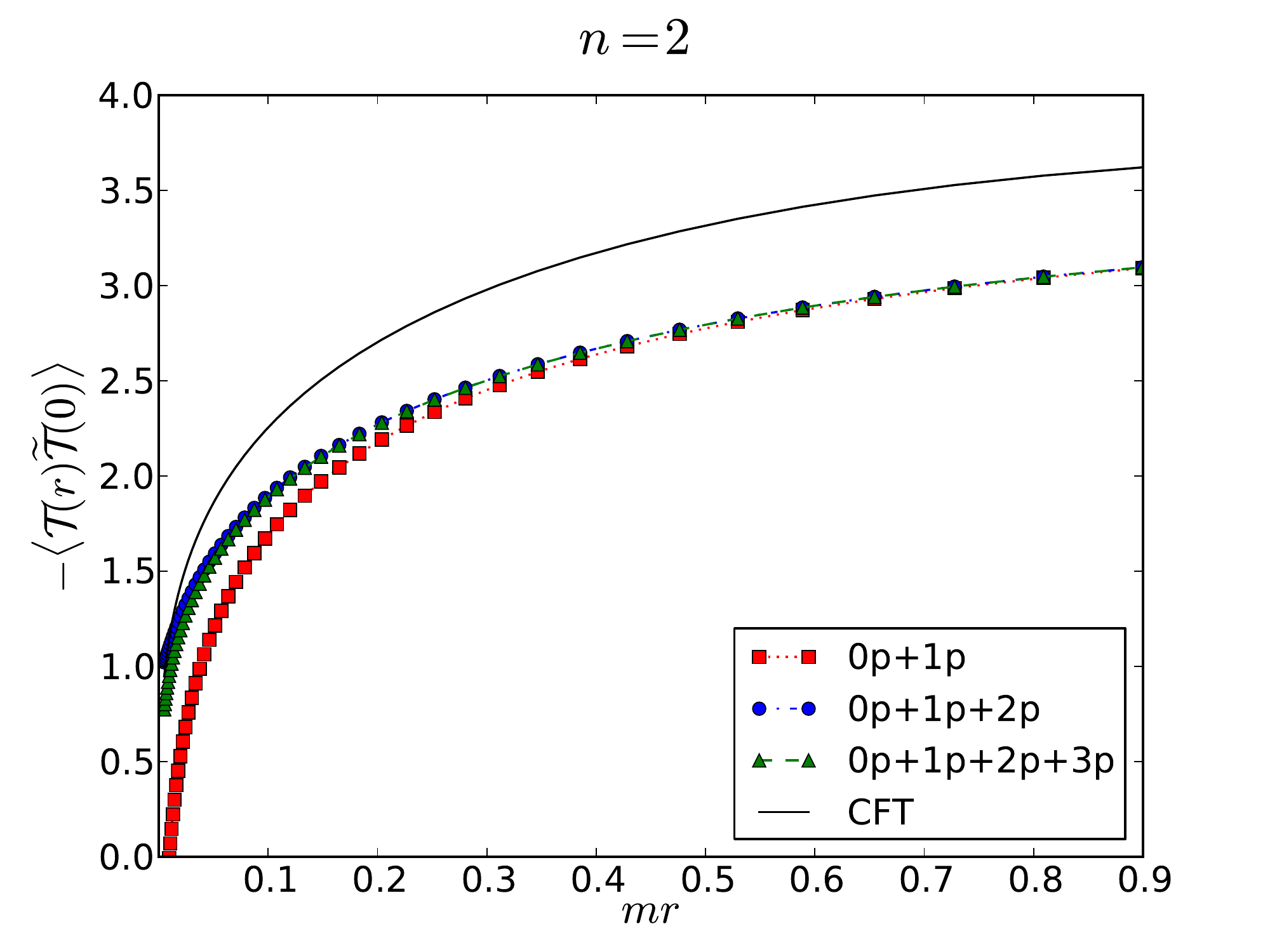}
  \includegraphics[width=7.5cm]{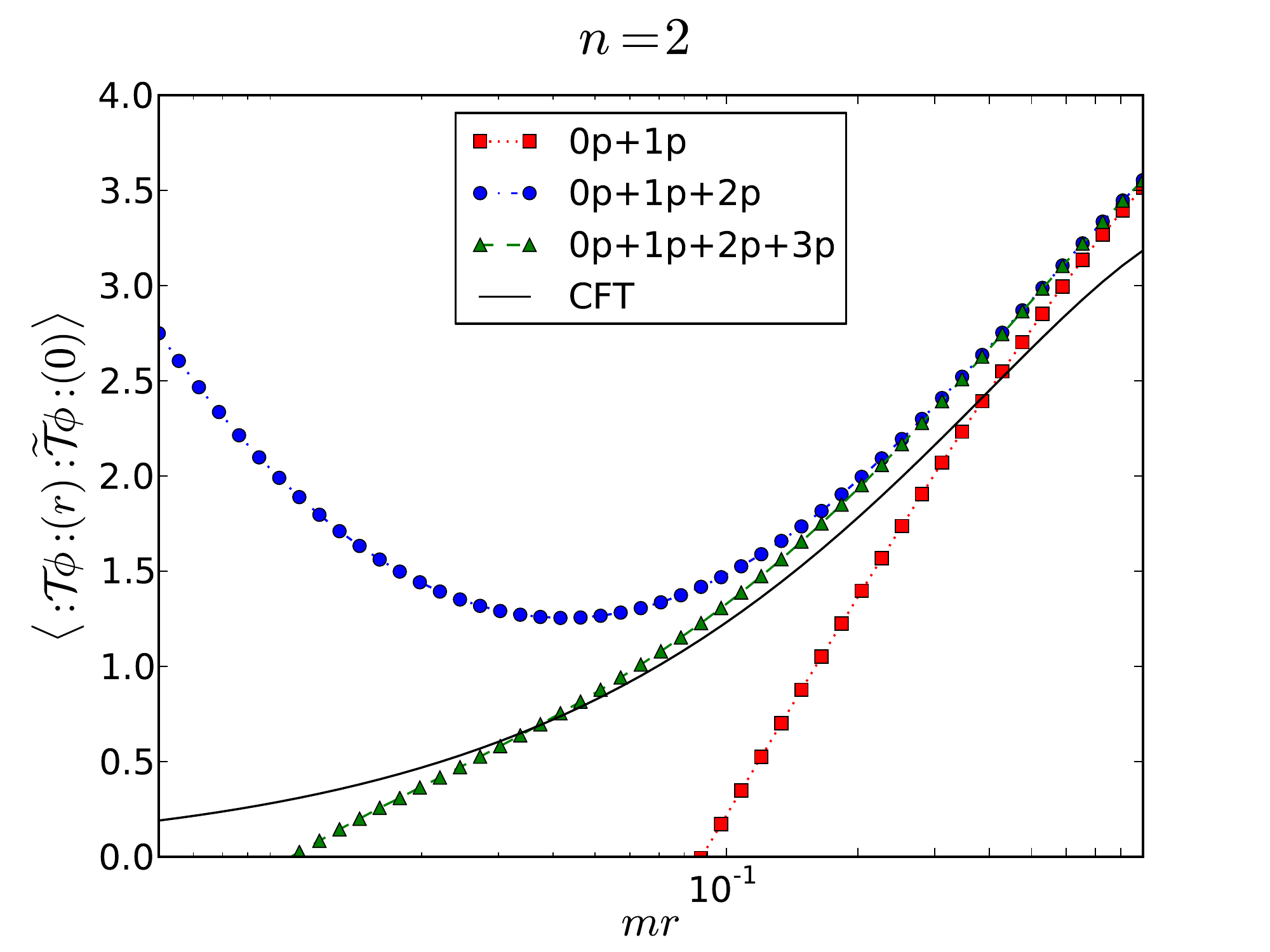} 
 \includegraphics[width=7.5cm]{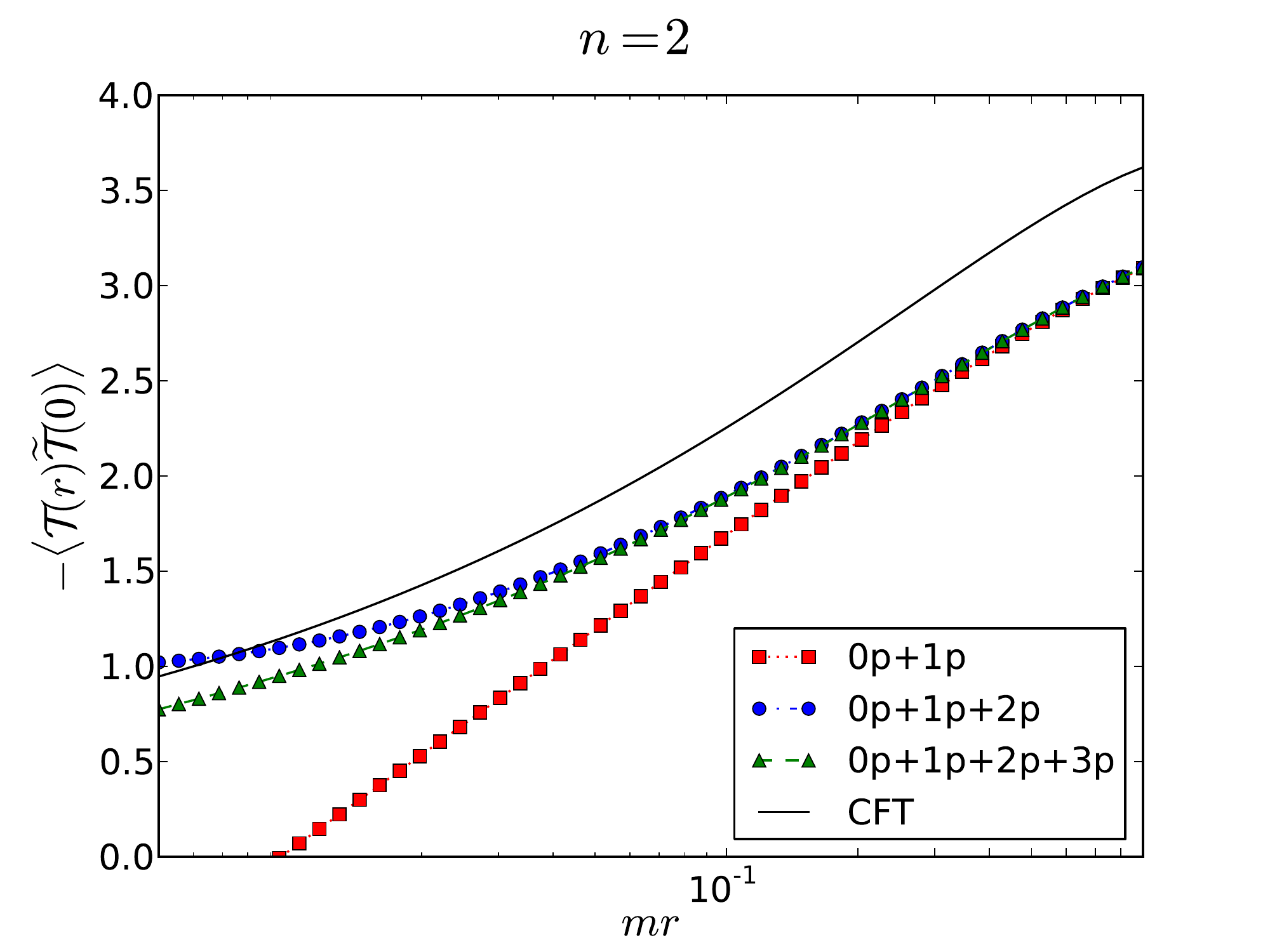}
 \end{center} 
 \caption{Zeroth order perturbed CFT versus form factor computation of the two-point function $\bra \Tp(r) \bTp(0) \ket$ and $-\bra \TT(r)\tilde{\TT}(0)\ket$. Squares, circles and triangles represent contributions up to one-, two- and three-particles to the form factor expansion. For each correlator we present results both in linear and logarithmic scale. As expected, we see that the form factor result (triangles) and the CFT computation (solid line) are in relatively good agreement for small values of $mr$ but quickly drift apart for larger values of $mr$. The range of agreement is seen more clearly by using a logarithmic scale, where we can directly compare the slopes of the form factor and CFT curves.} 
 \label{cute1} 
 \end{figure}
 
The four point function of the Lee-Yang model has been studied in \cite{DF,BKT,LYCardy}. Following \cite{LYCardy} we can write the four point function as in (\ref{fourpoint}) with
    \beq
    \kappa(\eta)= \left|\eta
    \right|^{\frac{4}{5}} (|F_1(\eta)|^2 + C^2 |F_2(\eta)|^2),
    \eeq
   where 
   \beq
  F_1(\eta)={\,}_2F_1\left(\frac{3}{5},\frac{4}{5},\frac{6}{5};\eta\right), \qquad F_2(\eta)=\eta^{-\frac{1}{5}}{\,}_2F_1\left(\frac{3}{5},\frac{2}{5},\frac{4}{5};\eta\right)\quad \text{and} \qquad
   C=\tilde{C}_{\phi\phi}^{\phi}.
   \eeq
   A simple calculation then gives
   \beq
   \kappa(2) = \lim_{y\to\infty}
	|y|^{4\Delta} \bra \phi(-1)\phi(0)\phi(1)\phi(y)\ket=-3.1802...
   \eeq
Plugging these values in (\ref{constants}) as well as the expectation value (\ref{ratio01}) we obtain
\beq
{\bra\TT(r) \tilde{\TT}(0)\ket}=r^{\frac{11}{10}}\left(1-(4.6566...) (mr)^{-\frac{4}{5}} \right)+\cdots,\label{for1}
\eeq
and
\beqa
 {\bra\Tp(r) \bTp(0)\ket} =
	r^{\frac{3}{2}}\left(1
	-(6.2515...) (mr)^{-\frac{2}{5}}+ (8.5055...) (mr)^{-\frac{4}{5}}\right)+\cdots,\label{for}
    \eeqa
which gives an approximation of the two-point function at zeroth order in perturbed CFT. 

In order to compare this with the form factor expansion, we need to fix the vacuum expectation values of $\TT$ and $\Tp$. Recall that we used the CFT normalization to set the coefficients of $r^{\frac{11}{10}}$ and $r^{\frac{3}{2}}$, respectively equal 1. This in principle uniquely fixes the expectation value. Although the resulting expectation value is not known explicitly, we may use the form factor expansion (\ref{vev}) to estimate it. 
From CFT the leading behaviours should be
\beq
\frc{\bra\TT(r) \tilde{\TT}(0)\ket}{\bra\TT\ket^2}
	\stackrel{{r \rightarrow 0}}\sim
	\frc{C_{\TT \tilde{\TT}}^{\Phi_{1,2}}\bra\phi\ket^2}{\bra\TT\ket^2}
	r^{4(\Delta-\Delta_{\TT})}=-\frac{(4.6566...) r^{\frac{11}{10}}}{\bra\TT\ket^2} (mr)^{-\frac{4}{5}}\label{vev3}
\eeq
and
\beq
	\frc{\bra\Tp(r) \bTp(0)\ket}{\bra\Tp\ket^2}
	\stackrel{{r \rightarrow 0}}\sim
	\frc{C_{\Tp \bTp}^{\Phi_{1,2}}\bra\phi\ket^2}{\bra\Tp\ket^2}
	r^{4(\Delta-\Delta_{\Tp})}=\frac{(8.5055...) r^{\frac{3}{2}}}{\bra\Tp\ket^2} (mr)^{-\frac{4}{5}}.\label{vev2}
\eeq
We observe from Figure \ref{cute1} that the truncated form factor expansions of the two-point functions potentially change sign (pass by the value 0) only at short distances, at positions that become smaller as more particles are added. Since they approach the CFT form at short distances, this implies that the full two-point function never becomes zero. Since the ratios $\frc{\bra\TT(r) \tilde{\TT}(0)\ket}{\bra\TT\ket^2}$ and $\frc{\bra\Tp(r) \bTp(0)\ket}{\bra\Tp\ket^2}$ tend to unity at large distances, they are then positive for all values of $r$. Hence, we find $\bra\TT\ket^2<0$ and $\bra\Tp\ket^2>0$.

In fact, from (\ref{vev3}) and (\ref{vev2}), we have that
\beq
\bra\TT\ket^2 = -\frc{(4.6566...)m^{-\frac{11}{10}}}{K_{\TT}},\qquad
	\bra\Tp\ket^2 = \frc{(8.5055...)m^{-\frac{3}{2}}}{K_{\Tp}}.
\eeq
The constants $K_\TT$ and $K_{\Tp}$ as expressed in \eqref{vev} are necessarily positive, and the fact that two-point functions never become zero is related to the convergence of the series \eqref{vev}. A numerical evaluation of (\ref{vev}) including up to three-particle form factors yields
\beq
	K_{\TT}\approx {1.35236},\quad K_{\Tp}\approx {1.95908}.
\eeq
Therefore
\beq
\bra\TT\ket^2\approx -3.443\, m^{-\frac{11}{10}} \quad \text{for} \quad n=2, \label{vev4}
\eeq 
\beq
\bra\Tp\ket^2\approx 4.342 \, m^{-\frac{3}{2}} \quad \text{for} \quad n=2. \label{vev4}
\eeq 
Employing these (approximate) values in our form factor expansion we can now compare it to the functions (\ref{for1}) and (\ref{for}). The results are depicted in Figures~\ref{cute1}.                                                                                                                                                                             
\subsection{The cases $n=3$ and $n=4$}
For the fields $\TT$ and $\tilde{\TT}$ it is also possible to compute the two-point function in the zeroth order approximation for $n=3, 4$. It is given by
\beqa
{\bra\TT(r) \tilde{\TT}(0)\ket} = 
	r^{-4\Delta_{\TT}}\left(1
	+ 3 \tilde{C}_{\TT \tilde{\TT}}^{\Phi_1} r^{2\Delta} \bra\phi\ket+
	3\tilde{C}_{\TT \tilde{\TT}}^{\Phi_{1,2}}r^{4\Delta}
	\bra\phi\ket^2+ \tilde{C}_{\TT \tilde{\TT}}^{\Phi_{1,2,3}}\bra \phi\ket^3 r^{6\Delta}\right)+\cdots,
\eeqa
for $n=3$ (where we have used the numerical coefficients $n/{\cal S}_{1,2} = 3$ and $n/{\cal S}_{1,2,3}=1$), with
\beq
\tilde{C}_{\TT \tilde{\TT}}^{\Phi_1}=0,\quad \tilde{C}_{\TT \tilde{\TT}}^{\Phi_{1,2}}=3^{-6\Delta}, \quad \tilde{C}_{\TT \tilde{\TT}}^{\Phi_{1,2,3}}=3^{-9\Delta} \tilde{C}_{\phi \phi}^\phi,
\eeq
giving
\beq
{\bra\TT(r) \tilde{\TT}(0)\ket}=r^{\frac{88}{45}}\left(1-(17.2221...) (mr)^{-\frac{4}{5}}+ (26.2893...) (mr)^{-\frac{6}{5}} \right)+\cdots \quad \text{for}\quad n=3;
\eeq
whereas for $n=4$ we have
\beqa
{\bra\TT(r) \tilde{\TT}(0)\ket} &=& 
	r^{-4\Delta_{\TT}}\left(1
	+ 4 \tilde{C}_{\TT \tilde{\TT}}^{\Phi_1} r^{2\Delta} \bra\phi\ket+
	(4\tilde{C}_{\TT \tilde{\TT}}^{\Phi_{1,2}}+2\tilde{C}_{\TT \tilde{\TT}}^{\Phi_{1,3}})r^{4\Delta}
	\bra\phi\ket^2\right. \nonumber\\
    && \left. + 4\tilde{C}_{\TT \tilde{\TT}}^{\Phi_{1,2,3}}r^{6\Delta}\bra \phi\ket^3+ \tilde{C}_{\TT \tilde{\TT}}^{\Phi_{1,2,3,4}}r^{8\Delta}\bra \phi \ket^4\right) \cdots,
\eeqa
with
\begin{figure}[h!]
 \begin{center} 
 \includegraphics[width=7.5cm]{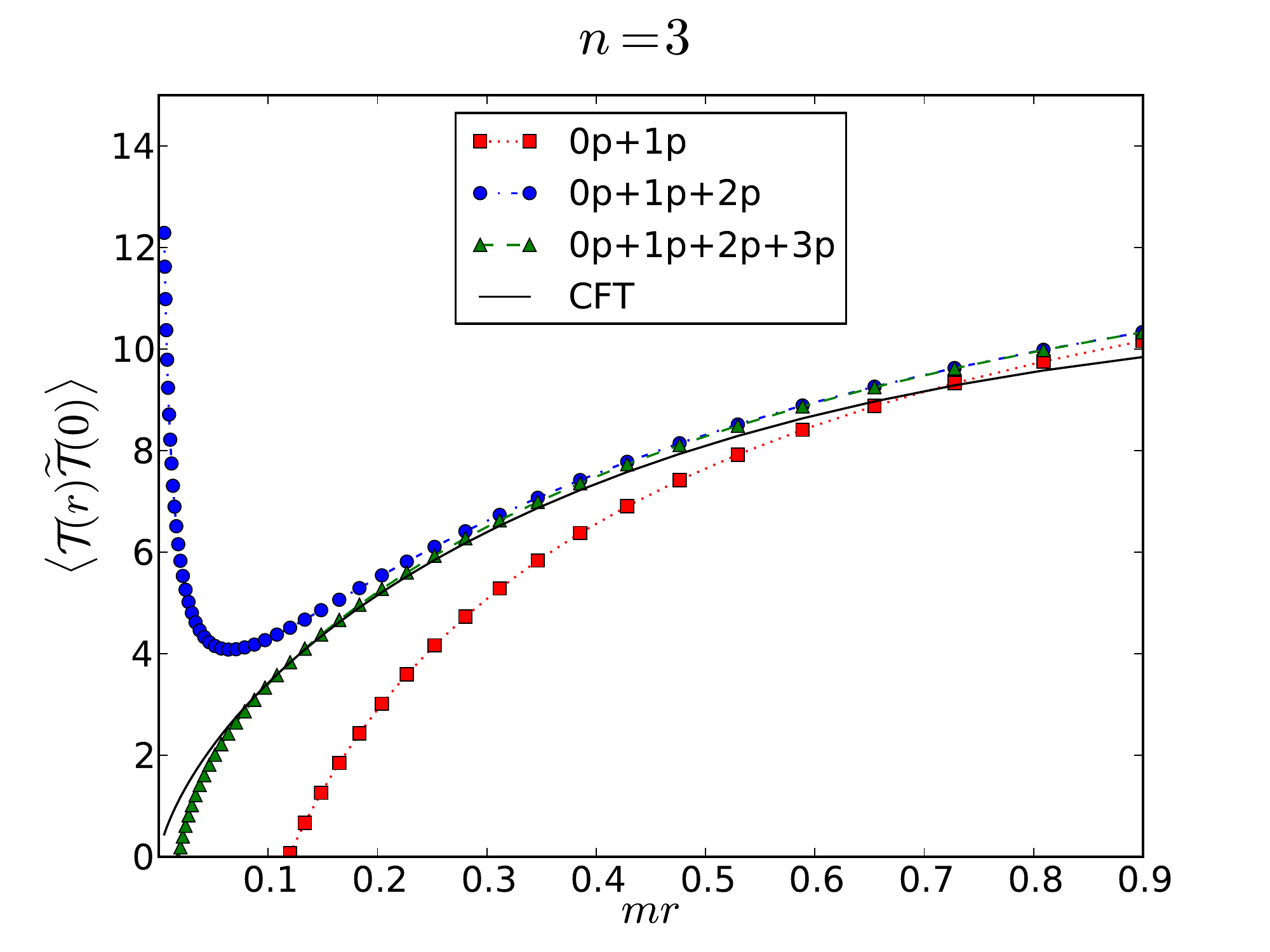} 
 \includegraphics[width=7.5cm]{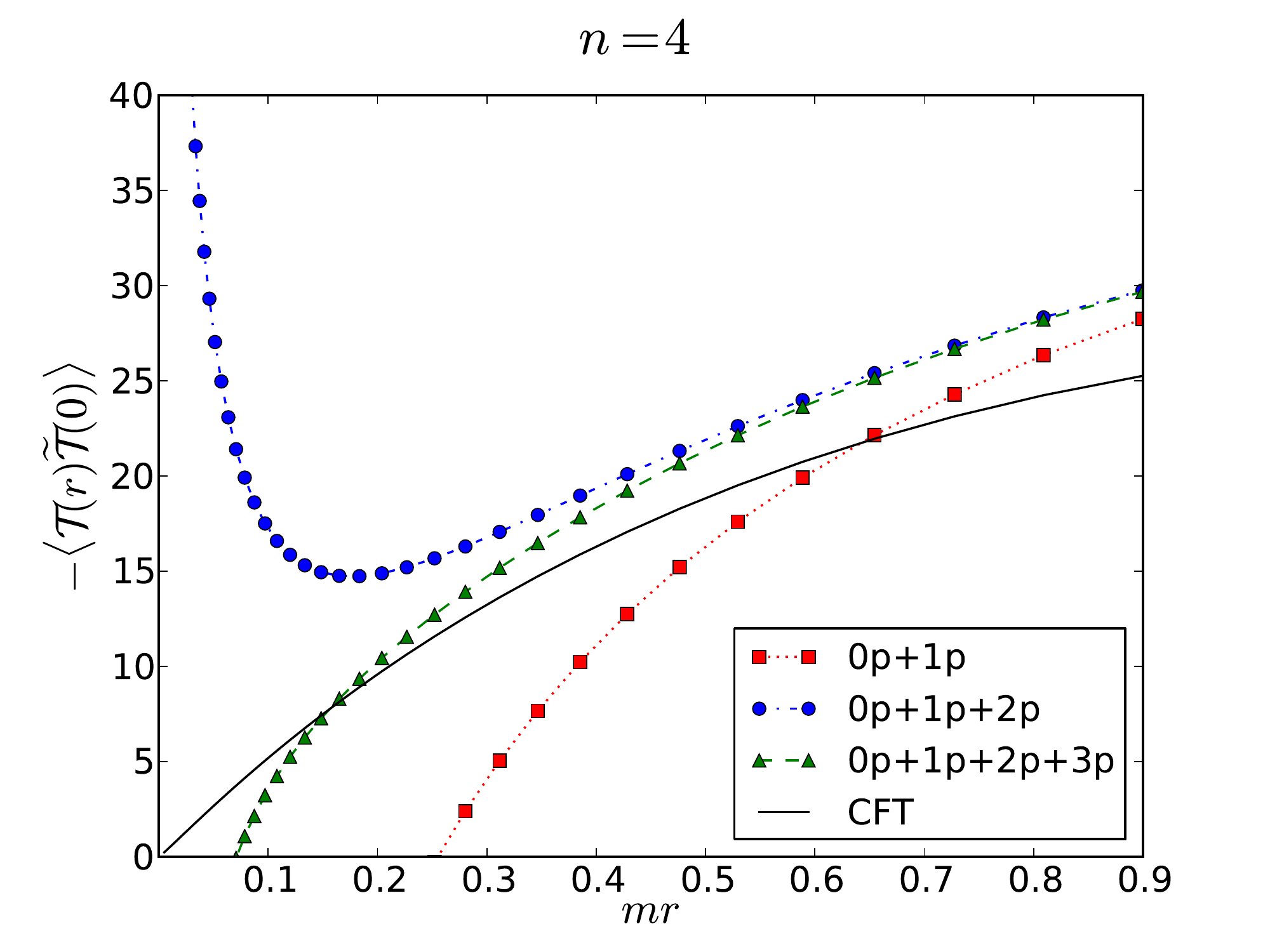}
  \includegraphics[width=7.5cm]{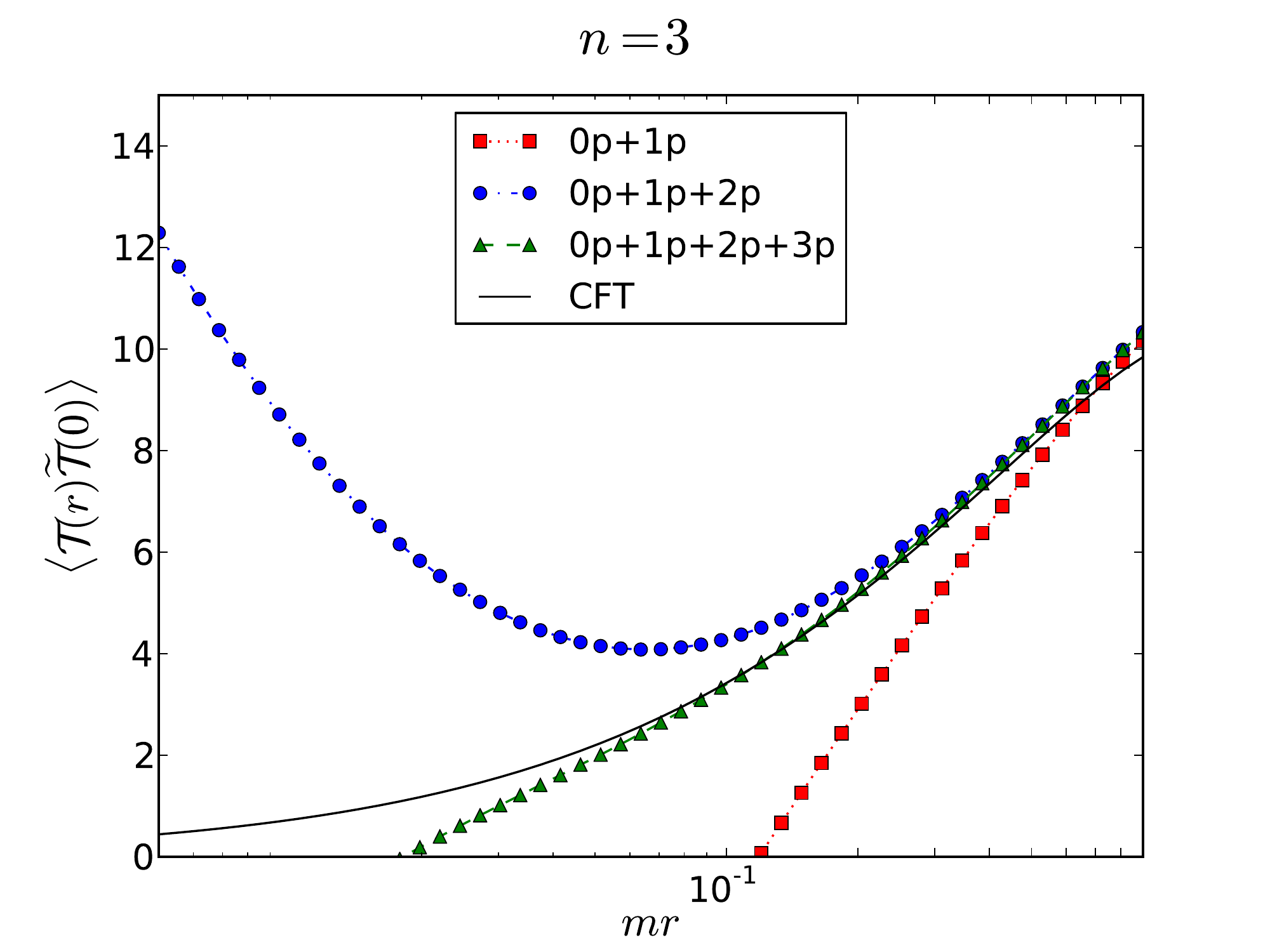} 
 \includegraphics[width=7.5cm]{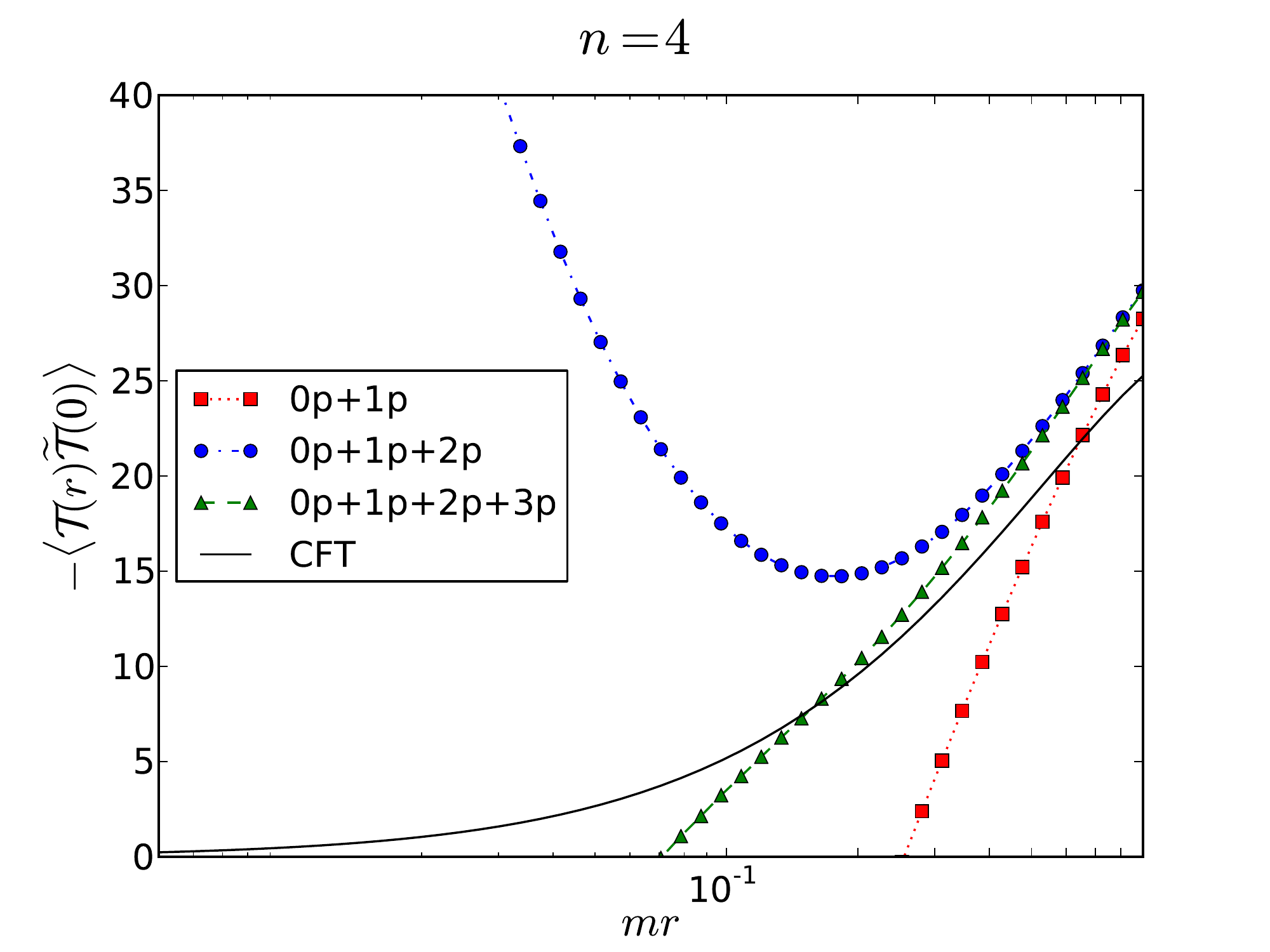}
 \end{center} 
 \caption{Zeroth order perturbed CFT versus three-particle form factor computation of the two-point function $\bra \TT(r) \tilde{\TT}(0) \ket$ for $n=3$ and $-\bra \TT(r)\tilde{\TT(0)}\ket$ for $n=4$. Squares, circles and triangles represent contributions up to one-, two- and three-particles to the form factor expansion. For each value of $n$ we present the same results both in linear and logarithmic scale. As expected, we see that the form factor result (triangles) and the CFT computation (solid line) are in relatively good agreement for small values of $mr$ but quickly drift apart for larger values of $mr$.}
 \label{cute2} 
 \end{figure}
 
\beqa
\tilde{C}_{\TT \tilde{\TT}}^{\Phi_1}&=& 0,\quad \tilde{C}_{\TT \tilde{\TT}}^{\Phi_{1,2}}=4^{-5\Delta}, \quad\tilde{C}_{\TT \tilde{\TT}}^{\Phi_{1,3}}=4^{-6\Delta},\nonumber\\
\tilde{C}_{\TT \tilde{\TT}}^{\Phi_{1,2,3}}&=&4^{-8\Delta}\tilde{C}_{\phi \phi}^\phi,\quad \tilde{C}_{\TT \tilde{\TT}}^{\Phi_{1,2,3,4}}=4^{-8\Delta}\bra \phi(i)\phi(-1)\phi(-i)\phi(1)\ket,
\eeqa
where we can again easily compute
\beq
\bra \phi(i)\phi(-1)\phi(-i)\phi(1)\ket=-5.53709...
\eeq
This gives
\beq
{\bra\TT(r) \tilde{\TT}(0)\ket}=r^{\frac{11}{4}}\left(1-(40.7927...) (mr)^{-\frac{4}{5}}+ (133.7569...) (mr)^{-\frac{6}{5}} -(120.0647...) (mr)^{-\frac{8}{5}}\right)+\cdots,
\eeq
for $n=4$. 

Like for the $n=2$ case, it is possible to compare these results to a form factor expansion once the expectations values of $\TT$ and $\Tp$ have been obtained by using (\ref{vev}). In the three-particle approximation we find
\beq
K_{\TT}= 2.02966\quad  \text{and}\quad
K_{\Tp}=  2.60713 \quad \text{for} \quad n=3,
\eeq
\beq
K_{\TT}= 2.89127 \quad \text{and}\quad
K_{\Tp}= 3.48758 \quad \text{for} \quad n=4,
\eeq
giving
\beq
\bra \TT \ket^2 = 12.953\quad \text{for}\quad n=3,
\eeq
and
\beq
\bra \TT \ket^2 = -41.5266 \quad \text{for}\quad n=4.
\eeq
\subsection{Summary and discussion}
In this section we have studied the two-point functions $\bra \Tp(r) \bTp(0) \ket$ and $\bra \TT(r)\t\TT(0)\ket$ by using two well-known approaches: a form factor expansion (up to 3 particles) and perturbed CFT (at zeroth order) calculation. Examining Figures~\ref{cute1} and \ref{cute2} we can say that agreement between both approaches is good in terms of the range of values that the correlators take but not particularly good if we compare the slope and precise values the functions take at particular points.

This level of agreement (and disagreement) is not entirely surprising given the expected range of validity of each approach: the form factors approach is eminently a large $mr$ expansion and although considering contributions up to three particles should provide a relatively good description for small values of $mr$ we do not expect it to be very precise for very short distances. Conformal perturbation theory works best near criticality, that is for very small values of $mr$, exactly where form factors should be less accurate. Besides, we have carried out perturbed CFT at zeroth order so the expectation is that this should really only be accurate for very small values of $mr$. Finally, a numerical comparison between CFT and form factors is only possible if the form factor normalization constant (that is the vacuum expectation value of the field) is known. In our case we can only access these expectation values approximately through yet again a form factor expansion. This introduces a further error (the vacuum expectation values obtained this way are smaller in absolute value than their exact values) which results in an overall shift of the form factor points.  

Overall the results we obtain are not dissimilar to Zamolodchikov's results \cite{Z} for the two-point function $\bra \phi(r)\phi(0)\ket$ in Lee-Yang. Agreement with CFT was slightly better in \cite{Z} as first order corrections in perturbed CFT were also included and the exact value of $\bra \phi\ket^2$ was known from an independent thermodynamic Bethe ansatz computation.

Despite the many limitations described above, it is still the case that agreement between form factor numerics and zeroth order perturbed CFT is better for some particular correlators than for others. We do not have a good physical explanation as to why this should be the case but it appears to depend on the particular functional form of the perturbed CFT curve obtained for each case, that is the relative weight of the various contributing terms and the region of values of $r$ where the term with the lowest power of $r$ is leading.



\section{Entanglement entropy from form factors}
\label{6}
\subsection{R\'enyi entropy from form factors with $mr\ll 1$}
\begin{figure}[h!]
 \begin{center} 
 \includegraphics[width=7.9cm]{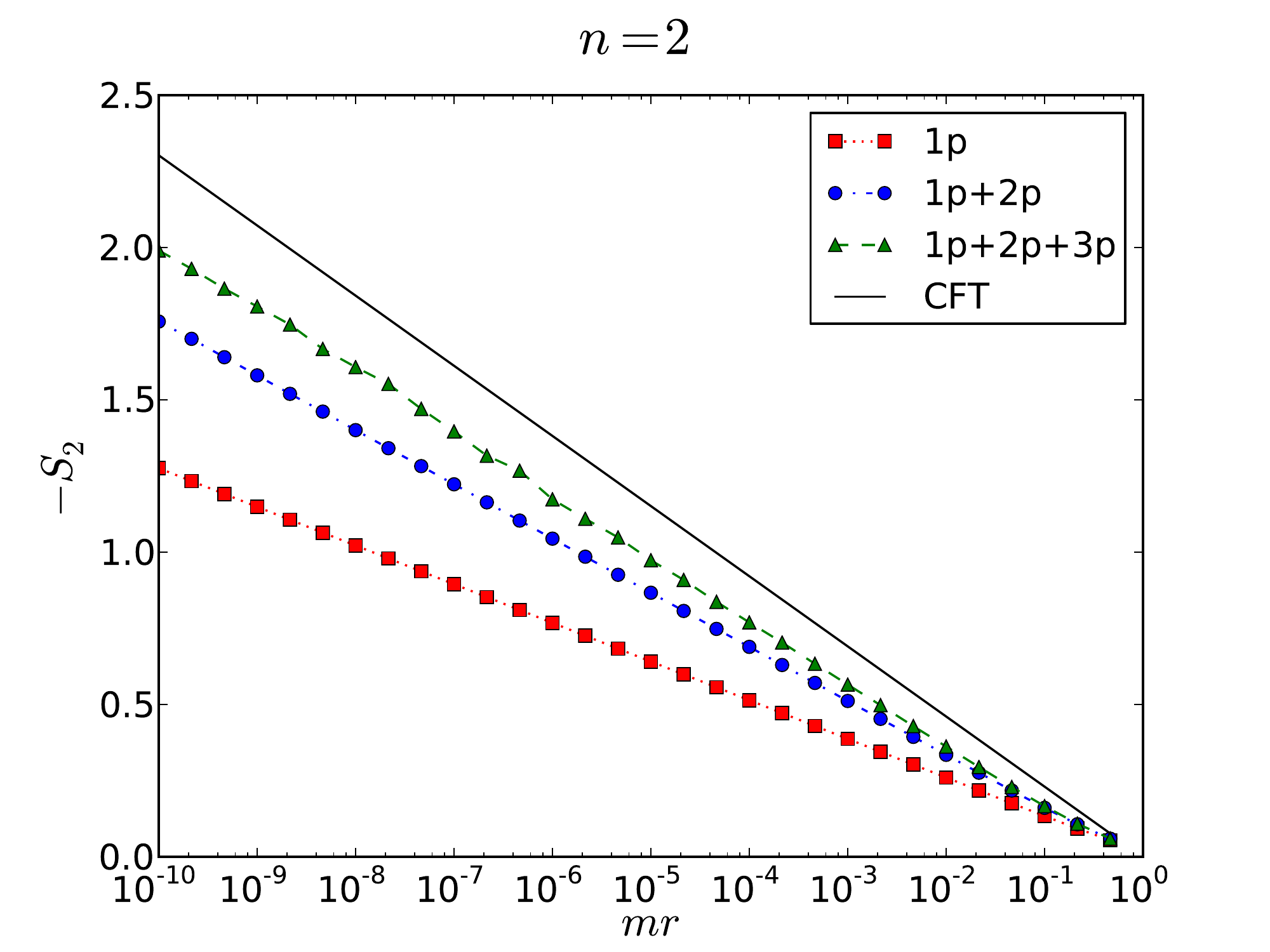} 
  \includegraphics[width=7.9cm]{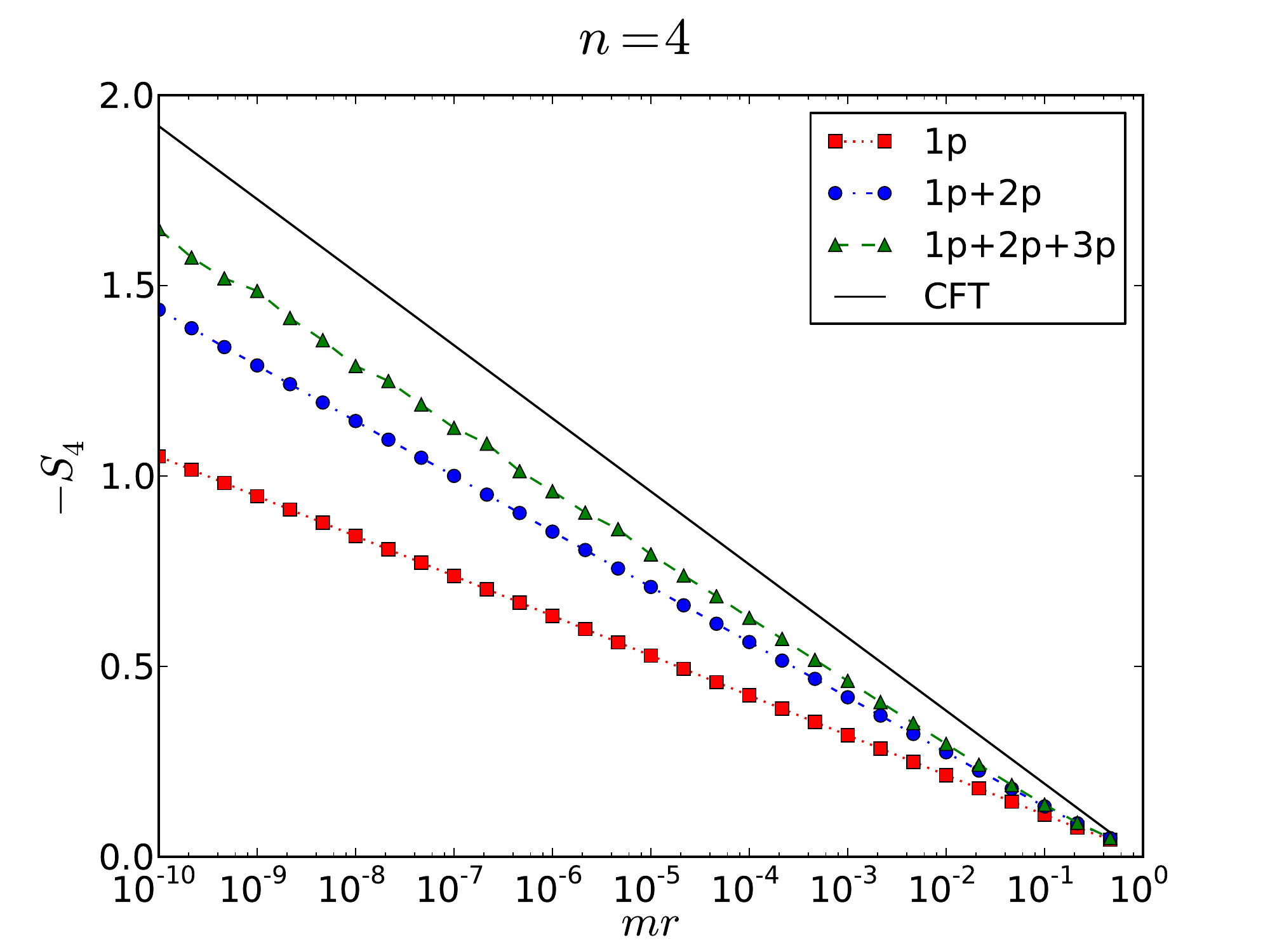} 
   \includegraphics[width=7.9cm]{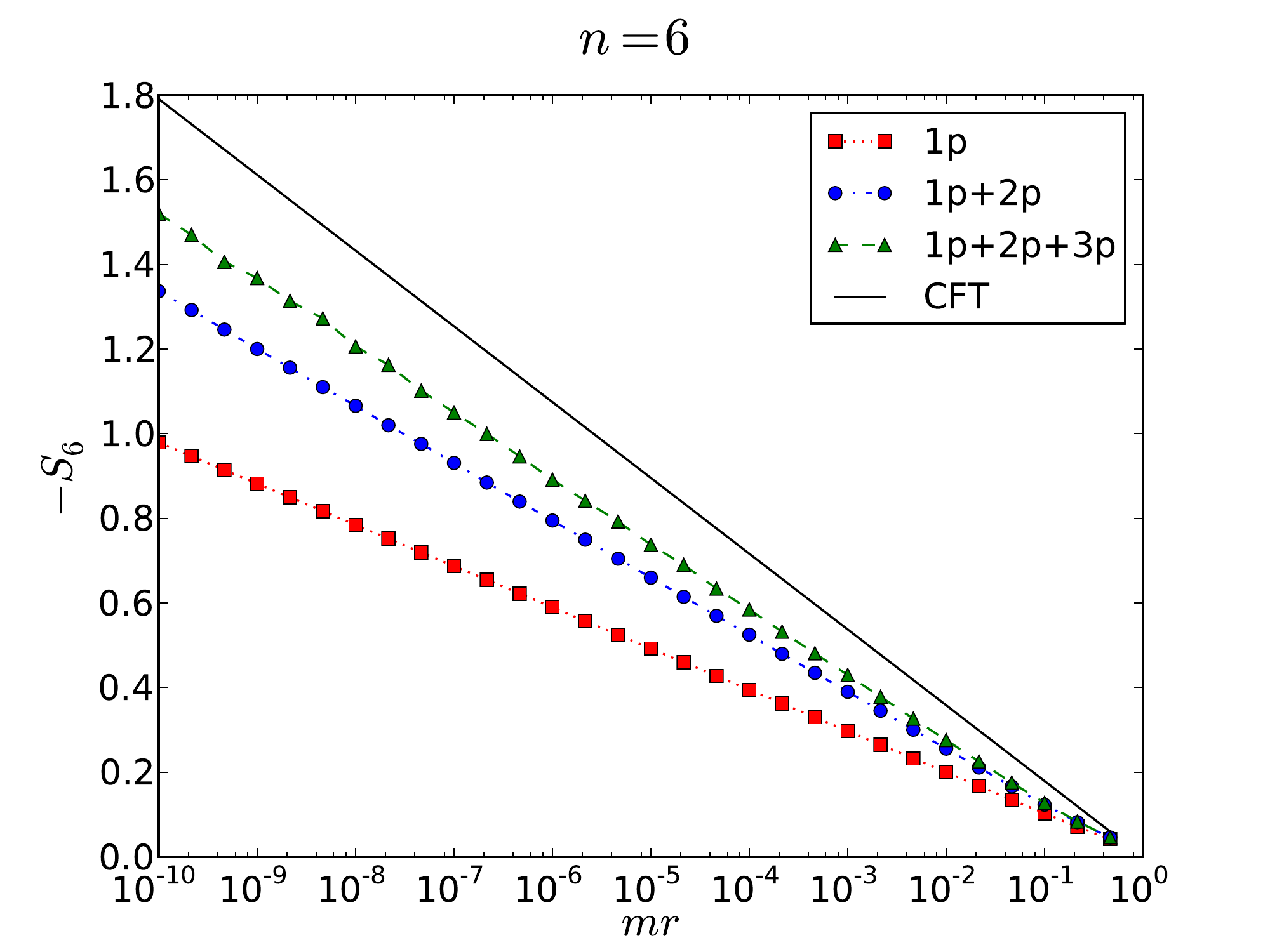} 
    \includegraphics[width=7.9cm]{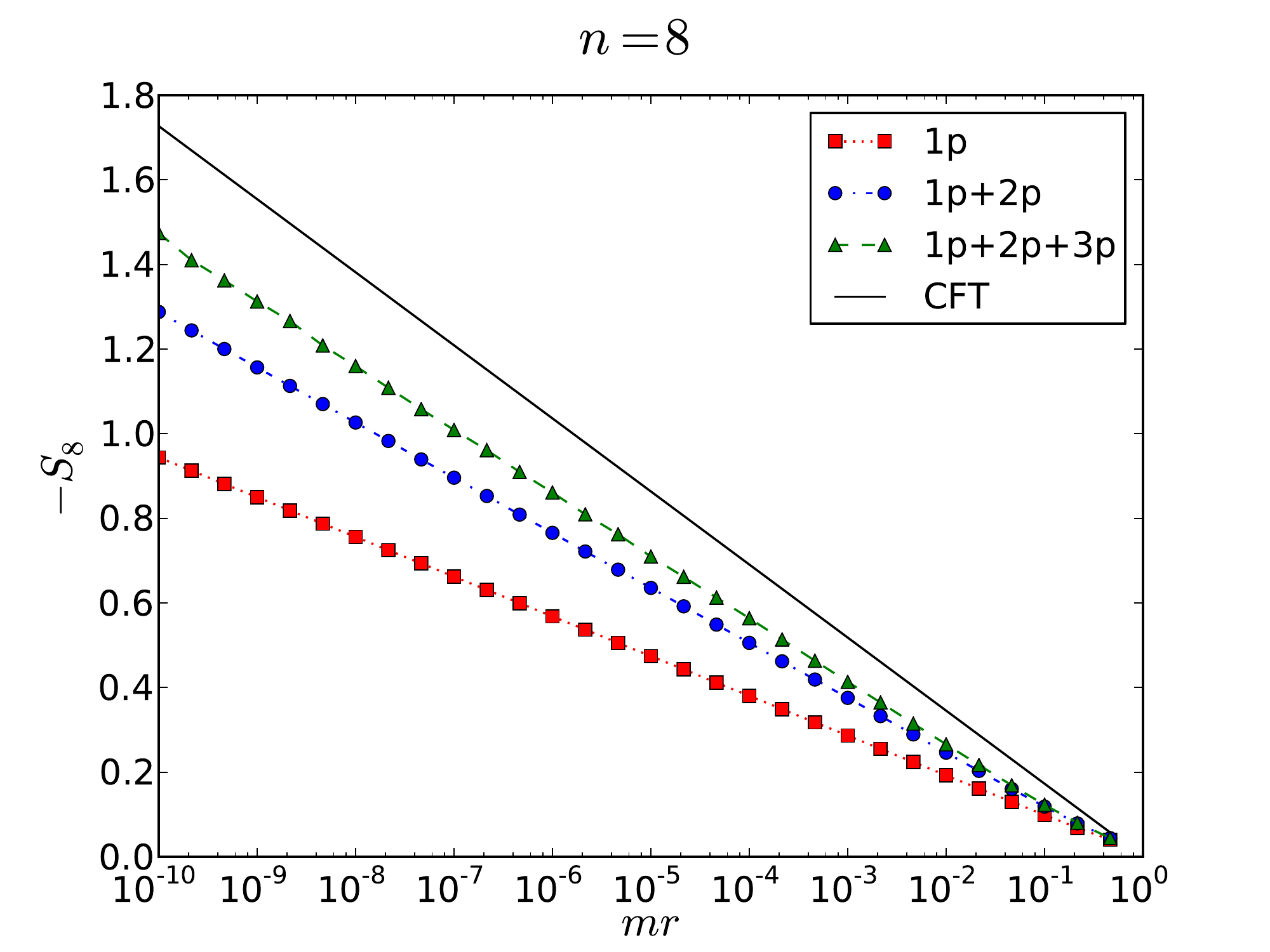} 
 \end{center} 
 \caption{The R\'enyi entropy (\ref{renyi}) with (\ref{newt}), subtracting the non-universal, positive, infinite additive contribution $\frc{1}{1-n}\log{\cal Z}_n - \frc{c_{\rm eff}}6 \lt(1+\frc1n\rt)\log(m\varepsilon)$, and evaluated in logarithmic scale using form factors. The form factor contributions up to one-, two- and three-particles are considered both for the correlators $\bra \Tp(r) \bTp(0)\ket$ and $\bra \phi(r) \phi(0) \ket$. The solid line represents the CFT prediction $\frac{c_{\text{eff}}(n+1)}{6n}\log(mr)$ (note that for $mr<1$, this is negative). All graphs show a clear logarithmic divergence at $mr=0$ (as expected). Additional form factor contributions (also as expected) improve agreement with CFT.} 
 \label{cuty} 
 \end{figure}

In the previous section we established that although the form factor expansion is only rapidly convergent for $mr \gg 1$, it does still provide a good estimate of the short distance behaviour of correlators. An alternative way of testing this results is by performing a computation of the R\'enyi entropy as defined in (\ref{newt}). This involves also the computation of $\bra \phi(r) \phi(0) \ket$, which was first obtained in \cite{Z} (and which can be obtained from $\bra \Tp(r)\bTp(0)\ket$ by setting $n=1$). Figure~\ref{cuty} shows the results of such a computation for $n=2, 4, 6$ and 8.

\subsection{Bi-partite entanglement entropy of large subsystems}

In this section we will used the form factors previously obtained to study the bi-partite entanglement entropy of the Lee-Yang model paying special attention to the region $mr>>1$.

\subsubsection{Saturation}

Using \eqref{newt}, the entanglement entropy of non-unitary theories is
\beq
S(r)=-\lim_{n\rightarrow 1} \frac{d}{d n} \left[{\cal Z}_n\varepsilon^{\frac{c_{\text{eff}}}{6}\left(n-\frac{1}{n}\right)}
\frac{\langle :\mathcal{T}\phi:\rangle^2}{\langle \phi \rangle^{2n}}\frac{A(r,n)}{B(r)^n}\right]
\eeq
where we have used the short-hand notation
\beq
A(r,n):=\langle :\mathcal{T}\phi: \rangle ^{-2}\langle :\mathcal{T}\phi:(r) :\mathcal{T}\phi:(0)\rangle, \qquad
B(r):=\langle \phi \rangle^{-2}\langle \phi(r) \phi(0) \rangle.
\eeq
Observe that 
\beq
\lim_{n\rightarrow 1} A(r,n)=B(r)
\eeq
and that $\lim_{mr\to\infty} A(r,n)= \lim_{mr\to\infty}B(r) = 1$.

The expression above can be written as
\beq 
S(r)=-\frac{c_{\text{eff}}}{3}\log(m\epsilon)+ U -\lim_{n\rightarrow 1} \frac{d}{d n} \frac{A(r,n)}{B(r)^n}. 
\eeq
The constant $\ep$ is a convenient short-distance cutoff related to $\varep$ by $\frac{c_{\text{eff}}}{3}\log(m\epsilon) = \frac{c_{\text{eff}}}{3}\log(m\varepsilon) + \lim_{n\to1}\frc{d}{dn}\lt({\cal Z}_n - (\tilde{C}_{\phi\phi}^\phi)^n/{\tilde{C}_{\Tp\Tp}^{\Phi_{1,\ldots,n}}}\rt)$. The dimensionless, universal saturation constant is
\beqa
U&=&-\lim_{n\rightarrow 1} \frac{d}{dn}\left(
m^{-\frc{c_{\rm eff}}6\lt(n-\frc1n\rt)}
\frac{\langle :\mathcal{T}\phi:\rangle^2 \,\lt(\tilde{C}_{\phi\phi}^\phi\rt)^{n}}{
\langle \phi \rangle^{2n}\,\tilde{C}_{\Tp\Tp}^{\Phi_{1,\ldots,n}}}
\right)\label{u} \n
&=&-\lim_{n\rightarrow 1} \frac{d}{dn}\left(
\frc{K_{\phi}^n}{K_{\Tp}}\rt)
\eeqa
where $K_{\cal O}$ was defined in \eqref{Kointro} (equivalently \eqref{Ko}). The meaning of $U$ is clear from a comparison of the small- and large-distance behaviour of the entanglement entropy:
\beqa
	S(r)&\sim& -\frc{c_{\rm eff}}3 \log(m\ep) + U + o(1) \qquad (mr\to\infty)\n
    &\sim& \frc{c_{\rm eff}}3 \log(r/\ep)+ o(1)\qquad\qquad (mr\to 0).
\eeqa
These easily generalize to the R\'enyi entanglement entropy at arbitrary $n$, giving the saturation behaviour \eqref{Snsl} with the universal saturation constant $U_n$ expressed in \eqref{Un}, and in particular $U=U_1$.

\subsubsection{Leading order correction to saturation}

It is easy to see that
\beq
\lim_{n\rightarrow 1} \frac{d}{d n} \frac{A(r,n)}{B(r)^n}= \lim_{n\rightarrow 1} \frac{A'(r,n)-A(r,n)\log B(r)}{B(r)^{n}}=\frac{A'(r,1)}{B(r)}-\log B(r)
\eeq
where $A':=d A/ d n$. We now need to compute the objects above, all of which are given in terms of various two-point functions and their limits at $n=1$. As we have seen in the introduction, both the two-point functions $A(r,n)$, $B(r)$ and the logarithm $\log B(r)$ admit expressions in terms of form factors. Here we only want to investigate the first and second order corrections to saturation of the entanglement so we will consider only up to the two-particle contribution to the form factor expansion. By doing so we have that
\beq
A(r,n)= 1 +A_1(r,n)+A_2(r,n)+\cdots
\eeq
where
\beq
A_1(r,n)=- n\int_{-\infty}^\infty \frac{d \theta}{2\pi}\left|\frac{F_1^{:\mathcal{T}\phi:|1}}{\langle :\mathcal{T}\phi: \rangle}\right|^2e^{-rm \cosh \theta}=-\frac{n}{\pi}\left|\frac{F_1^{:\mathcal{T}\phi:|1}}{\langle :\mathcal{T}\phi: \rangle} \right|^2 K_0(mr)
\eeq
is the one-particle contribution, and 
\beqa
A_2(r,n)&=& \frac{1}{2}\sum_{i,j=1}^n\int_{-\infty}^\infty 
\int_{-\infty}^\infty \frac{d \theta_1 d\theta_2}{(2\pi)^2}\left|\frac{F_2^{:\mathcal{T}\phi:|ij}(\theta_1,\theta_2)}{\langle :\mathcal{T}\phi: \rangle}\right|^2e^{-rm \cosh \theta_1-rm \cosh\theta_2}\nonumber\\
&=&n\sum_{j=1}^n\int_{-\infty}^\infty 
 \frac{d \theta}{(2\pi)^2}\left|\frac{F_2^{:\mathcal{T}\phi:|11}(\theta,2\pi i(j-1))}{\langle :\mathcal{T}\phi: \rangle}\right|^2 K_0\left(2mr \cosh\frac{\theta}{2}\right)
\eeqa
is the two-particle contribution. Similarly,
\beq
B(r)= 1+B_1(r)+B_2(r)+\cdots
\eeq
with 
\beq
B_1(r)=- \int_{-\infty}^\infty \frac{d \theta}{2\pi}\left|\frac{F_1^{\phi}}{\langle\phi \rangle}\right|^2 e^{-rm \cosh \theta}= -\frac{1}{\pi}\left|\frac{F_1^{\phi}}{\langle\phi \rangle}\right|^2 K_0(mr)=-\frac{ 2} {3^{1/2}\pi f(\frac{2\pi i}{3},1)^2}K_0(mr), \label{B1}
\eeq
where the ratio $\frac{F_1^{\phi}}{\langle\phi \rangle}$ was given in (\ref{ratio01}), and 
\beqa
B_2(r)&=& \frac{1}{2}\int_{-\infty}^\infty 
\int_{-\infty}^\infty \frac{d \theta_1 d\theta_2}{(2\pi)^2}\left|\frac{F_2^{\phi}(\theta_1-\theta_2)}{\langle \phi \rangle}\right|^2e^{-rm \cosh \theta_1-rm \cosh\theta_2}\nonumber\\
&=&   
\int_{-\infty}^\infty \frac{d \theta}{(2\pi)^2}\left|\frac{F_2^{\phi}(\theta)}{\langle \phi \rangle}\right|^2 K_0\left(2mr \cosh\frac{\theta}{2}\right).
\eeqa
The form factor $F_2^{\phi}(\theta)$ was given by Zamolodchikov in \cite{Z} and can be written as
\beq
F_2^{\phi}(\theta)=\frac{\pi m^2}{8}\frac{F_{\text{min}}(\theta,1)}{f(i\pi,1)}.
\eeq
Finally
\beq
\log B(r)=B_1(r)+B_2(r)-\frac{1}{2}B_1(r)^2+\cdots
\eeq
Thus, we find that
\beq
\lim_{n\rightarrow 1} \frac{d}{d n} \frac{A(r,n)}{B(r)^n}=A_1'(r,1)-B_1(r)+ A_2'(r,1)-B_2(r)+\frac{1}{2}B_1(r)^2-B_1(r)A_1'(r,1)+\cdots
\eeq 
where the first two terms will give the next-to-leading order contribution to the entanglement entropy (i.e. the leading correction to its saturation value) and the remaining terms give the next-to-next-to leading order contibution. We will now analyse this expression in more detail.

In appendix~\ref{appendixb} we show that $A_1'(r,1)$ is given by
\beq
A_1'(r,1)=B_1(r)+\frac{2}{f(\frac{2\pi i}{3},1)^2}\left(\frac{1}{\pi\sqrt{3}}-\frac{13}{108}\right)K_0(mr).
\eeq
We also need $A_2'(r,1)$ which is given by
\beq
A_2'(r,1)=\frac{1}{8}K_0(2mr).
\eeq
This simple result was established in \cite{entropy} for all integrable quantum field theories and even beyond integrability \cite{next}.
Then, the expression above simplifies to
\beqa
\lim_{n\rightarrow 1} \frac{d}{d n} \frac{A(r,n)}{B(r)^n}&=&\frac{2}{f(\frac{2\pi i}{3},1)^2}\left(\frac{1}{\pi\sqrt{3}}-\frac{13}{108}\right)K_0(mr)+\frac{1}{8}K_0(2mr)\nonumber\\
&& -\frac{4}{3f(\frac{2\pi i}{3},1)^4} 
\int_{-\infty}^\infty \frac{d \theta}{(2\pi)^2}\left(\left|{F_{\text{min}}(\theta,1)}\right|^2-1\right) K_0\left(2mr \cosh\frac{\theta}{2}\right)\nonumber\\ &&
-\frac{13} {3^3 \sqrt{3}\pi f(\frac{2\pi i}{3},1)^4}K_0(mr)^2+\cdots
\eeqa 
Thus, the von Neumann entropy of the Lee-Yang model takes the form
\beqa
S(r)&=&-\frac{2}{15}\log(m\epsilon)+ U -\frac{2}{f(\frac{2\pi i}{3},1)^2}\left(\frac{1}{\pi\sqrt{3}}-\frac{13}{108}\right)K_0(mr)-\frac{1}{8}K_0(2mr)\nonumber\\
&& +\frac{4}{3f(\frac{2\pi i}{3},1)^4} 
\int_{-\infty}^\infty \frac{d \theta}{(2\pi)^2}\left(\left|{F_{\text{min}}(\theta,1)}\right|^2-1\right) K_0\left(2mr \cosh\frac{\theta}{2}\right)\nonumber\\ &&
+\frac{13} {3^3 \sqrt{3}\pi f(\frac{2\pi i}{3},1)^4}K_0(mr)^2+\cdots\nonumber\\
&=& -\frac{2}{15}\log(m\epsilon)+ U-a K_0(mr)-  \frac{b e^{-2mr}}{\sqrt{2mr}}-\frac{c e^{-2mr}}{{2mr}}+O(e^{-2mr}(2mr)^{-3/2})
\eeqa
where $U$ is the model-dependent constant (\ref{u}) and
\beq
a:=\frac{2}{f(\frac{2\pi i}{3},1)^2}\left(\frac{1}{\pi\sqrt{3}}-\frac{13}{108}\right)=0.0769782...
\eeq
\beq
b:=\sqrt{\frac{\pi}{2}}\left(\frac{1}{8}-\frac{4}{3f(\frac{2\pi i}{3},1)^4} 
\int_{-\infty}^\infty \frac{d \theta}{(2\pi)^2}\left(\left|{F_{\text{min}}(\theta,1)}\right|^2-1\right) \right)=0.326234...
\eeq
and
\beq
c:=-\frac{13} {3^3 \sqrt{3} 2 f(\frac{2\pi i}{3},1)^4}=-0.0512159...
\eeq

In contrast to results found for unitary theories \cite{entropy, next}, the results above suggest that the leading and next-to-leading order correction to saturation of the entropy of large blocks are strongly model-dependent. In particular, the leading correction is proportional to the constant $a$ which clearly depends on specific features of the model under consideration (that is, the one-particle form factor). This term is directly related to the one-particle form factor and in particular to its value and the value of its derivative at $n=1$. The fact that both these quantities are non-zero for $\Tp$ 
is special for this field -- they would have been zero if we had used $\TT$ -- and we are tempted to conclude that this phenomenon is related to the non-unitary nature of the model. It would be interesting to test these results numerically for example by studying the spin chain model considered in \cite{gehlen1,gehlen2,meandreas,BCDLR}.

\section{Conclusions and outlook}
\label{7}

In this paper we have provided an in-depth study of the two-point functions of twist fields in the massive Lee-Yang model and their application to the computation of the bi-partite entanglement entropy. The main tools used for our study are branch-point twist fields and the relationship between their correlation functions in replica theories and the bi-partite entanglement. For massive unitary theories this connection was established and explored in \cite{entropy}, and the present work addresses the problem for a massive non-unitary model for the first time. Representing the EE using correlation functions of twist fields indeed provides the only known method so far for performing computations of the bi-partite entanglement in massive QFT models. The non-unitarity of the theory has important consequences for the computation of entanglement and several stark differences are found with respect to the unitary case. 

The twist field $\TT$ and its conjugate $\tilde{\TT}$ considered in \cite{entropy} {\it are not} the right operators to consider in non-unitary models when performing entropy computations. Instead, as first proposed in \cite{BCDLR} for CFT, one must consider the two-point function of suitably normalized composite fields $\Tp$ and $\bTp$ introduced in \cite{ctheorem}, defined as leading contributors to the OPE $\TT(x) \phi(0)$ and $\tilde{\TT}(x)\phi(0)$ where $\phi$ is the lowest-dimension primary field of the model.

In the present work we find the exact form factors of $\TT$ and $\Tp$ up to three particles. Remarkably we find that the form factor equations together with the requirement of clustering are sufficient to entirely fix all form factors and to provide in a natural way two families of solutions corresponding to the two twist fields $\TT$ and $\Tp$. We also give numerical evidence that the resulting correlation functions agree, at short distances, with CFT results. This is done by numerically evaluating truncated form factor expansions, or the logarithm thereof, of correlation functions at short distances, and comparing with a zeroth order perturbed CFT computation of the twist field two-point function, in the spirit of Zamolodchikov's work \cite{Z}. The CFT computation also provides some of the first general results regarding OPEs of the composite twist fields $\Tp$ and $\bTp$. Finally these results are used to compute the R\'enyi and von Neumann entropy, in particular in the limit of large blocks, which is well described by the form factor expansion. For large blocks we find that the corrections to saturation are strongly model-dependent for non-unitary theories. This is in contrast to the very universal form of the leading correction found for unitary models \cite{entropy,next}, only depending on the particle spectrum.

It would be interesting to compare the present results about entanglement entropy with a numerical evaluation in the quantum Ising model with imaginary transverse magnetic field spin first considered by von Gehlen \cite{gehlen1,gehlen2}, whose near-critical universal region is expected to be described by the Lee-Yang QFT \cite{Fisher, LYCardy}. This would provide the first strong evidence beyond criticality supporting the conjecture that the composite fields $\Tp$ and $\bTp$ are the correct fields for representing branch points, or conical singularities, in non-unitary models.

The ideas in the present work and in \cite{BCDLR}, in particular the form (\ref{newt}) of the twisted replica partition function in non-unitary QFT, lead us to speculate a relation between correlators of composite fields in non-unitary QFT and correlators of physical fields in its ``unitary counterpart''. Following ideas from the field of ${\tt PT}$-symmetric quantum mechanics \cite{sgh,pt}, given that the non-unitary theory (i.e.~with a non-Hermitian hamiltonian) considered has a real energy spectrum, we may infer that there must be a similarity transformation which maps the Hamiltonian and correlators of the Lee-Yang model to the Hamiltonian and  correlators of some unknown unitary theory. It is tempting to propose that the operation of taking composites with the lowest-dimension field $\phi$ implements, up to normalization, such a similarity transformation. That is, we may identify the correlators of local fields in the resulting unitary theory, denoted by $\bra \bra {\mathcal{O}}(r) {\mathcal{O}}(0) \ket\ket$, with correlators of composite fields in the non-unitary model, as 
\beq
\bra \bra {\mathcal{O}}(r) {\mathcal{O}'}(0) \ket\ket:=\frac{\bra :{\mathcal{O}\phi}:(r) :{\mathcal{O}'\phi}:(0)\ket}{\bra\phi(r)\phi(0) \ket}.
\eeq
It would be very interesting to test this and related ideas further.

Several future directions of research follow naturally from this work: a more detailed study of the OPEs of twist fields for arbitrary $n$ is desirable, not only to better understand the properties of replica CFTs but also as building blocks for perturbed CFT computations for larger values of $n$. A systematic understanding of higher particle form factor solutions is still missing as is the study of the entanglement entropy of excited states and multipartite regions in massive (unitary or not) QFT.

{\bf Acknowledgements:} D.~Bianchini is grateful to City University London for a University Studentship. We thank G.~Takacs for bringing references \cite{Smir,takacs} to our attention and F.~Ravanini for giving us the opportunity to present some preliminary results of this work at the 9th Bologna Workshop on CFT and Integrable Models (Bologna, September 2014). 
We thank also E.~Levi and F.~Ravanini for a previous collaboration which has inspired this work. Finally, we thank E.~Levi for his comments on the manuscript.

\appendix

\section{Definition of the field $\Tp$}
\label{appendixc}
Consider
\beq
	\Tp(y) = A \lim_{y\to x} |x-y|^a \sum_{j=1}^n \TT (y) \phi_j(x).
\eeq
The power $a$ is fixed by requiring that the limit exist, and the normalization $A$ is determined by requiring conformal normalization of the resulting field. These, as well as structure constants studied in Appendix \ref{appendixd}, may be evaluated by using standard methods of CFT \cite{FMS}. Correlation functions with twist field insertions at $y_1$ and $y_2$ in the $n$-copy model are interpreted as correlation functions on a $n$-sheeted Riemann surface ${\cal M}_{n,y_1,y_2}$ with branch points in place of the twist fields, and conformal uniformization to the sphere is used. For the uniformization step, one makes use of the
conformal map
\beq
	g: {\cal M}_{n,y_1,y_2} \to \hC\setminus\{0,\infty\},\quad
	g(z) = \lt(\frc{z-y_1}{z-y_2}\rt)^{1/n}, \label{manifold}
\eeq
with
\beq
	\frac{\p g}{\p z} :=\partial g= \frc1n \frc{y_2-y_1}{(z-y_1)(z-y_2)} \lt(\frc{z-y_1}{z-y_2}\rt)^{1/n}.
\eeq
In order to compute $A$ and $a$, we compute the following ratio of correlation functions:
\beqa
	\lefteqn{\frc{\bra \Tp (x_1) \bTp (x_2)\ket}{\bra \TT(x_1)\b\TT(x_2)\ket}}
	&&\n
	&=& |A|^2\lim_{x_i\to y_i}
	|x_1-y_1|^a |x_2-y_2|^a
	\sum_{j_1,j_2=1}^n \frc{\bra \TT(y_1)\TT(y_2)
	\phi_{j_1}(x_1)\phi_{j_2}(x_2)\ket}{\bra \TT(y_1)\b\TT(y_2)\ket}.
    \eeqa
The new ratio of correlators involved is interpreted as a correlator of $\phi_{j_1}(x_1) \phi_{j_2}(x_2)$ on the Riemann surface ${\cal{M}}_{n,y_1,y_2}$, and can be computed by using the conformal map above to relate them to correlators in the complex plane. Thus
    \beqa
    \lefteqn{\frc{\bra \Tp (x_1) \bTp (x_2)\ket}{\bra \TT(x_1)\b\TT(x_2)\ket}}
	&&\n
	&=& |A|^2\lim_{y_i\to x_i}
	|x_1-y_1|^a |x_2-y_2|^a
	|\p g(x_1)|^{2\Delta} |\p g(x_2)|^{2\Delta} \sum_{j_1,j_2=1}^n
	\bra \phi(e^{\frac{2\pi i j_1}{n}}g(x_1)) \phi(e^{\frac{2\pi i j_2}{n}} g(x_2))\ket \n
	&=& |A|^2 n^{-4\Delta} \lim_{y_i\to x_i}
	|x_1-y_1|^{a+2\Delta(\frac{1}{n}-1)}
    |x_2-y_2|^{a-2\Delta(\frac{1}{n}+1)} \sum_{j_1,j_2=1}^n
	\lt|e^{\frac{2\pi i j_1}{n}}g(x_1) - e^{\frac{2\pi i j_2}{n}} g(x_2)\rt|^{-4\Delta}\n
	&=& |A|^2 n^{2-4\Delta} \lim_{y_i\to x_i}
	|x_1-y_1|^{a+2\Delta(\frac{1}{n}-1)} |x_2-y_2|^{a-2\Delta(\frac{1}{n}+1)}
	\lt|\frc{x_2-y_1}{x_2-y_2}\rt|^{-\frac{4\Delta}{n}}\n
	&=& |A|^2 n^{2-4\Delta} |x_1-x_2|^{-\frac{4\Delta}{n}}
	\lim_{y_i\to x_i}
	\lt(|x_1-y_1| |x_2-y_2|\rt)^{a+2\Delta(\frac{1}{n}-1)}.
\eeqa
Hence we must set
\beq
	a = 2\Delta\lt(1-\frc1n\rt),\quad A = n^{2\Delta-1}.
\eeq
We find the known result \cite{ctheorem} that the dimension of $\Tp$ is
\beq
	\Delta_{\Tp} = \Delta_{\TT} +\frac{\Delta}{n},
\eeq
and we have the correct normalization
\beq
	\Tp(x_1) \bTp(x_2) \sim {\bf 1} |x_1-x_2|^{-4\Delta_{\Tp}}.
\eeq

\section{Large $n$ expansion of the one-particle form factor contribution}\label{appendixa}

The one particle contribution to the powers $x_{\TT}$ and $x_{\Tp}$ as defined in section \ref{4} listed in the tables 1 and 2 may be computed exactly as is simply given by the function (see eq.~(\ref{del}))
\beq
\frac{n}{\pi} |F_1^{\mathcal{O}|1}|^2.
\eeq
for $\mathcal{O=\TT}$ or $\mathcal{O=\Tp}$. It is easy to show that this function admits a large $n$ expansion in powers of $1/n$ starting with a term linear in $n$. Recall the expressions (\ref{identification}). Combining those with (\ref{rationfs}) we can rewrite the expectation values as
\beq
 \frac{n}{\pi}\left|\frac{F_1^{\mathcal{O}|1}}{\langle \mathcal{\mathcal{O}}\rangle}\right|^2={\frac{\sin \frac{\pi}{3n}}{2\pi\sin\frac{\pi}{6n} \sin \frac{\pi}{2n}f(i\pi,n)}}
 \left(\cos\left(\frac{\pi}{3n}\right)\pm 2 \sin ^2\left(\frac{\pi}{6 n}\right) \right)^2, \label{identifications}
\eeq
For large $n$ we find that
\begin{equation}
  {\frac{\sin \frac{\pi}{3n}}{2\pi\sin\frac{\pi}{6n} \sin \frac{\pi}{2n}}}
 \left(\cos\left(\frac{\pi}{3n}\right)+ 2 \sin ^2\left(\frac{\pi}{6 n}\right) \right)^2=\frac{2n}{\pi^2}+ \frac{1}{18n}+\frac{13\pi^2}{9720 n^3}+ \mathcal{O}(n^{-5}),\label{easy1}
\end{equation}
and
\begin{equation}
 {\frac{\sin \frac{\pi}{3n}}{2\pi\sin\frac{\pi}{6n} \sin \frac{\pi}{2n}}}
 \left(\cos\left(\frac{\pi}{3n}\right)-2 \sin ^2\left(\frac{\pi}{6 n}\right) \right)^2=\frac{2n}{\pi^2}- \frac{7}{18n}+\frac{173\pi^2}{9720 n^3}+ \mathcal{O}(n^{-5}),\label{easy1}
\end{equation}
We now study the expansion of $f(i \pi,n)^{-1}$ for $n$ large. It is possible to show that $f(i\pi,n)^{-1}=\prod_{k=0}^\infty f_k(n)$ with $f_k(n)$ given by the following expression
\begin{eqnarray}
 \frac{\Gamma \left(k n+\frac{1}{2}\right) \Gamma \left(k n+\frac{7}{6}\right) \Gamma \left(k n+\frac{4}{3}\right)
   \Gamma \left((k+1)n-\frac{1}{2}\right) \Gamma \left((k+1)n+\frac{1}{6}\right) \Gamma \left((k+1)n+\frac{1}{3}\right)} {\Gamma \left(k n+\frac{2}{3}\right) \Gamma \left(k n+\frac{5}{6}\right) \Gamma \left(k n+\frac{3}{2}\right) \Gamma \left((k+1)n-\frac{1}{3}\right) \Gamma \left((k+1)n-\frac{1}{6}\right) \Gamma \left((k+1)n+\frac{1}{2}\right)}.
\end{eqnarray}
It is easy to see that the leading contribution for $n$-large comes from the $n$-independent part of the $k=0$ term in the product. We have that
\begin{eqnarray}
f_0(n) &=& \frac{1}{2} \frac{\Gamma(\frac{5}{6})\Gamma(\frac{2}{3})}{\Gamma(\frac{4}{3})\Gamma(\frac{7}{6})}\left(1-\frac{1}{36n^2}-\frac{1}{36n^3}+ \mathcal{O}(n^{-4})\right), \\
  f_1(n) &=&  1-\frac{5}{144 n^2}+\frac{7}{288 n^3}+\mathcal{O}(n^{-4}),\\
  f_2(n) &=& 1-\frac{13}{1296 n^2}+\frac{19}{7776 n^3}+\mathcal{O}(n^{-4}),\\
  f_3(n)&=& 1-\frac{25}{5184 n^2}+\frac{37}{62208 n^3}+\mathcal{O}(n^{-4}),\\
  f_4(n) &= & 1-\frac{41}{14400 n^2}+\frac{61}{288000 n^3}+\mathcal{O}(n^{-4}),\\
  f_5(n) &=& 1-\frac{61}{32400 n^2}+\frac{91}{972000 n^3}+\mathcal{O}(n^{-4}),
\end{eqnarray}
and so on. 
This gives
\begin{eqnarray}
    \frac{n}{\pi}\left|\frac{F_1^{:\mathcal{T}\phi:|1}}{\langle :\mathcal{T}\phi: \rangle}\right|^2 \approx &=&\frac{1}{2} \frac{\Gamma(\frac{5}{6})\Gamma(\frac{2}{3})}{\Gamma(\frac{4}{3})\Gamma(\frac{7}{6})}\left(\frac{2n}{\pi^2}+\frac{1}{n}\left(-\frac{197}{2400\pi^2}+\frac{1}{18}\right)+\mathcal{O}(n^{-3})\right)\nonumber\\
    &=&(0.186944...)n-\frac{(0.0435792...)}{n} + \mathcal{O}(n^{-3}).
\end{eqnarray}
\begin{eqnarray}
    \frac{n}{\pi}\left|\frac{F_1^{\mathcal{T}|1}}{\langle \mathcal{T} \rangle}\right|^2 \approx &=&\frac{1}{2} \frac{\Gamma(\frac{5}{6})\Gamma(\frac{2}{3})}{\Gamma(\frac{4}{3})\Gamma(\frac{7}{6})}\left(\frac{2n}{\pi^2}-\frac{1}{n}\left(\frac{197}{ 2400\pi^2}+\frac{7}{18}\right)+\mathcal{O}(n^{-3})\right)\nonumber\\
    &=&(0.186944...)n-\frac{(0.366434...)}{n} + \mathcal{O}(n^{-3}).
\end{eqnarray}
where the coefficient of $1/n^2$ is not exact but has been obtained by considering contributions up to $f_5$ (the next term would have given a correction of $2\times 10^{-3}$ to the coefficient).

The expansion above indeed shows that the structure of the one-particle form factor contribution closely matches what is expected from CFT since
\beq
-4x_{\Tp}=\frac{22}{30}\left(n-\frac{1}{n}\right)+\frac{4}{5n}-\frac{2n}{5}=\frac{n}{3}+\frac{1}{15n}=(0.33333...)n + \frac{(0.06666...)}{n},
\eeq
and
\beq
-4 x_{\mathcal{T}}=\frac{22}{30}\left(n-\frac{1}{n}\right)-\frac{2n}{5}=\frac{n}{3}-\frac{11}{15 n}=(0.33333...)n - \frac{(0.73333...)}{n}.
\eeq
Indeed, comparing coefficients we find that the one-particle contribution provides 56\% of the coefficient of $n$ for both $\TT$ and $\Tp$, 65\% of the $1/n$ coefficient for the field $\Tp$, and 49\% of the same coefficient for the field $\TT$. In short, the one-particle form factor provides a very substantial contribution to the two-point function of twist fields, both for short and long distances.

\section{Computation of three-particle form factors}
\label{appendixe}

There are two recursive relations for the symmetric polynomial $Q_3$:
\begin{eqnarray}
Q_3(\alpha x_0,x_0,x_1) &=& x_0^2 P_1(x_0,x_1)Q_1(x_1)\nonumber\\
&=& {F_1^{\mathcal{O}|1} C_1(n)}x_0^2(x_1-\alpha^2x_0)(x_1-\alpha^{-1}x_0)(x_1-\beta x_0)(x_1-\alpha\beta^{-1}x_0),\\
Q_3(x_0\beta^{-\frac{1}{2}},x_0\beta^{\frac{1}{2}},x_1) &=& x_0^2 U_2(x_0,x_1)Q_2(x_0,x_1) \nonumber\\ 
&=&- H_1(n) x_0^2\left(\langle\mathcal{O}\rangle C_0(n) x_0x_1 + \frac{\alpha (F_1^{\mathcal{O}|1})^2}{\langle\mathcal{O}\rangle}\left((1+\alpha^2)x_0 x_1-\alpha (x_0^2+x_1^2)\right)\right)\nonumber\\
&& \qquad \qquad \times (x_1-\beta^{-2}x_0)(x_1-\beta^2x_0).
\end{eqnarray}
The recursive equations imply that $Q_3$ is a symmetric polynomial of degree 6. Thus its most general form in terms of elementary symmetric polynomials $\sigma_i = \sigma_i^{(3)}$ is
\begin{equation}
Q_3(x_1,x_2,x_3) = A_1 \sigma_1^3\sigma_3 + A_2 \sigma_1^2 \sigma_2^2 + A_3\sigma_1\sigma_2\sigma_3 +A_4\sigma_2^3 + A_5 \sigma_3^2+A_6 \sigma_1^6 + A_7 \sigma_1^4\sigma_2 . \label{q3}
\end{equation}
The constants $A_i$ with $i=1, \dots, 7$ are found to be:
\begin{eqnarray}
A_6&=& A_7=0, \quad
A_1= A_4=\frac{\alpha F_1^{\mathcal{O}|1} (C_0(n) \cos ^2\frac{\pi }{3 n}-\alpha \langle \mathcal{O}\rangle^{-1} F_1^{\mathcal{O}|1} H_1(n) \cos^2 \frac{\pi}{2n})}{\sin
   \frac{\pi }{6 n} \sin \frac{5 \pi }{6
   n}}, \nonumber\\
A_2 &=&\frac{\alpha F_1^{\mathcal{O}|1}(\alpha \langle \mathcal{O}\rangle^{-1} F_1^{\mathcal{O}|1} H_1(n) -C_0(n)) }{4\sin \frac{\pi }{6 n} \sin \frac{5
   \pi }{6 n}},\nonumber\\
A_3 &=&-\frac{\alpha C_0(n) F_1^{\mathcal{O}|1} \left( 5 \cos \frac{\pi }{6 n}+4 \cos \frac{\pi }{2 n}+2 \cos \frac{5 \pi }{6 n}+6 \cos \frac{7 \pi }{6 n}+\cos \frac{11
   \pi }{6 n}+\cos \frac{13 \pi }{6 n}-\cos \frac{5 \pi }{2 n}\right)}{4 \cos \frac{\pi }{2 n} \sin \frac{\pi }{6 n} \sin \frac{5 \pi }{6 n}} \nonumber\\ 
   && + \frac{\alpha^2 \langle \mathcal{O}\rangle^{-1} (F_1^{\mathcal{O}|1})^2 H_1(n)\left(11 \cos \frac{\pi }{2 n}+6 \cos \frac{3 \pi }{2 n}+\cos\frac{5 \pi }{2 n}\right)}{4\cos \frac{\pi }{2 n} \sin \frac{\pi }{6 n} \sin \frac{5 \pi }{6 n}}\nonumber\\
A_5&=&\frac{\alpha  C_0(n) F_1^{\mathcal{O}|1}\left(2 \cos \frac{\pi }{n}+1\right)^2 \left(\cos \frac{\pi }{6 n}+\cos \frac{\pi }{2 n}+2 \cos \frac{7
   \pi }{6 n}-\cos \frac{3 \pi }{2 n}+\cos \frac{11 \pi }{6 n}-\cos \frac{13 \pi }{6 n}\right) }{4 \cos \frac{\pi }{2 n}\sin \frac{\pi }{6 n}\sin \frac{5 \pi }{6 n}}\nonumber\\
   &&  -\frac{\alpha^2  \langle \mathcal{O}\rangle^{-1} (F_1^{\mathcal{O}|1})^2 H_1(n)\left(2 \cos \frac{\pi }{n}+1\right)^3 }{4\sin \frac{\pi }{6 n}\sin\frac{5 \pi }{6 n}}.
\end{eqnarray}

\section{Conformal structure constants of twist fields}
\label{appendixd}
In this appendix we present detailed computations of the conformal structure constants $\tilde{C}_{\TT{\tilde{\TT}}}^{\mathcal{O}}$ and $\tilde{C}_{\Tp \bTp}^{\mathcal{O}}$ for different choices of the local field $\mathcal{O}$. These structure constants are used in section \ref{PCFT} in zeroth order perturbed CFT computations. The general strategy relies upon the fact that correlation functions of twist fields in CFT may be computed in two different ways: on the one hand we may treat the twist fields as standard local fields in the $n$-copy model on the manifold $\mathcal{M}_{n,x_1,x_2}$ (as defined in \ref{manifold}); on the other hand we may conformally map the correlation function to the complex plane by using the map (\ref{manifold}) thus expressing it in terms of correlation functions of other local fields (e.g.~with no twist field insertions). Below we present many examples of this approach.

\subsection{Structure constants involving the fields $\TT$ and $\tilde{\TT}$}
\subsubsection{The structure constant $\tilde{C}_{\TT \tilde{\TT}}^{\Phi_{1}} $}
The CFT structure constant $\tilde{C}_{\TT \tilde\TT}^{\Phi_1}$ may be computed as follows. We may select out the term proportional to $\Phi_1=\sum_{j=1}^n \phi_j$ in the OPE of $\TT (x_1) \tilde{\TT} (x_2)$ by evaluating the three point function below, where a single field $\phi_1$ is inserted:
\beqa
\frac{\bra \TT(x_1) \tilde{\TT}(x_2) \phi_1(x_3)\ket}{\bra \TT(x_1) \tilde{\TT}(x_2)\ket} &\stackrel{x_1\to x_2}\sim& \tilde{C}_{\TT \tilde{\TT}}^{\Phi_{1}} |x_1-x_2|^{2\Delta}\sum_{j=1}^n\bra \phi_j(x_2) \phi_1(x_3)  \ket\nonumber\\
&=&\tilde{C}_{\TT \tilde{\TT}}^{\Phi_{1}} |x_1-x_2|^{2\Delta}\bra \phi(x_2) \phi(x_3)  \ket\nonumber\\
&=&\tilde{C}_{\TT \tilde{\TT}}^{\Phi_{1}} |x_1-x_2|^{2\Delta}|x_2-x_3|^{-4\Delta}
\eeqa
On the other hand, we identify the ratio of correlators on the left-hand side as a correlator of $\phi_1(x_3)$ on the manifold $\mathcal{M}_{n,x_1,x_2}$, and use the conformal map $g$ to relate this to the one-point function $\bra \phi(g(x_3)\ket$ on $\mathbb{R}^2$. Since this one-point function is zero in the complex plane, we have that, in general
\beq
\tilde{C}_{\TT \tilde{\TT}}^{\Phi_{1}}=0.
\eeq
\subsubsection{The structure constant $\tilde{C}_{\TT \tilde{\TT}}^{\Phi_{1,k}} $}

The third term in the OPE $\TT(x_1)\tilde{\TT}(x_2)$ contains bilinears of the fields $\phi_j$, with the constraint that they be cyclically symmetric. The only possibility are the fields $\Phi_{1,k}$ defined earlier, with $k=2,\ldots, [n/2]+1$; the restriction on $k$ is to avoid over-counting, as $\Phi_{1,k} = \Phi_{1,n-k+2}$.
The coupling $\tilde{C}_{\Tp \bTp}^{\Phi_{1,k}}$ may be computed exactly as in the previous subsection.
We consider, for some $k\in\{2,\ldots,[n/2]+1\}$, the following ratio of correlators, which we evaluate using the OPEs in order to extract the structure constant:
\begin{eqnarray}
\frac{\bra \TT(x_1) \tilde{\TT}(x_2) \phi_1(x_3)\phi_k(x_4)\ket}{\bra \TT(x_1) \tilde{\TT}(x_2)\ket} &\stackrel{x_1\to x_2}\sim&{|x_1-x_2|^{4\Delta} \sum_{j=2}^{\left[\frac{n}{2}\right]+1}\tilde{C}_{\TT \tilde{\TT}}^{\Phi_{1,j}}\bra\Phi_{1,j}(x_2)\phi_1(x_3)\phi_k(x_4)\ket}\nonumber\\
&=&\tilde{C}_{\TT \tilde{\TT}}^{\Phi_{1,k}} |x_1-x_2|^{4\Delta} \bra\Phi_{1,k}(x_2)\phi_1(x_3)\phi_k(x_4)\ket \n
&=&
\tilde{C}_{\TT \tilde{\TT}}^{\Phi_{1,k}} |x_1-x_2|^{4\Delta}(|x_2-x_3||x_2-x_4|)^{-4\Delta}.
\eeqa
In the last step, we have used the fact that, by definition, every independent bilinear in $\Phi_{1,k}$ occurs with coefficient 1. We can then evaluate this explicitly by conformally mapping to the complex plane:
\beqa
&& \frac{\bra \TT(x_1) \tilde{\TT}(x_2) \phi_1(x_3)\phi_k(x_4)\ket}{\bra \TT(x_1) \tilde{\TT}(x_2)\ket}  = |\partial g(x_3)|^{2\Delta}|\partial g(x_4)|^{2\Delta} \bra \phi(e^{\frac{2\pi i}{n}}g(x_3))\phi(e^{\frac{2 \pi i k}{n}} g(x_4)) \ket \nonumber\\
&=& \frac{n^{-4\Delta}|x_2-x_1|^{4\Delta}\bra \phi(e^{\frac{2\pi i}{n}} g(x_3))\phi(e^{\frac{2 \pi i k}{n}} g(x_4)) \ket}{|x_3-x_1|^{2\Delta(1-\frac{1}{n})}|x_3-x_2|^{2\Delta(1+\frac{1}{n})}|x_4-x_1|^{2\Delta(1-\frac{1}{n})}|x_4-x_2|^{2\Delta(1+\frac{1}{n})}}\nonumber\\
&=& \frac{n^{-4\Delta}|x_2-x_1|^{4\Delta}\left|e^{\frac{2\pi i}{n}}\lt(\frac{x_3-x_1}{x_3-x_2}\rt)^{\frac{1}{n}}-e^{\frac{2\pi i k}{n}}\lt(\frac{x_4-x_1}{x_4-x_2}\rt)^{\frac{1}{n}}\right|^{-4\Delta}}{|x_3-x_1|^{2\Delta(1-\frac{1}{n})}|x_3-x_2|^{2\Delta(1+\frac{1}{n})}|x_4-x_1|^{2\Delta(1-\frac{1}{n})}|x_4-x_2|^{2\Delta(1+\frac{1}{n})}}\nonumber\\
&\stackrel{x_1\to x_2}\sim & \frac{n^{-4\Delta}|x_2-x_1|^{4\Delta}|e^{\frac{2\pi i}{n}}-e^{\frac{2\pi i k}{n}}|^{-4\Delta}}{|x_3-x_2|^{4\Delta}|x_4-x_2|^{4\Delta}}
\eeqa
(where the power functions are on their principal branch), thus, comparing both formulae we find
\beq
\tilde{C}_{\TT \tilde{\TT}}^{\Phi_{1,k}}= n^{-4\Delta}|1-e^{\frac{2\pi i (k-1)}{n}}|^{-4\Delta}.
\eeq
\subsubsection{The structure constant $\tilde{C}_{\TT \tilde{\TT}}^{\Phi_{1,k, j}}$}
We consider now the next correction to the OPE of $\TT$ and $\tilde{\TT}$, involving the fields $\Phi_{1,k,j}$ with $k>j>1$. Again, the ranges of $k$ and $j$ must be further restricted in the OPE in order not to overcount the fields. We do not need to discuss this in general; we just note that in both cases $n=3$ and $n=4$ there is a single field to count, $\Phi_{1,2,3} = \phi_1\phi_2\phi_3$ (for $n=3$) and $\Phi_{1,2,3}=\phi_1\phi_2\phi_3 + \phi_2\phi_3\phi_4 + \phi_3\phi_4\phi_1 + \phi_4\phi_1\phi_2$ (for $n=4$). As usual we first consider the consequence of the OPE,
\beqa
\lefteqn{\frac{\bra \TT(x_1) \tilde{\TT}(x_2) \phi_1(x_3)\phi_k(x_4) \phi_j(x_5)\ket}{\bra \TT(x_1) \tilde{\TT}(x_2)\ket}} && \nonumber\\
&\stackrel{x_1\to x_2}\sim& \tilde{C}_{\TT \tilde{\TT}}^{\Phi_{1,k,j}} |x_1-x_2|^{6\Delta} \bra \Phi_{1,k,j}(x_2)\phi_1(x_3) \phi_k(x_4) \phi_j(x_5) \ket\n
&=& \tilde{C}_{\TT \tilde{\TT}}^{\Phi_{1,k,j}} |x_1-x_2|^{6\Delta}|x_2-x_3|^{-4\Delta} |x_2-x_4|^{-4 \Delta} |x_2-x_5|^{-4\Delta}.
\eeqa
We then perform the calculation of the correlation function by mapping to the sphere,
\beqa
&& \frac{\bra \TT(x_1) \tilde{\TT}(x_2) \phi_1(x_3)\phi_k(x_4) \phi_j(x_5)\ket}{\bra \TT(x_1) \tilde{\TT}(x_2)\ket} \nonumber\\
&=& |\partial g(x_3)|^{2\Delta}|\partial g(x_4)|^{2\Delta} |\partial g(x_5)|^{2\Delta} \bra \phi(e^{\frac{2\pi i}{n}}g(x_3))\phi(e^{\frac{2\pi i k}{n}}g(x_4)) \phi(e^{\frac{2\pi i j}{n}}g(x_5))\ket\nonumber\\
&=& \frac{n^{-6\Delta} |x_2-x_1|^{6\Delta}\bra \phi(e^{\frac{2\pi i}{n}}g(x_3))\phi(e^{\frac{2\pi i k}{n}}g(x_4)) \phi(e^{\frac{2\pi i j}{n}}g(x_5))\ket}{(|x_3-x_1||x_4-x_1||x_5-x_1|)^{2\Delta(1-\frac{1}{n})}(|x_3-x_2||x_4-x_2||x_5-x_2|)^{2\Delta(1+\frac{1}{n})}} \nonumber\\
& = &
\frac{\tilde{C}_{\phi \phi}^{\phi}|x_2-x_1|^{6\Delta} n^{-6\Delta}}{(|x_3-x_1||x_4-x_1||x_5-x_1|)^{2\Delta(1-\frac{1}{n})}(|x_3-x_2||x_4-x_2||x_5-x_2|)^{2\Delta(1+\frac{1}{n})}}\nonumber\\
&\times& \left(|g(x_3)-e^{\frac{2\pi i (k-1)}{n}}g(x_4)| |g(x_3)-e^{\frac{2\pi i (j-1)}{n}}g(x_5)||g(x_4)-e^{\frac{2\pi i (j-k)}{n}}g(x_5)|\right)^{-2\Delta}\nonumber\\
&\stackrel{x_1\to x_2}\sim &
\frac{n^{-6\Delta} \tilde{C}_{\phi \phi}^{\phi}|x_2-x_1|^{6\Delta} |(1-e^{\frac{2\pi i (k-1)}{n}})(1-e^{\frac{2\pi i (j-1)}{n}})(1-e^{\frac{2\pi i (j-k)}{n}})|^{-2\Delta}}{|x_3-x_2|^{4\Delta}|x_4-x_2|^{4\Delta}|x_5-x_2|^{4\Delta}},
\eeqa
thus
\beq
\tilde{C}_{\TT \tilde{\TT}}^{\Phi_{1,k,j}}=n^{-6\Delta} \tilde{C}_{\phi \phi}^{\phi}|(1-e^{\frac{2\pi i (k-1)}{n}})(1-e^{\frac{2\pi i (j-1)}{n}})(1-e^{\frac{2\pi i (j-k)}{n}})|^{-2\Delta}.
\eeq
\subsubsection{The structure constant $\tilde{C}_{\TT \tilde{\TT}}^{\Phi_{1,k,j,p}}$}
This may be computed as before with the final results involving now a four point function of fields $\phi$: 
\beq
\tilde{C}_{\TT \tilde{\TT}}^{\Phi_{1,k,j,p}}=n^{-8\Delta}\bra\phi(e^{\frac{2\pi i}{n}}) \phi(e^{\frac{2\pi i k}{n}})\phi(e^{\frac{2\pi i j}{n}})\phi(e^{\frac{2\pi i p}{n}}) \ket.
\eeq
\subsection{Structure constants involving the fields $\Tp$ and $\bTp$}
Computations for the fields $\Tp$ and $\bTp$ are very similar to those performed in the previous subsection, once the representation (\ref{newfield}) is used. Below we provide some examples.
\subsubsection{The structure constant $\tilde{C}_{\Tp \bTp}^{\Phi_1}$}
As before, we compute first
\beqa
	\frc{\bra \Tp (x_1) \bTp (x_2)\phi_1(x_3)\ket}{\bra \TT(x_1)\b\TT(x_2)\ket}
	&\stackrel{x_1\to x_2}\sim& \tilde{C}_{\Tp \bTp}^{\Phi_1} |x_1-x_2|^{2\Delta-4\Delta_{\Tp}} \frc{
	\bra \phi_1(x_2)\phi_1(x_3)\ket}{\bra\TT(x_1) \TT(x_2)\ket} \n
	&=& \tilde{C}_{\Tp \bTp}^{\Phi_1} |x_1-x_2|^{2\Delta(1-\frac{2}{n})}
	|x_2-x_3|^{-4\Delta},
\eeqa
and we may compute the same three point function by using the conformal map (\ref{manifold}) together with the definition (\ref{newfield})
\beqa
	\lefteqn{\frc{\bra \Tp (x_1) \bTp (x_2)\phi_1(x_3)\ket}{\bra \TT(x_1)\b\TT(x_2)\ket}}
	&&\n
	&=& n^{4\Delta-2}\lim_{y_i\to x_i}
	|x_1-y_1|^{2\Delta(1-\frac{1}{n})} |x_2-y_2|^{2\Delta(1-\frac{1}{n})}
	\sum_{j_1,j_2=1}^n \frc{\bra \TT(x_1)\b\TT(x_2)
	\phi_{j_1}(y_1)\phi_{j_2}(y_2)\phi_1(x_3)\ket}{
	\bra \TT(x_1)\b\TT(x_2)\ket} \n
	&=& n^{4\Delta-2}\lim_{y_i\to x_i}
	|x_1-y_1|^{2\Delta(1-\frac{1}{n})} 
    |x_2-y_2|^{2\Delta(1-\frac{1}{n})}
	|\p g(y_1)|^{2\Delta} |\p g(y_2)|^{2\Delta} |\p g(x_3)|^{2\Delta}\n 
	&\times& \sum_{j_1,j_2=1}^n
	\bra \phi(e^{\frac{2\pi i j_1}{n}}g(y_1)) \phi(e^{\frac{2\pi i j_2}{n}}g(y_2))
	\phi(e^{\frac{2\pi i}{n}}g(x_3)) \ket \n
	&=& \tilde{C}_{\phi\phi}^{\phi}n^{4\Delta-2} n^{-4\Delta} \lim_{y_i\to x_i}
	|x_2-y_2|^{4\Delta(1-\frac{1}{n})} |\p g(x_3)|^{2\Delta}\n 
	&& \sum_{j_1,j_2=1}^n
	\lt(|e^{\frac{2\pi i j_1}{n}}g(y_1)-e^{\frac{2\pi i j_2}{n}}g(y_2)|\,|
	e^{\frac{2\pi i j_1}{n}}g(y_1) - 
    e^{\frac{2\pi i}{n}}g(x_3)|\,|e^{\frac{2\pi i j_2}{n}}g(y_2) - 
    e^{\frac{2\pi i}{n}}g(x_3)|\rt)^{-2\Delta} \n
	&=& \tilde{C}_{\phi\phi}^\phi \lim_{y_i\to x_i}
	 |x_2-y_2|^{4\Delta(1-\frac{1}{n})} |\p g(x_3)|^{2\Delta}
	\lt(|g(y_2)|^2\,|g(x_3)|\rt)^{-\frac{4\Delta}{n}} \n
	&=& \tilde{C}_{\phi\phi}^{\phi} |x_1-x_2|^{-\frac{4\Delta}{n}}\lim_{y_i\to x_i}
	 |\p g(x_3)|^{2\Delta}
	|g(x_3)|^{-2\Delta} \n
	&=& \tilde{C}_{\phi\phi}^{\phi} n^{-2\Delta} |x_1-x_2|^{2\Delta(1-\frac{2}{n})} |x_3-x_1|^{-2\Delta} |x_3-x_2|^{-2\Delta}.
\eeqa
Hence we conclude
\beq
	\tilde{C}_{\Tp\bTp}^\phi = n^{-2\Delta} \tilde{C}_{\phi\phi}^{\phi}.
\eeq
\subsubsection{The structure constant $\tilde{C}_{\Tp \bTp}^{\Phi_{1,k}}$}
Again we use
\beqa
\lefteqn{\frc{\bra \Tp (x_1) \bTp (x_2)\phi_1(x_3)\phi_k(x_4)\ket}{\bra \TT(x_1)\b\TT(x_2)\ket}} && \nonumber\\
	&\stackrel{x_1\to x_2}\sim& C_{\Tp \bTp}^{\Phi_{1,k}} |x_1-x_2|^{4\Delta(1-\frac{1}{n})}
	\bra \phi_1(x_2)\phi_1(x_3)\ket
	\bra \phi_k(x_2)\phi_k(x_4)\ket \n
	&=&
	C_{\Tp \bTp}^{\Phi_{1,k}} |x_1-x_2|^{4\Delta(1-\frac{1}{n})}
	\lt(|x_2-x_3|\,|x_2-x_4|\rt)^{-4\Delta}.
\eeqa
We then calculate:
\beqa
	\lefteqn{\frc{\bra \Tp (x_1) \bTp (x_2)\phi_1(x_3)\phi_k(x_4)\ket}{\bra \TT(x_1)\b\TT(x_2)\ket}} &&\n
	&=& n^{4\Delta-2}\lim_{y_i\to x_i}
	|x_1-y_1|^{2\Delta(1-\frac{1}{n})} |x_2-y_2|^{2\Delta(1-\frac{1}{n})} 
	|\p g(y_1)|^{2\Delta} |\p g(y_2)|^{2\Delta} |\p g(x_3)|^{2\Delta}
    |\p g(x_4)|^{2\Delta}\n 
	&\times& \sum_{j_1,j_2=1}^n
	\bra \phi(e^{\frac{2\pi i j_1}{n}}g(y_1)) 
    \phi(e^{\frac{2\pi i j_2}{n}}g(y_2))
	\phi(e^{\frac{2\pi i}{n}}g(x_3))\phi(e^{\frac{2\pi i k}{n}}g(x_4)) \ket \n
	&=& n^{4\Delta-2} n^{1-4\Delta} \lim_{y_i\to x_i}
	|x_2-y_2|^{-\frac{4\Delta}{n}} |\p g(x_3)|^{2\Delta}
	|\p g(x_4)|^{2\Delta}\n 
	&\times& \sum_{j=1}^n
	\bra \phi(0) \phi(g(y_2))
	\phi(e^{-\frac{2\pi i (j-1)}{n}}g(x_3))\phi(e^{\frac{2\pi i (k-j)}{n}}g(x_4)) \ket \n
	&=& |x_1-x_2|^{-\frac{4\Delta}{n}}\lim_{y_i\to x_i}
	|\p g(x_3)|^{2\Delta}
	|\p g(x_4)|^{2\Delta} |g(x_4)|^{-4\Delta}
	\kappa\lt(1-e^{\frac{2\pi i (k-1)}{n}}\frc{g(x_4)}{g(x_3)}\rt)\n
	&=& n^{-4\Delta} |x_1-x_2|^{4\Delta(1-\frac{1}{n})}
	(|x_3-x_1||x_4-x_2|)^{-2\Delta(1-\frac{1}{n})}(|x_3-x_2||x_4-x_1|)^{-2\Delta(1+\frac{1}{n})}\n &\times&
	\kappa\lt(1-e^{\frac{2\pi i (k-1)}{n}}
	\lt(\frc{(x_4-x_1)(x_3-x_2)}{(x_4-x_2)(x_3-x_1)}\rt)^{\frac{1}{n}}\rt)\n
	&\stackrel{x_1\to x_2}\sim &
	n^{-4\Delta} |x_1-x_2|^{4\Delta(1-\frac{1}{n})}
	|x_3-x_2|^{-4\Delta}
	|x_4-x_2|^{-4\Delta}
	\kappa\lt(1-e^{\frac{2\pi i (k-1)}{n}}\rt),
\eeqa
whence we conclude
\beq
	\tilde{C}_{\Tp\bTp}^{\Phi_{1,k}} = n^{-4\Delta} \kappa\lt(1-e^{\frac{2\pi i (k-1)}{n}}\rt),
\eeq
where $\kappa$ is a model-dependent function which characterizes the four-point function
\beq
	\bra\phi(x_1)\phi(x_2)\phi(x_3)\phi(x_4)\ket
	= \kappa(\eta) |x_1-x_4|^{-4\Delta} |x_2-x_3|^{-4\Delta},
	\quad \eta = \frc{x_{12}x_{34}}{x_{13}x_{24}}.
\eeq
\section{Computation of $A_1'(r,1)$}
\label{appendixb}
We have seen that
\beq
A_1(r,n)=-\frac{n}{\pi}\left|\frac{F_1^{:\mathcal{T}\phi:|1}}{\langle :\mathcal{T}\phi: \rangle} \right|^2 K_0(mr).
\eeq
The $n$-dependence of this expression is contained on the one-particle form factor, so we need to compute
\beq
\lim_{n \rightarrow 1}\frac{d}{d n} \left(n \left|\frac{F_1^{:\mathcal{T}\phi:|1}}{\langle :\mathcal{T}\phi: \rangle} \right|^2\right)=-\left(\frac{F_1^{\phi}}{\langle \phi \rangle} \right)^2-2\frac{F_1^{\phi}}{\langle \phi \rangle} \lim_{n \rightarrow 1}\frac{d}{d n} \left(\frac{F_1^{:\mathcal{T}\phi:|1}}{\langle :\mathcal{T}\phi: \rangle}\right),
\eeq
where we have used the fact that $\frac{F_1^{:\mathcal{T}\phi:|1}}{\langle :\mathcal{T}\phi: \rangle}$ has zero real part. We now compute the derivative above employing the formula (\ref{identification})
\beqa
\lim_{n \rightarrow 1}\frac{d}{d n} \left(\frac{F_1^{:\mathcal{T}\phi:|1}}{\langle :\mathcal{T}\phi: \rangle}\right)
&=&\frac{i\sqrt{2}}{3^{\frac{1}{4}}f(\frac{2\pi i}{3},1)}\left(-1+{\pi}{3^{-\frac{3}{2}}}\right)-\frac{i\sqrt{2}}{3^{\frac{1}{4}}f(\frac{2\pi i}{3},1)^2} \lim_{n\rightarrow 1} \frac{\partial {f(\frac{2\pi i}{3},n)}}{\partial n}.
\eeqa
The derivative above can be computed from the integral representation (\ref{eq}) to
\beq
 \lim_{n\rightarrow 1} \frac{d {f(\frac{2\pi i}{3},n)}}{d n}=-2 f(\frac{2\pi i}{3},1)\int_{0}^{\infty} \frac{\sinh\frac{t}{3}\sinh\frac{t}{6}\cosh\frac{2t}{3}}{\sinh^2 t \cosh\frac{t}{2}} dt =\left(\frac{11 \pi}{72 \sqrt{3}}-\frac{1}{2}\right)f(\frac{2\pi i}{3},1).
\eeq
Simplifying we obtain,
\beqa
A_1'(r,1)=B_1(r)+\frac{2}{f(\frac{2\pi i}{3},1)^2}\left(\frac{1}{\pi\sqrt{3}}-\frac{13}{108}\right)K_0(mr),
\eeqa
with $B_1(r)$ defined in (\ref{B1}).

\bibliographystyle{phreport}

\end{document}